\documentclass[aps, prx, reprint, superscriptaddress, longbibliography]{revtex4-1}
\usepackage[dvipdfmx]{graphicx}
\usepackage{amsmath}
\usepackage{amssymb}
\usepackage{dcolumn}
\usepackage{color}
\usepackage{bm}
\usepackage{braket}
\usepackage{txfonts}

\newcommand{\up}{\uparrow}
\newcommand{\down}{\downarrow}

\begin{document}
\title{Quasiparticles of Decoherence Processes in Open Quantum Many-Body Systems: Incoherentons}
\author{Taiki Haga}
\email[]{taiki.haga@omu.ac.jp}
\affiliation{Department of Physics and Electronics, Osaka Metropolitan University, Sakai-shi, Osaka 599-8531, Japan}
\author{Masaya Nakagawa}
\affiliation{Department of Physics, University of Tokyo, 7-3-1 Hongo, Bunkyo-ku, Tokyo 113-0033, Japan}
\author{Ryusuke Hamazaki}
\affiliation{Nonequilibrium Quantum Statistical Mechanics RIKEN Hakubi Research Team, RIKEN Cluster for Pioneering Research (CPR), RIKEN iTHEMS, Wako, Saitama 351-0198, Japan}
\author{Masahito Ueda}
\affiliation{Department of Physics, University of Tokyo, 7-3-1 Hongo, Bunkyo-ku, Tokyo 113-0033, Japan}
\affiliation{RIKEN Center for Emergent Matter Science (CEMS), Wako, Saitama 351-0198, Japan}
\affiliation{Institute for Physics of Intelligence, University of Tokyo, 7-3-1 Hongo, Bunkyo-ku, Tokyo 113-0033, Japan}
\date{\today}

\begin{abstract}
The relaxation dynamics of an open quantum system is determined by the competition between the coherent Hamiltonian dynamics of a system and the dissipative dynamics due to interactions with environments.
It is therefore of fundamental interest to understand the transition from the coherent to incoherent regimes.
We find that hitherto unrecognized quasiparticles -- incoherentons -- describe this coherent-to-incoherent transition in eigenmodes of a Liouvillian superoperator that governs the dynamics of an open quantum many-body system.
Here, an incoherenton is defined as an interchain bound state in an auxiliary ladder system that represents the density matrix of a system.
The Liouvillian eigenmodes are classified into groups with different decay rates that reflect the number of incoherentons involved therein.
We also introduce a spectral gap -- quantum coherence gap -- that separates the different groups of eigenmodes.
We demonstrate the existence of incoherentons in a lattice boson model subject to dephasing, and show that the quantum coherence gap closes when incoherentons are deconfined, which signals a dynamical transition from incoherent relaxation with exponential decay to coherent oscillatory relaxation.
Furthermore, we discuss how the decoherence dynamics of quantum many-body systems can be understood in terms of the generation, localization, and diffusion of incoherentons.
\end{abstract}

\maketitle

\section{Introduction}

Understanding the role of environments on quantum coherence presents a key challenge in quantum physics \cite{Breuer, Weiss, Rivas}.
The concomitant decoherence of quantum superposition of a system places a major obstacle in the development of quantum technologies \cite{Pellizzari-95, Balasubramanian-09, Lanyon-11, Paik-11}.
Moreover, there has been a surge of interest in nonequilibrium dynamics of open quantum many-body systems owing to experimental progress in atomic, molecular, and optical (AMO) systems, enabling one to control not only the Hamiltonian of a quantum system but also its coupling to an environment \cite{Kasprzak-06, Bloch-08-1, Bloch-08-2, Diehl-08, Syassen-08, Baumann-10, Barreiro-11, Schauss-12, Ritsch-13, Carusotto-13, Daley-14}.

The dynamics of an open quantum system can, in general, be described by a quantum master equation for its density matrix.
In particular, in a typical AMO system, the weak coupling and the separation of time scales between the system and an environment allow the dynamics of the density matrix to be described by a Markovian quantum master equation \cite{Rivas}.
The superoperator that generates the time evolution of the density matrix is referred to as the Liouvillian $\mathcal{L}$.
The relaxation dynamics of an open quantum system is fully characterized by the complex spectrum and eigenmodes of $\mathcal{L}$.
In general, the Liouvillian $\mathcal{L}$ consists of a coherent part describing the unitary time evolution governed by the Hamiltonian of the system and an incoherent part due to the coupling with the environment.
The competition between these contributions causes a transition from a coherent regime to an incoherent one.
Such a coherent-to-incoherent transition, a phenomenon found in many quantum systems \cite{Leggett-87, Mak-91, Egger-97, Chin-06, Matsuo-08, Wang-08, Kast-13, Magazzu-15, Leegwater-96, Nazir-09}, is detrimental to quantum technologies, including quantum computation.
However, it is a formidable task to understand how decoherence proceeds in open quantum many-body systems because of exponentially large Hilbert-space dimensions.
It is highly desirable to establish an effective description of the competition between coherent and incoherent dynamics in many-body systems.
In this regard, it should be recalled that the concepts of spectral gaps and quasiparticles play a pivotal role in quantum many-body physics.
In isolated systems, quantum phase transitions in the ground state are characterized by the closing of the spectral gap \cite{Sachdev}, and the low-energy behavior is governed by quasiparticle excitations, which allow an effective description of complex many-body systems \cite{Wen}.

\begin{figure}
	\centering
	\includegraphics[width=0.45\textwidth]{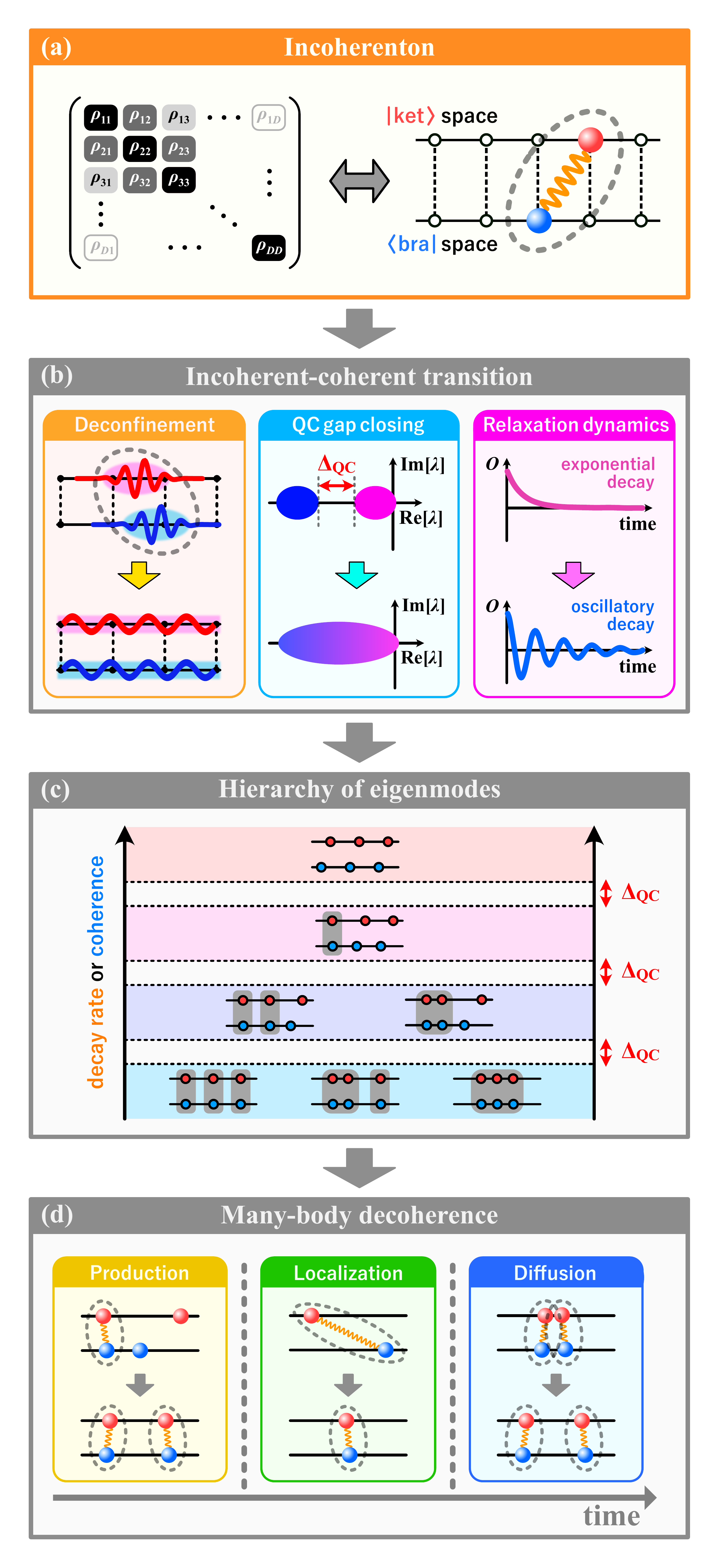}
	\caption{Quasiparticle description of relaxation processes in terms of incoherentons.
		(a) An incoherenton is a bound state between the degrees of freedom of ket and bra spaces of a density matrix.
		(b) The incoherent-coherent transition of eigenmodes can be characterized by the deconfinement of incoherentons (left panel) and the closing of the quantum coherence (QC) gap $\Delta_{\mathrm{QC}}$ (middle panel), where $\lambda$ is the eigenvalue of the Liouvillian.
		The QC gap closing causes a dynamical transition from incoherent exponential relaxation to coherent oscillatory relaxation (right panel).
		(c) The many-body eigenmodes are classified into groups according to the number of incoherentons involved.
		Each group of eigenvalues is separated from the others by the QC gaps.
		Since the system loses coherence over time, the more coherent the mode, the larger the decay rate.
		(d) The relaxation dynamics of open quantum many-body systems is effectively described by the production, localization, and diffusion of incoherentons.}
	\label{Fig-concept}
\end{figure}

In the present paper, we investigate spectral gaps and quasiparticles that characterize the physics of Markovian open quantum many-body systems described by a Liouvillian superoperator.
Note, however, that the incoherent-coherent transition often occurs far from the steady state, in which the conventional low-energy description in terms of quasiparticles is inapplicable.
We here discover quasiparticles, {\it ``incoherentons,"} that naturally describe the incoherent-coherent transition in Liouvillian eigenmodes of open quantum many-body systems.
As opposed to the conventional notion of quasiparticles, incoherentons are applicable to far-from-equilibrium regions.
Incoherentons are defined on a space of {\it operators} due to the matrix nature of the density matrix.
To show this, we use the fact that any density matrix of a system can be mapped to a vector in the tensor product space of bra and ket spaces.
Since this product space can be interpreted as the Hilbert space of a ladder system consisting of two chains of bra and ket spaces [see Fig.~\ref{Fig-concept}(a)], we call such a mapping the ladder representation of the density matrix.
In the ladder representation, the Liouvillian $\mathcal{L}$ is mapped to a non-Hermitian Hamiltonian $\tilde{\mathcal{L}}$ of the ladder system.
The coherent part of $\tilde{\mathcal{L}}$ governs the independent dynamics of particles in each chain, while the incoherent part acts as a non-Hermitian interaction between different chains.
Thus, depending on which contribution is dominant, the Liouvillian eigenmode either forms a scattering state extended over the entire ladder or an interchain bound state in which the degrees of freedom of the two chains are strongly correlated.
Since the existence of such a bound state implies the localization of matrix elements near diagonal components in the original matrix representation, we will refer to it as an incoherenton [see Fig.~\ref{Fig-concept}(a)].

\begin{table*}
	\begingroup
	\renewcommand{\arraystretch}{1.2}
	\begin{ruledtabular}
		\begin{tabular}{cccc}
			& Isolated quantum system & \multicolumn{2}{l}{\hspace{80pt} Open quantum system} \\ \hline
			Operator & Hamiltonian $H$ & \multicolumn{2}{l}{\hspace{100pt} Liouvillian $\mathcal{L}$} \\ \hline
			Transition & Quantum phase transition & Dissipative phase transition & Deconfinement of incoherentons
			\\ \hline
			State & Ground states (GS) & Steady states (SS) & Non-steady eigenmodes
			\\ \hline
			Spectral gap & 
			\begin{tabular}{c}
				Energy gap $\Delta_E$
			\end{tabular} & 
			\begin{tabular}{c}
				Liouvillian gap $\Delta_L$
			\end{tabular} & 
			\begin{tabular}{c}
				Quantum coherence gap $\Delta_{\mathrm{QC}}$
			\end{tabular}
			\\ \hline
			Length scale &
			\begin{tabular}{c}
				Correlation length $\xi_g$ of GS
			\end{tabular} & 
			\begin{tabular}{c}
				Correlation length $\xi_s$ of SS
			\end{tabular} & 
			\begin{tabular}{c}
				Confinement length $\xi_{\mathrm{con}}$ of incoherentons
			\end{tabular}
			\\ \hline
			Dynamics & \multicolumn{2}{c}{Divergence of characteristic timescales} & Incoherent-coherent dynamical transition \\
		\end{tabular}
	\end{ruledtabular}
	\endgroup
	\caption{Quantum phase transition of the ground state (GS), dissipative phase transition of the steady state (SS), and deconfinement of incoherentons, which are characterized by the energy gap $\Delta_E$, the Liouvillian gap $\Delta_L$, and the QC gap $\Delta_{\mathrm{QC}}$, respectively.
	The closing of the QC gap is accompanied by a divergence of the confinement length of incoherentons.
	The critical slowing down of relaxation dynamics is characteristic of both quantum and dissipative phase transitions.
	The deconfinement of incoherentons signals a dynamical transition from incoherent exponential relaxation to coherent oscillatory one.}
	\label{tab-gaps}
\end{table*}

The concept of incoherentons provides several insights into the dynamics of open quantum many-body systems and allows us to discover a universal mechanism for incoherent-coherent transitions.
They are summarized below as well as in Figs.~\ref{Fig-concept}(b)--(d), and will be discussed in the following sections.
\begin{description}
\item[Deconfinement of incoherentons (Sec.~\ref{sec:incoherentons})]~\\
The incoherent-coherent transition of Liouvillian eigenmodes can be understood by the deconfinement of incoherentons.
Since the dissipation corresponds to chain-to-chain interactions in the ladder representation, the confinement length of an incoherenton increases with decreasing dissipation, and eventually a transition from bound to scattering states occurs at some critical strength of dissipation [see the left panel of Fig.~\ref{Fig-concept}(b)].

\item [Quantum coherence gap closing (Sec.~\ref{sec:incoherentons})]~\\
When the dissipation is sufficiently strong, a gap exists between groups of Liouvillian eigenvalues with different numbers of incoherentons.
We call the gap between such groups {\it quantum coherence (QC) gap} because it separates groups of eigenmodes with different degrees of quantum coherence.
The QC gap $\Delta_{\mathrm{QC}}$ closes at the deconfinement transition (see the middle panel of Fig.~\ref{Fig-concept}(b), where $\lambda$ denotes the eigenvalues of $\mathcal{L}$).

\item [Incoherent-coherent dynamical transition (Sec.~\ref{sec:incoherent_coherent_dynamical_transition})]~\\
The QC gap closing signals the onset of a dynamical transition from overdamped relaxation, where an inhomogeneous initial state relaxes exponentially to a uniform steady state, to underdamped relaxation, where the local density exhibits oscillatory behavior (see the right panel of Fig.~\ref{Fig-concept}(b), where $O$ is an appropriate observable).
We argue that this provides a hitherto unknown type of incoherent-coherent transitions in an extended lattice system.

\item [Hierarchy of eigenmodes (Secs.~\ref{sec:hierarchy}, \ref{sec:exact_many_body_solution}, and \ref{sec:effect_of_particle_loss})]~\\
The many-body eigenmodes are classified into groups with different decay rates characterized by the number of incoherentons involved therein [see Fig.~\ref{Fig-concept}(c), where the small gray boxes represent incoherentons].
Each group of eigenvalues is separated from the others by the QC gaps.
The more incoherentons an eigenmode involves, the slower it decays.

\item [Many-body decoherence (Sec.~\ref{sec:many_body_decoherence})]~\\
The number of incoherentons in the density matrix increases as relaxation proceeds, which means that the relaxation of a many-body state is accompanied by the production of incoherentons.
Furthermore, the late decoherence process is characterized by the localization and diffusion of incoherentons [see Fig.~\ref{Fig-concept}(d)].
\end{description}

Here, we highlight the distinction between the incoherent-coherent transition described in this work and conventional phase transitions in isolated and open quantum many-body systems.
Table \ref{tab-gaps} summarizes different types of transitions, spectral gaps, and characteristic length scales.
In isolated quantum systems (see the left column of Table \ref{tab-gaps}), the energy gap $\Delta_E$ of a Hamiltonian is defined as the energy difference between the ground state and the first excited state.
The correlation length $\xi_g$ of the ground state and $\Delta_E$ are related to each other by $\xi_g \sim v/\Delta_E$, where $v$ is the propagation velocity of low-energy excitations with wavelengths comparable to $\xi_g$.
Here and henceforth, the Planck constant $\hbar$ is set to unity.
At a quantum phase transition of the ground state, $\xi_g$ diverges, accompanied by the closing of $\Delta_E$ and the divergence of characteristic time scales of low-energy excitations \cite{Sachdev}.

In open quantum systems, a phase transition of the steady state, known as the dissipative phase transition \cite{Tomadin-11, Lee-11, Ludwig-13, Carr-13, Marcuzzi-14, Weimer-15, Maghrebi-16, Sieberer-16, Biondi-17}, is characterized by the Liouvillian gap $\Delta_L$, which is defined as the smallest absolute value of the real parts of nonzero Liouvillian eigenvalues (see the middle column of Table \ref{tab-gaps}).
The relation between the correlation length $\xi_s$ of the steady state and $\Delta_L$ is given by $\xi_s \sim v/\Delta_L$, where $v$ is the propagation velocity of excitations near the steady state.
The dissipative phase transition is characterized by the divergence of $\xi_s$ and the closing of $\Delta_L$ \cite{Kessler-12, Honing-12, Lee-13, Horstmann-13, Rota-18, Minganti-18, Ferreira-19}.
The longest timescale for the system to reach the steady state is expected to be inversely proportional to $\Delta_L$ \cite{Cai-13, Bonnes-14, Znidaric-15} (see, however, Refs.~\cite{Mori-20, Haga-21} for exceptions).
Thus, the closing of $\Delta_L$ leads to the divergence of the relaxation time.

The deconfinement of incoherentons together with the QC gap closing constitutes the third type of transition in quantum many-body systems (see the right column of Table \ref{tab-gaps}).
The relation between the confinement length $\xi_{\mathrm{con}}$ of incoherentons and the QC gap $\Delta_{\mathrm{QC}}$ is given by
\begin{equation}
	\xi_{\mathrm{con}} \sim \frac{\Gamma}{\Delta_{\mathrm{QC}}},
	\label{xi_con_Delta_QC}
\end{equation}
where $\Gamma$ is the decay rate of relevant eigenmodes and $\xi_{\mathrm{con}}$ is measured in units of the lattice spacing.
An important distinction of the deconfinement of incoherentons from other well-known transitions is that it is a transition of non-steady eigenmodes having finite lifetimes.
Thus, the deconfinement of incoherentons significantly alters the transient dynamics of open quantum systems, where incoherent-coherent transitions are expected to take place [see the right panel of Fig.~\ref{Fig-concept}(b)].

This paper is organized as follows.
Section~\ref{sec:preliminaries} details the ladder representation of the Liouvillian and introduces a system of hard-core bosons subjected to on-site dephasing, serving as a representative model for open quantum many-body systems.
In Sec.~\ref{sec:incoherentons}, the concept of incoherenton is introduced for the one-particle case.
We describe the deconfinement of incoherentons and the QC gap closing in terms of the prototypical model.
Section~\ref{sec:incoherent_coherent_dynamical_transition} demonstrates that the relaxation dynamics of particle density display an incoherent-coherent transition corresponding to the parameter at which the QC gap closes.
In Sec.~\ref{sec:hierarchy}, the concept of incoherentons is generalized to many-body systems.
By numerically diagonalizing the Liouvillian of the dephasing hard-core bosons, we demonstrate the deconfinement of incoherentons and the closing of the QC gap for the many-body case.
In Sec.~\ref{sec:exact_many_body_solution}, we obtain an exact many-body solution of the dephasing hard-core boson model with the Bethe ansatz method, which analytically confirms the existence of incoherentons and their deconfinement transitions.
In Sec.~\ref{sec:effect_of_particle_loss}, we discuss how the incoherenton framework can be applied in the presence of particle exchange with the environment, and demonstrate that the phenomenology of incoherentons remains intact for small loss and gain rates of particles.
Section~\ref{sec:many_body_decoherence} introduces a simple description of many-body decoherence via incoherentons, identifying three distinct decoherence regimes related to the production, localization, and diffusion of incoherentons.
In Sec.~\ref{sec:conclusion}, we summarize our results and discuss prospects for future work.
In Appendix \ref{appendix:Liouvillian_eigenmodes}, general properties of the Liouvillian spectrum and eigenmodes are summarized.
In Appendix \ref{appendix:single_particle_solution}, we present a thorough analysis of the Liouvillian spectrum and eigenmodes for the one-particle case without resorting to the Bethe ansatz.
In Appendix \ref{appendix:absence_of_incoherenton}, we show that incoherentons do not exist in continuous systems.
This fact implies that the spatial discreteness of lattice systems is crucial for the existence of incoherentons.
In Appendix \ref{appendix:incoherenton_correlation}, we discuss measuring incoherenton correlation functions in ultracold atomic systems.
In Appendix \ref{appendix:Bose_Hubbard_model}, we explore the Liouvillian spectra of a dephasing Bose-Hubbard model through numerical diagonalization and shows the deconfinement of incoherentons within this model.
In Appendix \ref{appendix:hard_core_bosons_with_next_nearest_neighbor_hopping}, we present the results for dephasing hard-core bosons with next-nearest-neighbor hopping. 
In Appendices \ref{appendix:Bose_Hubbard_model} and \ref{appendix:hard_core_bosons_with_next_nearest_neighbor_hopping}, we provide evidence supporting the universality of the incoherenton framework.

\section{Ladder representation of the Liouvillian}
\label{sec:preliminaries}

\subsection{Liouvillian superoperator}

We focus on Markovian open quantum lattice systems with bulk dissipation, in which the dissipation acts uniformly on every site.
Within the Born-Markov approximation \cite{Rivas}, the time evolution of the density matrix $\rho$ is described by a quantum master equation, which is generated by a Liouvillian superoperator $\mathcal{L}$ \cite{Lindblad-76, Gorini-76}:
\begin{equation}
	\frac{d \rho}{dt} = \mathcal{L}(\rho) := -i[H, \rho] + \sum_{\nu} \left( L_{\nu} \rho L_{\nu}^{\dag} - \frac{1}{2} \{ L_{\nu}^{\dag}L_{\nu}, \rho \} \right),
	\label{master_eq_1}
\end{equation}
where $[A,B] := AB-BA$, $\{A,B\} := AB+BA$, and $L_{\nu}$ is a Lindblad operator.
The quantum master equation \eqref{master_eq_1} is justified when the time scale of dynamics induced by the system-environment coupling is much longer than the characteristic time scale of the environment.
This condition is well satisfied for typical AMO systems such as trapped two-level atoms with spontaneous emission and an optical cavity with photon loss \cite{Daley-14, Sieberer-16, Ritsch-13}.
The index $\nu$ for the Lindblad operator $L_{\nu}$ denotes the lattice sites and the types of dissipation.
We assume that each $L_{\nu}$ has support on a finite number of sites.

The master equation \eqref{master_eq_1} can be rewritten as
\begin{equation}
	\frac{d \rho}{dt} = -i(H_{\mathrm{eff}} \rho - \rho H_{\mathrm{eff}}^{\dag}) + \sum_{\nu} L_{\nu} \rho L_{\nu}^{\dag},
	\label{master_eq_2}
\end{equation}
where the non-Hermitian effective Hamiltonian $H_{\mathrm{eff}}$ reads
\begin{equation}
	H_{\mathrm{eff}} := H - \frac{i}{2} \sum_{\nu} L_{\nu}^{\dag}L_{\nu}.
\end{equation}
It is convenient to define
\begin{equation}
	\mathcal{L}_H(\rho) := -i(H_{\mathrm{eff}} \rho - \rho H_{\mathrm{eff}}^{\dag}),
	\label{L_H}
\end{equation}
and
\begin{equation}
	\mathcal{L}_{\mathrm{jump}}(\rho) := \sum_{\nu} L_{\nu} \rho L_{\nu}^{\dag}.
	\label{L_jump}
\end{equation}
In the quantum trajectory description \cite{Daley-14}, where the dynamics of an open quantum system is described by stochastic trajectories of pure states, $\mathcal{L}_H$ describes a deterministic time evolution generated by the effective Hamiltonian $H_\mathrm{eff}$, and $\mathcal{L}_{\mathrm{jump}}$ describes quantum jump processes.

If the Liouvillian is diagonalizable, its eigenmodes $\rho_{\alpha}$ can be defined by
\begin{equation}
	\mathcal{L}(\rho_{\alpha}) = \lambda_{\alpha} \rho_{\alpha} \quad (\alpha=0, 1, ... , D^2-1),
	\label{eigen_eq}
\end{equation}
where $\lambda_{\alpha}$ is the $\alpha$th eigenvalue and $D$ is the dimension of the Hilbert space $\mathcal{H}$ of the system.
A steady state $\rho_{\mathrm{ss}}$ corresponds to an eigenmode with zero eigenvalue.
We arrange the eigenvalues $\{ \lambda_{\alpha} \}_{\alpha = 0,...,D^2-1}$ such that $0 = \mathrm{Re}[\lambda_0] \leq |\mathrm{Re}[\lambda_1]| \leq \cdots \leq |\mathrm{Re}[\lambda_{D^2-1}]|$.
General properties of the Liouvillian spectrum and eigenmodes are summarized in Appendix \ref{appendix:Liouvillian_eigenmodes}.
In terms of the Liouvillian eigenmodes, the time evolution of the density matrix is given by
\begin{equation}
	\rho(t) = \rho_{\mathrm{ss}} + \sum_{\alpha=1}^{D^2-1} c_{\alpha} e^{\lambda_{\alpha} t} \rho_{\alpha},
	\label{rho_expansion_t}
\end{equation}
where $c_{\alpha}$ is the coefficient of eigenmode expansion of the initial density matrix.
We have assumed that the steady state $\rho_0 = \rho_{\mathrm{ss}}$ is unique.
Equation \eqref{rho_expansion_t} implies that the relaxation dynamics of an open quantum system is fully characterized by the spectrum and eigenmodes of the Liouvillian.

\begin{figure}
	\centering
	\includegraphics[width=0.45\textwidth]{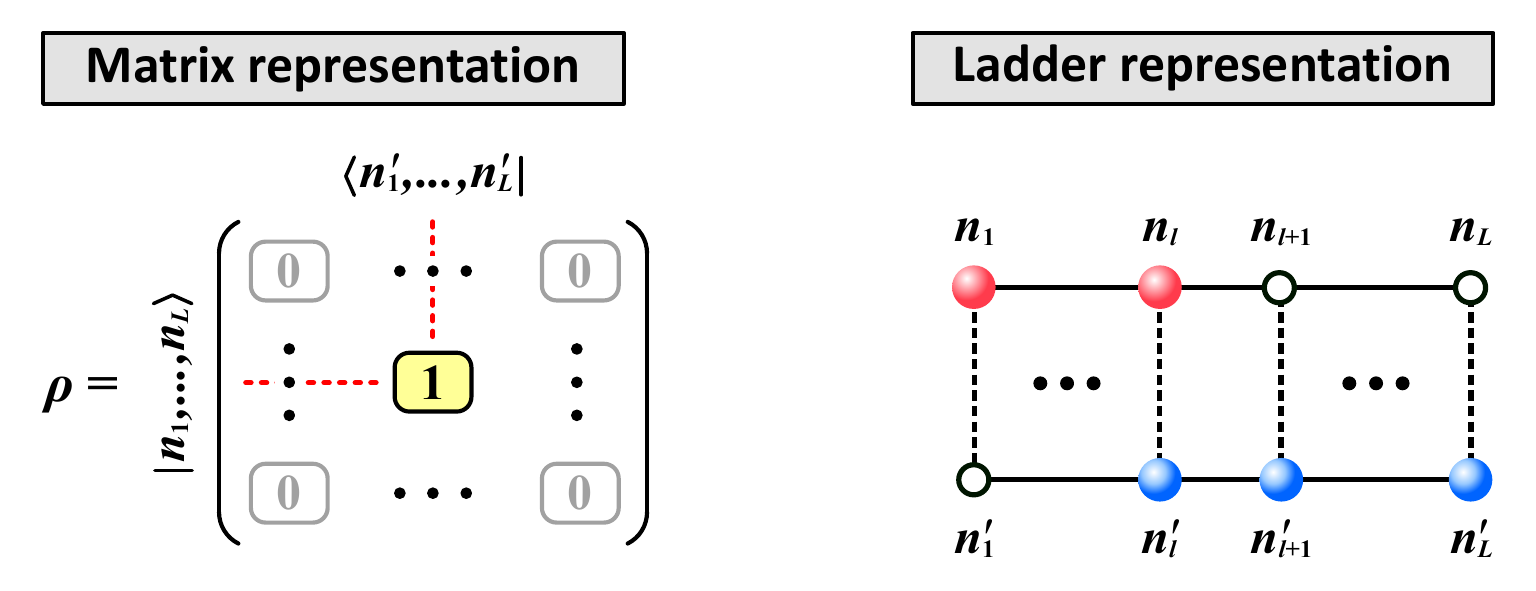}
	\caption{Ladder representation of the density matrix.
		A matrix $\rho = \sum_{i,j}\rho_{ij}\ket{i} \bra{j}$ is mapped to a vector $|\rho) = \sum_{i,j}\rho_{ij}\ket{i} \otimes \ket{j}$.
		In a one-dimensional tight-binding model with $L$ sites, a basis vector $\ket{i}$ of the Hilbert space can be written as $\ket{n_1,...,n_L}$, where $n_l=0,1,...$ is the occupation number of particles at site $l=1,...,L$.
		Similarly, $\bra{j}$ is represented as $\bra{n_1',...,n_L'}$.
		In the ladder representation, a basis vector $\ket{i}\otimes\ket{j}$ describes a state of a two-leg ladder.}
	\label{Fig-ladder}
\end{figure}

Let $\{ \ket{i} \}_{i=1,...,D}$ be an orthonormal basis set of $\mathcal{H}$ that specifies real-space configurations of particles or spins.
For example, one can consider the real-space Fock basis $\ket{n_1,...,n_L}$, where $n_l=0,1,...$ denotes the occupation number of particles at site $l$, and $L$ is the system size.
In terms of this orthonormal basis, the density matrix $\rho$ is written as
\begin{equation}
	\rho = \sum_{i,j=1}^D \rho_{ij} \ket{i} \bra{j},
	\label{rho_ket_bra}
\end{equation}
where $\rho_{ij}:=\braket{i|\rho|j}$.
Let us identify an operator $\ket{i} \bra{j}$ on $\mathcal{H}$ with a vector $\ket{i} \otimes \ket{j}$ in the tensor product space $\mathcal{H} \otimes \mathcal{H}$, the first (second) space of which will be referred to as the ket (bra) space \cite{Prosen-12, Znidaric-15, Shibata-19, Yoshioka-19, Hamazaki-22}.
Then, the density matrix \eqref{rho_ket_bra} is mapped onto the following vector:
\begin{equation}
	| \rho ) = \sum_{i,j=1}^D \rho_{ij} \ket{i} \otimes \ket{j},
	\label{rho_ladder}
\end{equation}
where we have used a round ket symbol $| ... )$ to emphasize that it belongs to $\mathcal{H} \otimes \mathcal{H}$ rather than $\mathcal{H}$.
It should be noted that, for one-dimensional cases, $\mathcal{H} \otimes \mathcal{H}$ can be considered as the Hilbert space of a ladder system composed of two chains (see Fig.~\ref{Fig-ladder}).
Thus, in the following, we refer to Eq.~\eqref{rho_ladder} as the ladder representation of the density matrix.

In the ladder representation, the Liouvillian $\mathcal{L}$ is mapped to a non-Hermitian Hamiltonian $\tilde{\mathcal{L}}$ of the ladder system.
The ladder representations of $\mathcal{L}_H$ and $\mathcal{L}_{\mathrm{jump}}$ are given by
\begin{equation}
	\tilde{\mathcal{L}}_H = - i H_\mathrm{eff} \otimes I + i I \otimes H_\mathrm{eff}^*, 
	\label{tilde_L_H}
\end{equation}
and
\begin{equation}
	\tilde{\mathcal{L}}_{\mathrm{jump}} = \sum_{\nu} L_{\nu} \otimes L_{\nu}^*, 
	\label{tilde_L_jump}
\end{equation}
where $I$ is the identity operator, and $H_\mathrm{eff}^*$ and $L_{\nu}^*$ are defined as $\braket{i|H_\mathrm{eff}^*|j} := \braket{i|H_\mathrm{eff}|j}^*$ and $\braket{i|L_{\nu}^*|j} := \braket{i|L_{\nu}|j}^*$.
The eigenmodes of $\tilde{\mathcal{L}}$ are given by
\begin{equation}
	\tilde{\mathcal{L}}|\rho_{\alpha}) = \lambda_{\alpha} |\rho_{\alpha}) \quad (\alpha=0, 1, ... , D^2-1),
	\label{eigen_eq_ladder}
\end{equation}
where $\rho_{\alpha}$ in Eq.~\eqref{eigen_eq} and $|\rho_{\alpha})$ in Eq.~\eqref{eigen_eq_ladder} are related to each other by Eqs.~\eqref{rho_ket_bra} and \eqref{rho_ladder}.

We comment on the diagonalizability of the Liouvillian.
Contrary to Hermitian operators, a non-Hermitian operator is not diagonalizable at exceptional points (EPs) \cite{Bender-98, Guo-09, Ruter-10, Heiss-12, El-Ganainy-18, Minganti-19, Ozdemir-19, Miri-19, Ashida-20, Bergholtz-21}.
While the set of EPs has zero measure in the parameter space (see, e.g., Sec.~2.6.1 in Ref.~\cite{Ashida-20}), the system can encounter an exceptional point when a certain parameter is continuously adjusted while keeping others fixed.
We note, however, that the diagonalizability of a Liouvillian is unimportant for our argument in this work.
An EP only indicates that the Liouvillian contains a Jordan block with size larger than one. 
In the most typical case of the lowest-order EP, two eigenvectors coalesce, and thus the Liouvillian involves a two-by-two Jordan block.
Nevertheless, the remaining eigenvectors, corresponding to one-by-one Jordan blocks, are unaffected, and it is worth studying their structure.
Consequently, even if the Liouvillian is not diagonalizable, our argument based on Liouvillian eigenmodes is applicable due to the predominance of one-by-one Jordan blocks in all eigenmodes.

\subsection{Example: hard-core bosons under dephasing}
\label{sec:hard_core_bosons}

\begin{figure}
	\centering
	\includegraphics[width=0.45\textwidth]{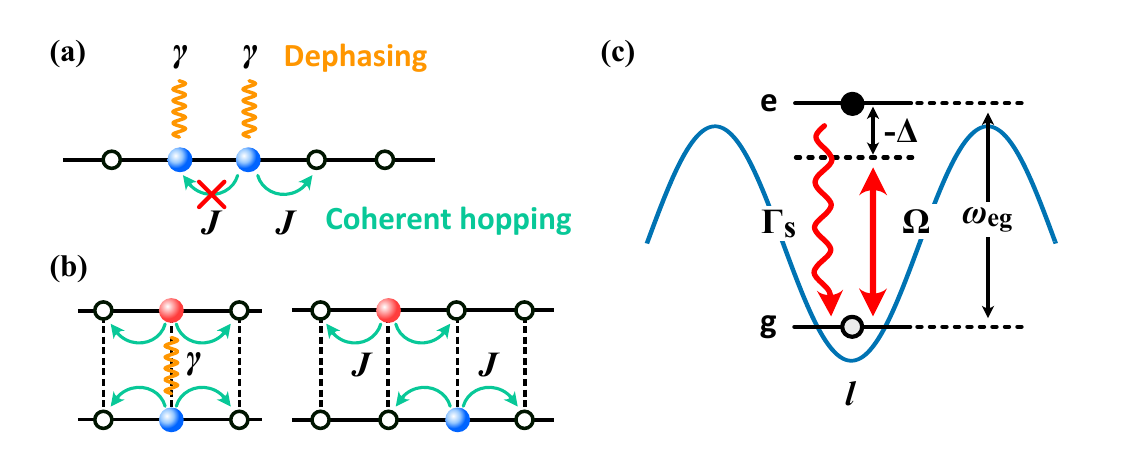}
	\caption{(a) Schematic illustration of a system of hard-core bosons with on-site dephasing.
		(b) Ladder representation of the model.
		While coherent hopping acts on individual particles, the on-site dephasing acts on a particle pair occupying the same rung (vertical dashed line).
		(c) Physical implementation of on-site dephasing for an atom in an optical lattice.
		The double arrow shows the Rabi coupling $\Omega$ induced by a laser with frequency $\omega_\mathrm{L}$.
		The wavy arrow shows spontaneous decay with rate $\Gamma_\mathrm{s}$.
		$\Delta=\omega_\mathrm{L}-\omega_\mathrm{eg}$ is the detuning of a laser, where $\omega_\mathrm{eg}$ is the excitation energy of the atom.}
	\label{Fig-model}
\end{figure}

We introduce a prototypical model of open quantum many-body systems, which will be analyzed in the following sections to demonstrate the concept of incoherentons.
The system is defined on a one-dimensional lattice with size $L$ under the periodic boundary condition.
The Hamiltonian of the system is given by
\begin{equation}
	H = -J \sum_{l=1}^{L} (b_{l}^{\dag} b_{l+1} + b_{l+1}^{\dag} b_{l}),
	\label{H_hard_core_boson}
\end{equation}
where $b^{\dag}_{l}$ and $b_{l}$ are the creation and annihilation operators of a boson at site $l$, and $J$ represents the tunneling amplitude.
We assume the hard-core condition $(b_l^{\dag})^2=0$, which prohibits more than two particles from occupying a single site.
The Lindblad operators for on-site dephasing are given by
\begin{equation}
	L_l = \sqrt{\gamma} b_{l}^{\dag} b_{l} \quad (l = 1, ..., L),
	\label{L_dephasing}
\end{equation}
where $\gamma$ denotes the strength of dephasing.
Note that the total particle number $N=\sum_{l=1}^L b_l^{\dag} b_l$ is conserved, i.e., $\mathrm{Tr}[N \mathcal{L}(\rho)]=0$ for any density matrix $\rho$.
The steady state of the corresponding master equation is the infinite-temperature state $\rho_{\mathrm{ss}}=D^{-1}I$, where $I$ is the identity operator, which is a consequence of the Hermiticity of the Lindblad operator $L_l$.
Figure \ref{Fig-model}(a) shows a schematic illustration of the model.

Let $n_l=0,1$ be the occupation number of the hard-core bosons at site $l$.
We define $2^L$ orthonormal basis vectors
\begin{equation}
	\ket{\{ n_l \}} = \prod_{l=1}^L (b_l^{\dag})^{n_l} \ket{\mathrm{v}},
\end{equation}
where $\ket{\mathrm{v}}$ is the vacuum state of the system.
In the ladder representation, an operator $\ket{\{ n_l \}} \bra{\{ m_l \}}$ on the Hilbert space of the system is interpreted as a state $\ket{\{ n_l \}} \otimes \ket{\{ m_l \}}$ of the ladder.
The Liouvillian is then rewritten as
\begin{align}
	\tilde{\mathcal{L}} &= i J \sum_{l=1}^{L} (b_{l,+}^{\dag} b_{l+1,+} + b_{l+1,+}^{\dag} b_{l,+} - b_{l,-}^{\dag} b_{l+1,-} - b_{l+1,-}^{\dag} b_{l,-}) \nonumber \\
	&\quad+ \gamma \sum_{l=1}^{L} b_{l,+}^{\dag} b_{l,+} b_{l,-}^{\dag} b_{l,-} - \gamma N,
	\label{Liouvillian_dissipative_boson_ladder}
\end{align}
where $b_{l,+(-)}$ represents the annihilation operator on the first (second) chain of the ladder. 
The on-site dephasing can be considered as an interchain interaction acting on a particle pair occupying the same rung.
Figure \ref{Fig-model}(b) shows a schematic illustration of the ladder representation.

This model can be realized with ultracold atomic gases in an optical lattice.
The on-site dephasing can be induced by the combined effect of coherent laser fields coupled to two internal atomic levels and spontaneous emission \cite{Pichler-10, Sarkar-14, Luschen-17, Bouganne-20}.
Suppose that a ground-state atom is excited by a laser field with frequency $\omega_\mathrm{L}$ and subsequently returns to its ground state through spontaneous emission with rate $\Gamma_\mathrm{s}$.
Figure \ref{Fig-model}(c) shows a level diagram of the atom excited by the laser.
The transition rate between the ground and excited states is characterized by the Rabi coupling $\Omega$, which is proportional to the intensity of the laser.
The detuning of the laser is given by $\Delta=\omega_\mathrm{L}-\omega_\mathrm{eg}$, where $\omega_\mathrm{eg}$ is the excitation energy of the atom.
When $|\Delta| \gg \Omega, \Gamma_\mathrm{s}$, the excited state can be adiabatically eliminated and one obtains the Lindblad master equation with an on-site dephasing $\gamma=\Gamma_\mathrm{s}\Omega^2/\Delta^2$ \cite{Pichler-10, Sarkar-14, Sieberer-16}.
The dephasing-type Lindblad operator given by Eq.~\eqref{L_dephasing} also appears in a master equation of ultracold atoms in an optical lattice driven by a stochastically fluctuating on-site potential \cite{Pichler-13}.

\section{Incoherent-coherent transition as deconfinement of incoherentons}
\label{sec:incoherentons}

\subsection{Incoherenton: an interchain bound state}

The non-Hermitian Hamiltonian $\tilde{\mathcal{L}}_H$ defined by Eq.~\eqref{tilde_L_H} independently acts on each chain of the ladder and does not create correlations between the bra space and the ket space.
If the Hamiltonian only contains kinetic energy terms which cause hopping of particles along each chain, $\tilde{\mathcal{L}}_H$ prefers plane-wave eigenmodes extended over each chain of the ladder.
On the other hand, $\tilde{\mathcal{L}}_{\mathrm{jump}}$ defined by Eq.~\eqref{tilde_L_jump} plays a role of an interchain interaction, e.g., see Eq.~\eqref{Liouvillian_dissipative_boson_ladder}.
Since each Lindblad operator $L_{\nu}$ has its support on a finite number of sites, $\tilde{\mathcal{L}}_{\mathrm{jump}}$ describes a local interaction between chains.
The interchain Hamiltonian $\tilde{\mathcal{L}}_{\mathrm{jump}}$ leads to the formation of an interchain bound state, in which the degrees of freedom in each chain are strongly correlated.
As a consequence, in the case of a one-particle system, the eigenmodes of $\tilde{\mathcal{L}}$ can be classified into the following two groups depending on which of the contributions from $\tilde{\mathcal{L}}_H$ and $\tilde{\mathcal{L}}_{\mathrm{jump}}$ is dominant:
\begin{enumerate}
	\item {\it Deconfined eigenmode}, where the intrachain kinetic energy dominates the interchain interaction and the eigenmodes are extended over the entire ladder.
	\item {\it Confined eigenmode}, where the interchain interaction dominates the intrachain kinetic energy and an interchain bound state is formed.
\end{enumerate}
In terms of this classification of eigenmodes, the interplay between the coherent and incoherent dynamics in open quantum systems is understood as a competition between the intrachain kinetic energy and the interchain interaction in Liouvillian eigenmodes. 
Here, the existence of an interchain bound state in the Lindblad ladder is nontrivial since the interchain interaction $\tilde{\mathcal{L}}_{\mathrm{jump}}$ has no clear notion of repulsiveness or attractiveness due to non-Hermiticity of $\tilde{\mathcal{L}}$.

\begin{figure}
	\centering
	\includegraphics[width=0.45\textwidth]{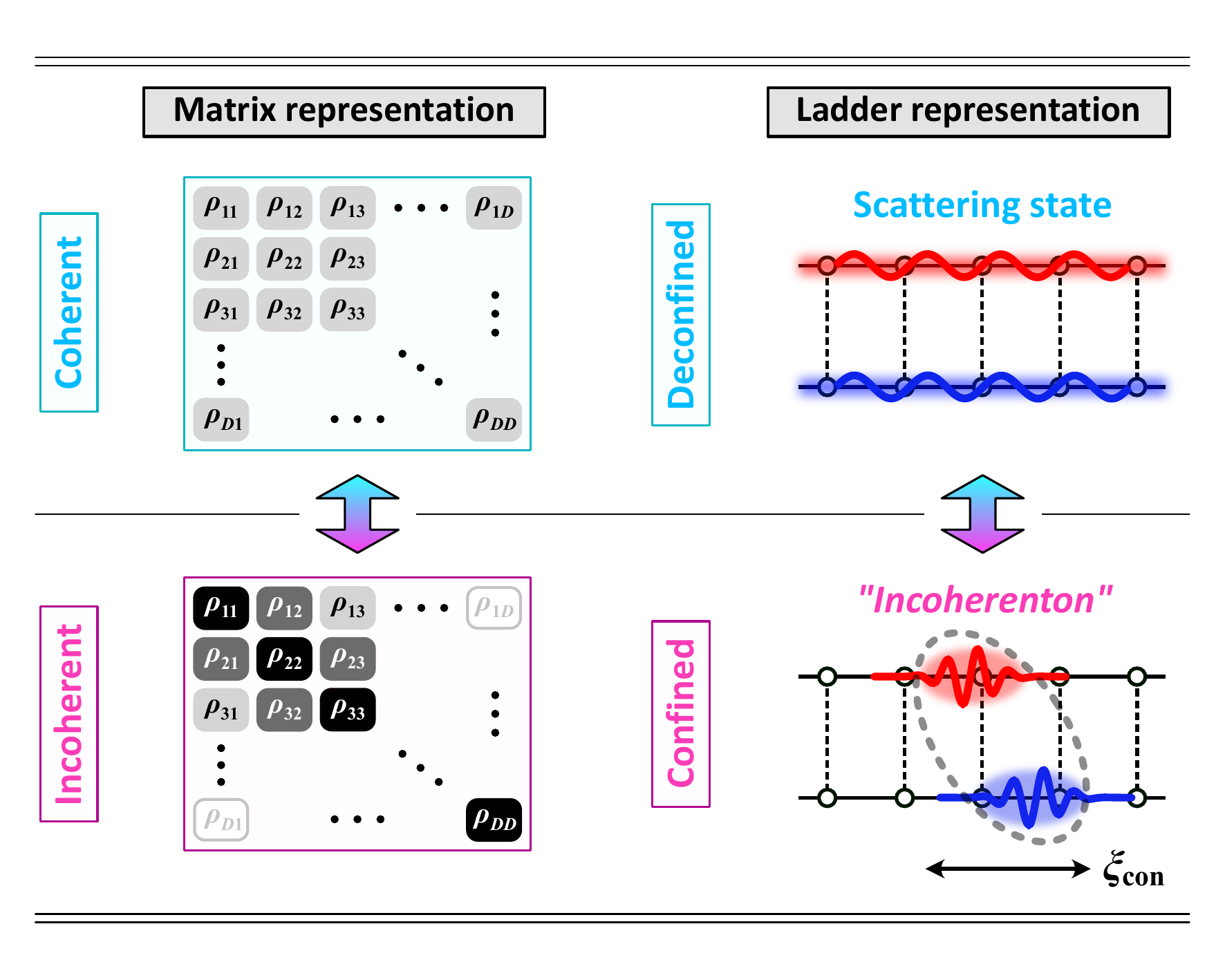}
	\caption{Schematic illustrations of coherent eigenmodes and incoherent eigenmodes for a one-particle system.
		We call an interchain bound state on the ladder as an incoherenton, for which the confinement length is denoted by $\xi_{\mathrm{con}}$.
		The gray scale shows the magnitude of each matrix elements, where the darker one shows the larger magnitude. 
		The coherent eigenmode has both diagonal and off-diagonal matrix elements that are comparable in magnitude, whereas the incoherent eigenmode has the predominant diagonal matrix elements.}
	\label{Fig-incoherenton}
\end{figure}

Deconfined and confined eigenmodes are schematically illustrated in the right column of Fig.~\ref{Fig-incoherenton}.
We call the interchain bound state in a confined eigenmode as an {\it ``incoherenton."}
For the one-particle case, an incoherenton is defined as follows.
In the ladder representation, an eigenmode can be written as
\begin{equation}
	|\rho_{\alpha}) =  \sum_{l,m=1}^L \rho_{\alpha, l m} b_{l,+}^{\dag} b_{m,-}^{\dag} |\mathrm{v}),
	\label{eigenmode_one_particle}
\end{equation}
where $b_{l,+(-)}^{\dag}$ is the creation operator of a particle at site $l$ on the first (second) chain of the ladder and $|\mathrm{v})$ is the vacuum state of the ladder.
An incoherenton is represented by matrix elements $\rho_{\alpha, l m}$ that decay exponentially with respect to the relative coordinate,
\begin{equation}
	|\rho_{\alpha, l m}| \sim e^{-|l-m|/\xi_{\mathrm{con}}} \quad (|l-m| \gg 1),
	\label{def_xi_con}
\end{equation}
where $\xi_{\mathrm{con}}$ is the confinement length of the incoherenton (see the right-bottom panel of Fig.~\ref{Fig-incoherenton}).
The divergence of $\xi_{\mathrm{con}}$ signals {\it deconfinement of an incoherenton}.
It should be noted that the critical values of control parameters at which the deconfinement transition occurs depend on the eigenmode under consideration.

In the matrix representation of the density matrix, the presence of an incoherenton implies the localization of the eigenmodes near diagonal matrix elements, and its deconfinement implies the delocalization over off-diagonal matrix elements.
Since the off-diagonal elements of the density matrix measure the degree of quantum coherence, we refer to the deconfined (confined) eigenmodes in the ladder representation as {\it coherent (incoherent) eigenmodes} in the matrix representation.
The left column of Fig.~\ref{Fig-incoherenton} illustrates the coherent and incoherent eigenmodes.
The confinement length $\xi_{\mathrm{con}}$ of an incoherenton in a confined (incoherent) eigenmode quantifies the characteristic length scale in which the quantum coherence in the eigenmode is retained.
We also call eigenvalues associated with these eigenmodes as coherent-mode eigenvalues or incoherent-mode eigenvalues.

\subsection{Deconfinement transition and quantum coherence gap}

We demonstrate the coexistence of the confined and deconfined eigenmodes for the one-particle case of the model introduced in Sec.~\ref{sec:hard_core_bosons}.
Let $\ket{l} = b_l^{\dag} \ket{\mathrm{v}}$ be the state in which the particle is located at site $l$.
Then, $\{ \ket{l} \}_{l=1,...,L}$ provides an orthonormal basis set of the Hilbert space of the one-particle sector.
In terms of this basis, an eigenmode of $\tilde{\mathcal{L}}$ is written as
\begin{equation}
	| \rho_{\alpha} ) = \sum_{l,m=1}^L \rho_{\alpha,lm} \ket{l} \otimes \ket{m} \quad (\alpha=0,1,...,L^2-1),
\end{equation}
where we assume the normalization $\sum_{l,m=1}^L |\rho_{\alpha,lm}|^2 =1$.
In the absence of coherent hopping ($J=0$), the matrix elements of $\tilde{\mathcal{L}}$ are given by 
\begin{equation}
	(\bra{l} \otimes \bra{m}) \tilde{\mathcal{L}} (\ket{l'} \otimes \ket{m'}) = \gamma \delta_{lm} \delta_{l'm'} \delta_{ll'} -\gamma \delta_{ll'} \delta_{mm'}.
	\label{L_dephasing_J=0}
\end{equation}
Thus, the action of $\tilde{\mathcal{L}}$ is decoupled into a ``diagonal'' subspace spanned by $\{ \ket{l} \otimes \ket{l} \}_{l=1,...,L}$ and an ``off-diagonal'' subspace spanned by $\{ \ket{l} \otimes \ket{m} \}_{l,m=1,...,L; l \neq m}$.
In the diagonal subspace, there is an $L$-fold degenerate eigenvalue $\lambda=0$, and in the off-diagonal subspace, there is an $(L^2-L)$-fold degenerate eigenvalue $\lambda=-\gamma$.

\begin{figure}
	\centering
	\includegraphics[width=0.45\textwidth]{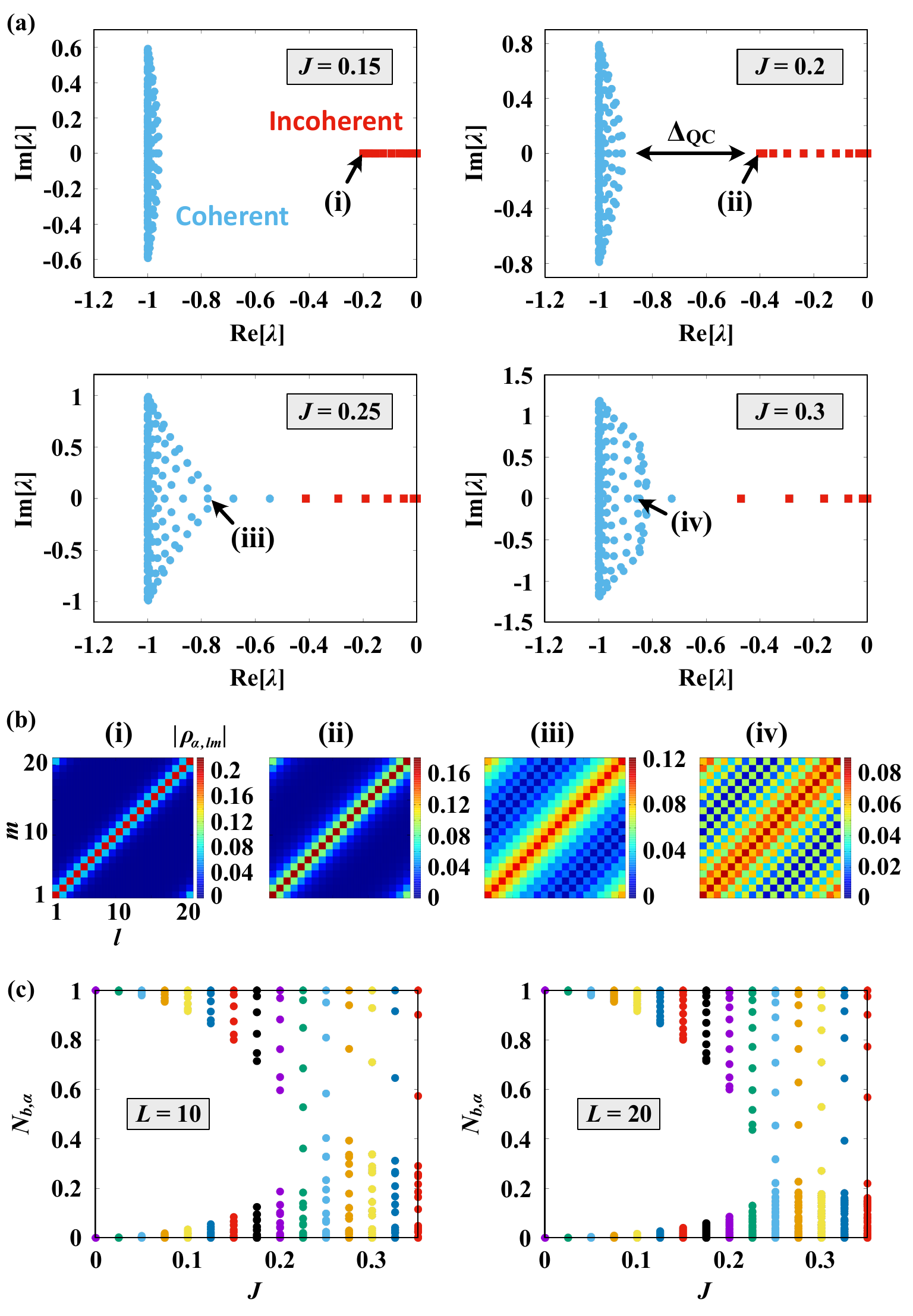}
	\caption{Liouvillian spectra and eigenmodes of the one-particle model under on-site dephasing subject to the periodic boundary condition.
		(a) Spectra with $\gamma=1$ and $J=0.15$, $0.2$, $0.25$, and $0.3$.
		The system size is $L=20$.
		The eigenvalues satisfying $S_{\alpha, \mathrm{off}}/S_{\alpha, \mathrm{diag}}<0.1$ are shown by red squares and the other eigenvalues are shown by blue circles.
		(b) Color plots of $|\rho_{\alpha, lm}|$ corresponding to the eigenvalues indicated by the arrows $\rm(\hspace{.1em}i\hspace{.1em})$--$\rm(i\hspace{-.08em}v\hspace{-.06em})$ in (a).
		(c) $\{N_{b,\alpha}\}_{\alpha=0,...,L^2-1}$ of eigenmodes with varying $J$ for system size $L=10$ (left) and $L=20$ (right).}
	\label{Fig-spec-1P}
\end{figure}

We next consider the cases of $J \neq 0$.
A detailed analysis of the one-particle eigenmodes is presented in Appendix \ref{appendix:single_particle_solution}.
In the presence of a nonzero coherent hopping, since $\tilde{\mathcal{L}}_H$ mixes the diagonal subspace with the off-diagonal one, the diagonal eigenmodes with eigenvalues near $0$ are no longer exactly diagonal.
However, when $J \ll \gamma$, the matrix elements $\rho_{\alpha,lm}$ of the eigenmodes are still localized near the diagonal elements as in Eq.~\eqref{def_xi_con}.
Figure \ref{Fig-spec-1P}(a) shows the Liouvillian spectra obtained by numerical diagonalization.
To distinguish the incoherent (confined) eigenmodes from the coherent (deconfined) ones, it is convenient to define
\begin{equation}
	S_{\alpha, \mathrm{diag}} := \sum_{|l-m|<L/4} |\rho_{\alpha, lm}|, \quad S_{\alpha, \mathrm{off}} := \sum_{|l-m| \geq L/4} |\rho_{\alpha, lm}|.
	\label{S_diag}
\end{equation}
The smaller $S_{\alpha, \mathrm{off}}/S_{\alpha, \mathrm{diag}}$ is, the stronger the localization of the eigenmode is.
In Fig.~\ref{Fig-spec-1P}(a),  the incoherent-mode eigenvalues whose eigenmodes satisfy $S_{\alpha, \mathrm{off}}/S_{\alpha, \mathrm{diag}}<0.1$ are shown by red squares, and the other eigenvalues by blue circles.

For a weak coherent hopping ($J=0.15$ or $0.2$), the Liouvillian spectrum consists of the incoherent-mode eigenvalues on the real axis and the coherent-mode eigenvalues accumulated around $\mathrm{Re}[\lambda] = -\gamma = -1$.
For $J=0$, these two types of eigenvalues are highly degenerate at $\lambda=0$ and $-\gamma$.
The presence of a nonzero $J$ lifts such degeneracy and leads to two elongated bands parallel to the real and imaginary axes.
Let us define the {\it quantum coherence (QC) gap} $\Delta_{\mathrm{QC}}$ as
\begin{equation}
	\Delta_{\mathrm{QC}} := \min_{\alpha, \ \beta} \big| \lambda_\alpha^\mathrm{(c)} - \lambda_\beta^\mathrm{(i)} \big|,
	\label{def_Delta_QC_1P}
\end{equation}
where $\{ \lambda_\alpha^\mathrm{(c)} \}$ and $\{ \lambda_\beta^\mathrm{(i)} \}$ are the coherent-mode eigenvalues (blue circles) and the incoherent-mode eigenvalues (red squares), respectively.
The QC gap $\Delta_{\mathrm{QC}}$ should not be confused with the Liouvillian gap $\Delta_L$.
While $\Delta_L=|\mathrm{Re}[\lambda_1]|$ is the gap between the steady state and the slowest decaying eigenmode, $\Delta_{\mathrm{QC}}$ is the gap between spectral bands of non-steady eigenmodes.

As $J$ increases, $\Delta_{\mathrm{QC}}$ decreases, and for $J \geq J_c=0.25$, the bands of the coherent-mode and incoherent-mode eigenvalues touch one another.
The arrows $\rm(\hspace{.1em}i\hspace{.1em})$--$\rm(i\hspace{-.08em}v\hspace{-.06em})$ in Fig.~\ref{Fig-spec-1P}(a) track an evolution of one eigenvalue that has the smallest real part in the incoherent-mode spectrum for $J<J_c$.
Figure \ref{Fig-spec-1P}(b) shows the color plots of $|\rho_{\alpha, lm}|$ corresponding to these eigenvalues.
For $\rm(\hspace{.1em}i\hspace{.1em})$ and $\rm(i\hspace{-.05em}i)$, $\rho_{\alpha, lm}$ is well localized near the diagonal elements.
In contrast, for $\rm(i\hspace{-.08em}v\hspace{-.06em})$, $\rho_{\alpha, lm}$ is delocalized over the off-diagonal elements.
Thus, in the ladder representation, the deconfinement transition of an incoherenton occurs at $J_c=0.25$.

From Eq.~\eqref{lambda_inc_1P} in Appendix \ref{appendix:single_particle_solution}, the incoherent-mode eigenvalue with the maximal $|\mathrm{Re}[\lambda]|$ is given in the limit of $L \to \infty$ by
\begin{equation}
	\lambda=-\gamma + \sqrt{\gamma^2 - 16 J^2}.
	\label{lambda_inc_1P_k=pi}
\end{equation}
Since the real parts of the coherent-mode eigenvalues are identical to $-\gamma$ in this limit (see Appendix \ref{appendix:single_particle_solution}), we have
\begin{equation}
	\Delta_{\mathrm{QC}} = \sqrt{\gamma^2 - 16 J^2}.
	\label{Delta_QC_1P}
\end{equation}
Thus, the critical value of $J$ at which $\Delta_{\mathrm{QC}}$ closes is given by
\begin{equation}
	J_c = \frac{\gamma}{4}.
	\label{J_c_1P}
\end{equation}
While Eq.~\eqref{lambda_inc_1P_k=pi} implies that a real-complex transition occurs at $J=J_c$ for the infinite system, the incoherent-mode eigenvalues for a finite system indicated by the arrows $\rm(\hspace{.1em}i\hspace{.1em})$--$\rm(i\hspace{-.08em}v\hspace{-.06em})$ in Fig.~\ref{Fig-spec-1P}(a) remain real for $J>J_c$.

For each eigenmode $|\rho_{\alpha})$, we define the fraction of an on-site bound pair in $|\rho_{\alpha})$ as
\begin{equation}
	N_{b,\alpha} := \sum_{l=1}^L \frac{(\rho_{\alpha}| n_{l,+}  n_{l,-} |\rho_{\alpha})}{(\rho_{\alpha}|\rho_{\alpha})},
	\label{def_N_b}
\end{equation}
where $n_{l,\pm}=b_{l,\pm}^{\dag} b_{l,\pm}$ is the number-density operator, and $N_{b,\alpha} \in [0,1]$.
For $J=0$, $N_{b,\alpha}=1$ for the incoherent eigenmodes and $N_{b,\alpha}=0$ for the coherent eigenmodes.
Figure \ref{Fig-spec-1P}(c) shows $\{N_{b,\alpha}\}_{\alpha=0,...,L^2-1}$ for different values of $J$ and the system sizes $L=10$ and $L=20$.
For $J<J_c=0.25$, there exists a gap between clusters of $N_{b,\alpha}$ around $0$ and $1$, and it closes at $J=J_c$.
The width of the cluster around $N_{b,\alpha}=0$ decreases in inverse proportion to $L$, which implies that the eigenmodes in this cluster are scattering states that extend over the entire system.
In contrast, the width of the cluster around $N_{b,\alpha}=1$ is independent of $L$ because the eigenmodes in this cluster are localized with a confinement length $\xi_{\mathrm{con}}$, which is independent of $L$.
Thus, in the limit of $L \to \infty$, $N_{b,\alpha}$ can be considered as an order parameter, which has a nonzero value for incoherent eigenmodes but vanishes for coherent eigenmodes.

The relation between $\Delta_{\mathrm{QC}}$ and the confinement length $\xi_{\mathrm{con}}$ that is maximized over all incoherentons is given by Eq.~\eqref{xi_con_Delta_QC}.
This relation may be interpreted as follows.
Let us denote the typical decay rate of coherent eigenmodes (without incoherenton) as $\Gamma_{\mathrm{coh}}$ and that of incoherent eigenmodes (with an incoherenton) as $\Gamma_{\mathrm{inc}}$. 
For on-site dissipation, in which each $L_{\nu}$ acts on a single lattice site, the decay rate of extended coherent eigenmodes is larger than that of localized incoherent eigenmodes because the dissipation suppresses off-diagonal elements of the density matrix.
Note that $\Gamma_{\mathrm{inc}}$ approaches $\Gamma_{\mathrm{coh}}$ as $\xi_{\mathrm{con}}$ diverges to infinity.
Thus, we assume $(\Gamma_{\mathrm{coh}}-\Gamma_{\mathrm{inc}})/\Gamma_{\mathrm{coh}} \sim 1/\xi_\mathrm{con}$, where $\xi_\mathrm{con}$ is measured in units of the lattice spacing.
While we have no general proof for this assumption, we can prove it for the one-particle model under on-site dephasing (see Appendix \ref{appendix:single_particle_solution}).
Since $\Delta_{\mathrm{QC}} \simeq \Gamma_{\mathrm{coh}}-\Gamma_{\mathrm{inc}}$, we obtain Eq.~\eqref{xi_con_Delta_QC}.

It should be noted that the examination of the one-particle spectra and eigenmodes provided above is essentially the same as the analysis in Sec.~III of Ref.~\cite{Znidaric-15}.
The author of this reference considered the XX spin chain with bulk dephasing, which is equivalent to dephasing hard-core bosons discussed in our study.
We note that Eq.~(8) and Figs.~3 and 4 in Ref.~\cite{Znidaric-15} correspond to Eq.~\eqref{lambda_inc_1P} and Fig.~\ref{Fig-spec-1P} in our study, respectively.
The primary distinction between Ref.~\cite{Znidaric-15} and our study is that the former emphasizes the most slowly decaying mode which determines the Liouvillian gap $\Delta_L$, whereas the latter focuses on the most rapidly decaying mode in the incoherent eigenmodes which governs the QC gap $\Delta_\mathrm{QC}$ [see the arrows in Fig.~\ref{Fig-spec-1P}(a)].

We draw attention to the connection between our findings and exceptional points (EPs) in non-Hermitian physics \cite{Bender-98, Guo-09, Ruter-10, Heiss-12, El-Ganainy-18, Minganti-19, Ozdemir-19, Miri-19, Ashida-20, Bergholtz-21}.
The eigenvalues typically exhibit a square-root dependence on the parameter near an EP.
As indicated by Eq.~\eqref{lambda_inc_1P_k=pi}, the critical value $J_c$, where the deconfinement of an incoherenton takes place, signifies an EP of the Liouvillian in the limit of infinite system size. 
Thus, we have identified a novel class of EPs associated with the transition between coherent and incoherent attributes of eigenmodes.
In addition, as detailed in Secs.~\ref{sec:hierarchy} and \ref{sec:exact_many_body_solution}, the deconfinement of incoherentons can be generalized to many-body cases.
The deconfinement of incoherentons offers a generic mechanism of producing EPs that have a significant consequences on the dynamics in open quantum many-body systems.

The discrete nature of a lattice system is essential for the formation of incoherentons.
In fact, for a free particle under a dephasing-type dissipation in continuous space, we can show the absence of such an interchain bound state (see Appendix \ref{appendix:absence_of_incoherenton}).
The creation of a bound state due to spatial discreteness in a lattice system has also been known in the conventional two-body problem with a repulsive interaction \cite{Winkler-06}; while in continuous space no bound state is allowed between particles with a repulsive interaction, on a lattice a bound state exists for an arbitrarily strong repulsive interaction.

\section{Incoherent-coherent dynamical transition}
\label{sec:incoherent_coherent_dynamical_transition}

We have shown that the QC gap $\Delta_{\mathrm{QC}}$ closes at a certain critical point.
Since the dynamics of open quantum systems are intimately related to the Liouvillian spectrum, it is natural to expect that such a change in the structure of the spectrum would significantly alter the transient dynamics to the steady state.
In this section, we demonstrate that the QC gap closing is accompanied by an incoherent-coherent dynamical transition from overdamped relaxation dominated by dissipation to underdamped relaxation dominated by unitary time evolution.

In the one-particle sector, $\{ \ket{l} = b_l^{\dag} \ket{\mathrm{v}} \}_{l=1,...,L}$ provides an orthonormal basis set of the Hilbert space.
We denote the corresponding matrix elements of the density matrix as $\rho_{lm}$, which satisfies a normalization condition $\sum_l \rho_{ll}=1$.
We note that the steady state $\rho_{\mathrm{ss}}$ is given by the infinite-temperature state, $(\rho_{\mathrm{ss}})_{lm} = L^{-1} \delta_{lm}$.
We consider the following initial state, whose particle density is modulated with wavenumber $k$,
\begin{equation}
	\rho_{lm}(0) = L^{-1} (1 + \Delta n \cos kl) \delta_{lm},
	\label{rho_ini}
\end{equation}
where $k= 2\pi s/L \ (s=-L/2+1,...,L/2)$, and $\Delta n$ represents the amplitude of the density modulation.
For $J=0$, the initial state given by Eq.~\eqref{rho_ini} shows no time evolution because the action of the Liouvillian to any diagonal density matrix vanishes.
In the presence of a nonzero $J$, the density matrix relaxes toward the uniform steady state $\rho_{\mathrm{ss}}$.
The perturbation with wavenumber $k$ to the steady state can selectively excite the incoherent eigenmodes with the same wavenumber $k$.
Thus, we expect that the decay rate of the particle density starting from the initial state \eqref{rho_ini} with each $k$ reflects the structure of the incoherent-mode spectrum.

\begin{figure}
	\centering
	\includegraphics[width=0.45\textwidth]{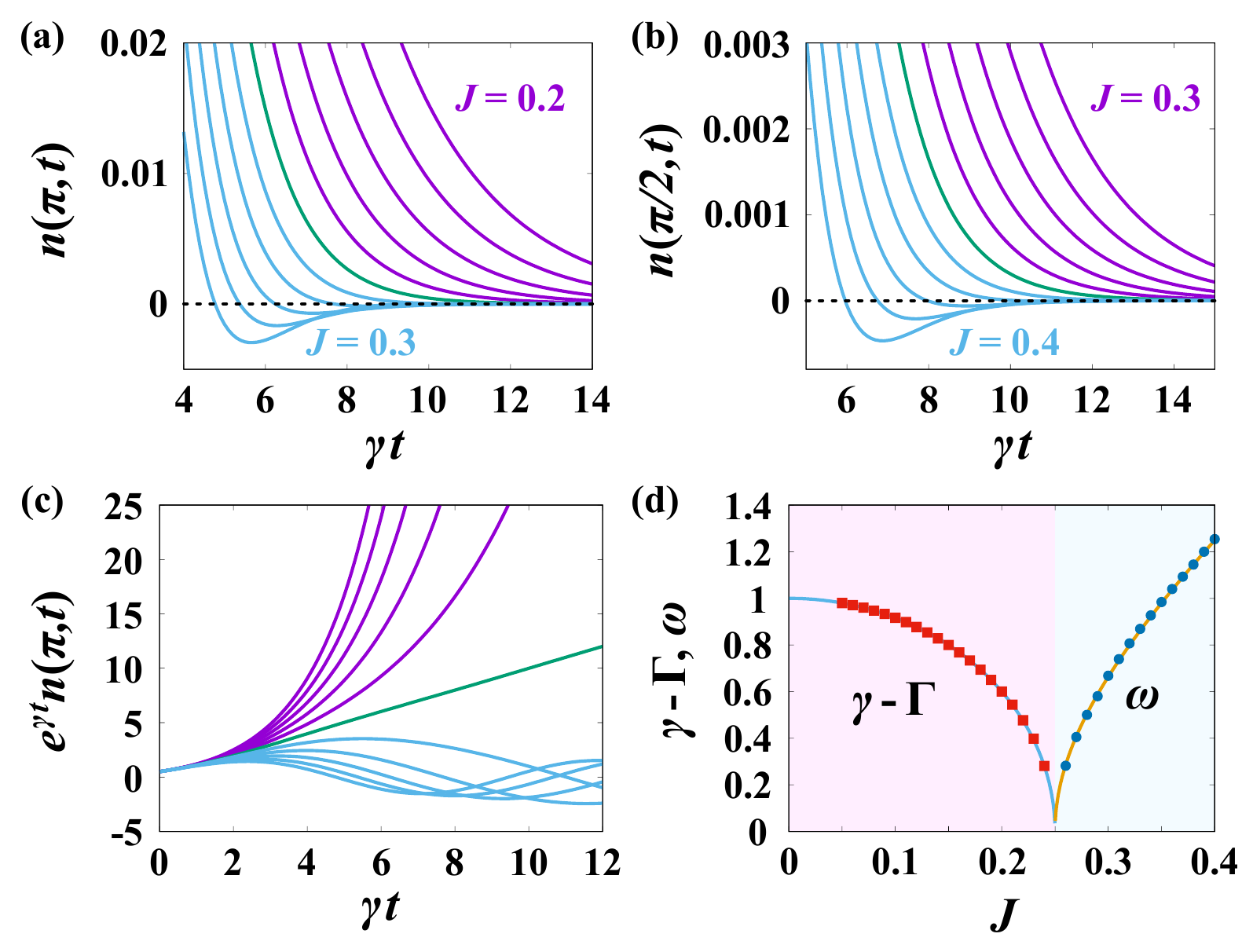}
	\caption{Time evolution of the amplitude of the density modulation.
		(a) $n(\pi,t)$ with $\gamma=1$ and system size $L=20$.
		The values of $J$ are increased from $0.2$ (top) to $0.3$ (bottom) in increments of $0.01$.
		The amplitude of the density modulation for the initial state is $\Delta n=0.5$.
		(b) $n(\pi/2,t)$ with $\gamma=1$ and system size $L=20$.
		The values of $J$ are increased from $0.3$ (top) to $0.4$ (bottom) in increments of $0.01$.
		(c) $e^{\gamma t} n(\pi,t)$ for the same values of $J$ as in (a).
		(d) $\gamma-\Gamma$ and $\omega$ as functions of $J$ obtained from $n(\pi,t)$.
		The solid curves represent $(\gamma^2-16J^2)^{1/2}$ for $J<0.25$ and $(16J^2-\gamma^2)^{1/2}$ for $J>0.25$.}
	\label{Fig-density-rel}
\end{figure}

From the density profile $n_l(t)=\rho_{ll}(t)$ at time $t$, we define 
\begin{equation}
	n(k,t) := \sum_{l=1}^L n_{l}(t) e^{-i kl},
	\label{def_n_k}
\end{equation}
which relaxes to zero in the limit of $t \to \infty$ for $k \neq 0$.
Figures \ref{Fig-density-rel} (a) and (b) show $n(\pi,t)$ and $n(\pi/2,t)$ for different values of $J$, respectively.
The decay rate is a decreasing function of $J$ and it vanishes at $J=0$.
From these figures, one finds that there exists a $k$-dependent critical value $J_c^{\mathrm{rel}}(k)$ below which $n(k,t)$ exhibits an exponential decay $e^{-\Gamma t}$ and above which it shows a damped oscillation $e^{-\gamma t} \cos \omega t$.
The critical value is estimated as $J_c^{\mathrm{rel}}(\pi) \simeq 0.25$ and $J_c^{\mathrm{rel}}(\pi/2) \simeq 0.35$.
An important observation is that $J_c^{\mathrm{rel}}(k)$ is close to the value of $J$ at which the incoherent eigenmode with wavenumber $k$ exhibits the deconfinement transition.
From Eq.~\eqref{lambda_inc_1P}, the incoherent-mode eigenvalue is written as
\begin{equation}
	\lambda_{\mathrm{inc}}(k)=-\gamma + \sqrt{\gamma^2 - 16 J^2 \sin^2 (k/2)}
\end{equation}
in terms of the wavenumber $k$.
This eigenvalue becomes complex at a critical value:
\begin{equation}
	J_c(k) = \frac{\gamma}{4|\sin(k/2)|}.
\end{equation}
We then obtain $J_c(\pi)=0.25\gamma$ and $J_c(\pi/2) \simeq 0.354\gamma$, which are close to $J_c^{\mathrm{rel}}(\pi)$ and $J_c^{\mathrm{rel}}(\pi/2)$, respectively.
It is reasonable that the real-complex transition of $\lambda_{\mathrm{inc}}(k)$ is accompanied by an incoherent-coherent dynamical transition from overdamped to underdamped relaxations starting from an incoherent initial state.
Recall that the QC closing occurs at $J=\min_k J_c(k)=\gamma/4$, where the minimum is attained at $k=\pi$.

Figure \ref{Fig-density-rel} (c) shows $e^{\gamma t} n(\pi,t)$ for different values of $J$.
For $J<0.25$, $n(\pi,t)$ decays exponentially with a rate $\Gamma<\gamma$.
As $J$ approaches $0.25$ from below, the decay rate $\Gamma$ approaches $\gamma$, and at the critical point $J=0.25$, one finds
\begin{equation}
	n(\pi,t) \sim t e^{-\gamma t}.
	\label{t_exp}
\end{equation}
Note that Eq.~\eqref{t_exp} exactly holds only in the limit of $L \to \infty$.
For a finite system, there is a crossover from an early stage described by Eq.~\eqref{t_exp} to a later stage showing an exponential decay $n(\pi,t) \sim e^{-\Gamma t}$ with $\Gamma<\gamma$.
The interval of this early stage diverges in the limit of $L \to \infty$.
It should be recalled that the polynomial correction \eqref{t_exp} to an exponential decay also appears at EPs of non-Hermitian systems \cite{Bender-98, Guo-09, Ruter-10, Heiss-12, El-Ganainy-18, Minganti-19, Ozdemir-19, Miri-19, Ashida-20, Bergholtz-21}.

For $J<0.25$, the decay rate $\Gamma$ is evaluated from the fitting of $n(\pi,t)$ by $a e^{-\Gamma t}$, and for $J>0.25$, the frequency $\omega$ is evaluated from the fitting of $e^{\gamma t} n(\pi,t)$ by $a \sin(\omega t + b)$.
Figure \ref{Fig-density-rel} (d) shows $\gamma-\Gamma$ for $J<0.25$ and $\omega$ for $J>0.25$.
The solid lines represent $\mathrm{Re}[\lambda_{\mathrm{inc}}(\pi)] + \gamma = (\gamma^2-16J^2)^{1/2}$ for $J<0.25$ and $\mathrm{Im}[\lambda_{\mathrm{inc}}(\pi)] = (16J^2-\gamma^2)^{1/2}$ for $J>0.25$.
We can confirm that $\Gamma$ and $\omega$ are well approximated by $-\mathrm{Re}[\lambda_{\mathrm{inc}}(\pi)]$ and $\mathrm{Im}[\lambda_{\mathrm{inc}}(\pi)]$, respectively. 
Thus, by measuring the decay rate and the frequency of the density relaxation starting from incoherent initial states with density modulation of various wavenumbers, one can reconstruct the incoherent-mode spectrum of the Liouvillian.

We close this section by stressing that our work provides a new type of incoherent-coherent dynamical transition in open quantum lattice systems.
Previous incoherent-coherent dynamical transition has mostly been studied with respect to the spin-boson model, where the expectation value of the spin variable shows a transition from overdamped to underdamped relaxation \cite{Leggett-87, Mak-91, Egger-97, Chin-06, Matsuo-08, Wang-08, Kast-13, Magazzu-15}.
An important distinction from the spin-boson model is that our model has spatially extended degrees of freedom, which play an essential role in the deconfinement of incoherentons.
Furthermore, it should be noted that the transition discussed here becomes sharp only in the limit of infinite system size.

\section{Hierarchy of eigenmodes}
\label{sec:hierarchy}

\begin{figure}
	\centering
	\includegraphics[width=0.45\textwidth]{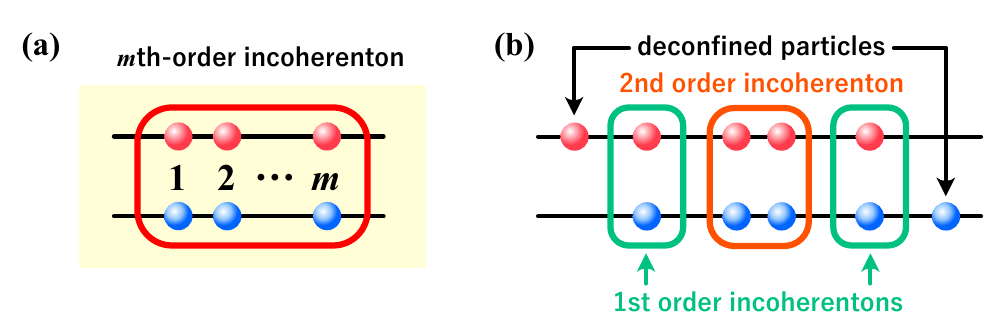}
	\caption{(a) Schematic illustration of an $m$th-order incoherenton.
		(b) Ladder representation of a typical Liouvillian eigenmode of a many-body system.}
	\label{Fig-many-body-incoherenton}
\end{figure}

In the following sections, we discuss the generalization of incoherentons and QC gap to many-body systems.
In these cases, incoherentons and deconfined particles, in general, coexist.
Furthermore, two or more particles can form a single bound state.
We refer to such a $2m$-particle composite incoherenton as an $m$th-order incoherenton [see Fig.~\ref{Fig-many-body-incoherenton}(a)].
An $m$th-order incoherenton can be represented by the $m$-particle reduced eigenmode,
\begin{equation}
	G^{(m)}_{\alpha; l_1, ..., l_m; l_{m+1}, ..., l_{2m}} := \mathrm{Tr}[b_{l_1} \cdots b_{l_m} \rho_\alpha b_{l_{m+1}}^{\dag} \cdots b_{l_{2m}}^{\dag}].
	\label{def_G_mP}
\end{equation}
If all particles form a single $m$th-order incoherenton, the $m$-particle reduced eigenmode is expected to behave as
\begin{equation}
	|G^{(m)}_{\alpha; l_1, ..., l_m; l_{m+1}, ..., l_{2m}}| \sim \prod_{i,j=1}^{2m} e^{-|l_i - l_j| / \xi_{\mathrm{con}}},
\end{equation}
where $\xi_{\mathrm{con}}$ is the confinement length.
Figure \ref{Fig-many-body-incoherenton} (b) shows the ladder representation of a typical Liouvillian eigenmode.
The eigenmodes of the $N$-body system can be classified according to how many $m$th-order incoherentons ($1 \leq m \leq N$) they contain.
In this section, we describe a typical scenario for the structure of eigenmodes and spectra, which is expected to be applicable to a broad class of dephasing-type dissipation.

\begin{figure}
	\centering
	\includegraphics[width=0.45\textwidth]{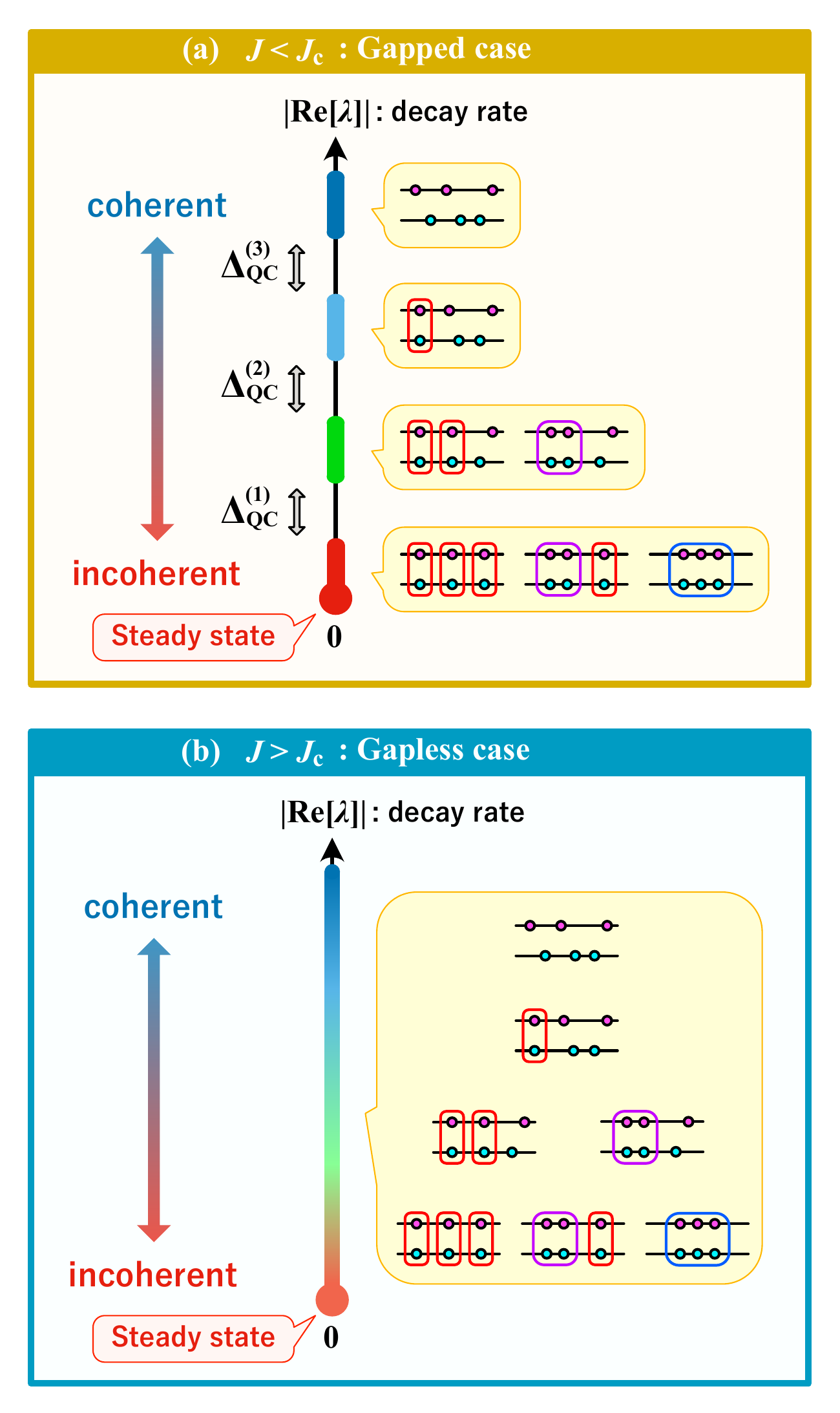}
	\caption{Hierarchy of eigenmodes for a three-particle case.
		The eigenmodes are ordered from top to bottom in decreasing order of their decay rates $|\mathrm{Re}[\lambda]|$.
		The lower end of the most incoherent band represents the steady state.
		For (a) with $J < J_c$, the groups of eigenmodes with different decay rates are separated by the QC gap $\Delta_{\mathrm{QC}}$.
		For (b) with $J > J_c$, the QC gap closes and all groups are continuously connected.}
	\label{Fig-hierarchy}
\end{figure}

Let $J$ be a parameter of the Hamiltonian that creates quantum coherence between lattice sites, such as the tunneling amplitude between adjacent sites, and let $\gamma$ be a parameter that describes the strength of dissipation.
The balance between $J$ and $\gamma$ characterizes the competition between the intrachain kinetic energy $\tilde{\mathcal{L}}_H$ and the interchain interaction $\tilde{\mathcal{L}}_{\mathrm{jump}}$.
Figure \ref{Fig-hierarchy} shows a schematic diagram of the Liouvillian eigenmodes.
The structure of the eigenmodes in the ladder representation is depicted in boxes, with circles and squares representing particles and incoherentons, respectively.
The eigenmodes are ordered from top to bottom in decreasing order of their decay rates $|\mathrm{Re}[\lambda]|$.
Note that eigenmodes with a larger number of incoherentons decay more slowly.
This is because the dissipation suppresses off-diagonal elements of the density matrix and thus coherent eigenmodes with less incoherentons decay faster.

\begin{figure}
	\centering
	\includegraphics[width=0.45\textwidth]{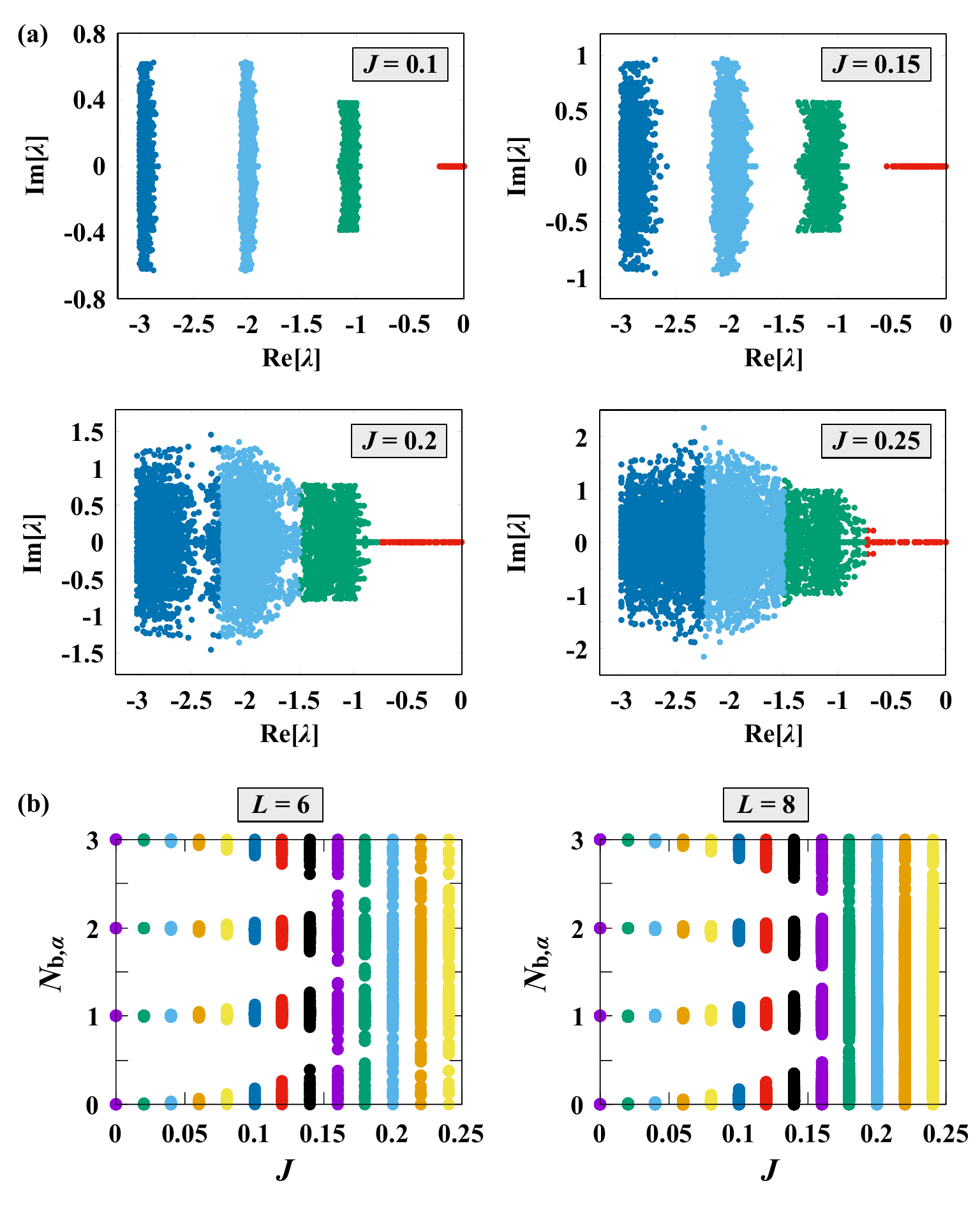}
	\caption{Liouvillian spectra and eigenmodes of the hard-core bosons under on-site dephasing.
		(a) Spectra with $\gamma=1$ and $J=0.1$, $0.15$, $0.2$, and $0.25$.
		The system size is $L=10$ and the particle number is $N=3$.
		The eigenvalues satisfying $0 \leq N_{b,\alpha} < 3/4$, $3/4 \leq N_{b,\alpha} < 3/2$, $3/2 \leq N_{b,\alpha} < 9/4$, and $9/4 \leq N_{b,\alpha} \leq 3$ are colored by blue, light blue, green, and red, respectively.
		(b) $\{N_{b,\alpha}\}_{\alpha=0,...,D^2-1}$ of eigenmodes with $\gamma=1$.
		The system sizes are $L=6$ and $8$, and the number of particles is $N=3$.}
	\label{Fig-spec-MP}
\end{figure}

We first consider the case where the dissipation is dominant over the coherent tunneling ($J \ll \gamma$).
In general, strong dissipation has the effect of suppressing the coherent evolution of quantum systems, known as the quantum Zeno effect \cite{Misra-77, Itano-90, Koshino-05}, which has recently attracted considerable attention in the context of AMO systems \cite{Zhu-14, Patil-15, Li-18, Froml-20, Popkov-21, Zhou-21}.
As shown in Fig.~\ref{Fig-hierarchy}(a), the eigenmodes of an $N$-particle system are divided into $N+1$ groups with different decay rates.
Each group of eigenmodes is characterized by the number of particles that do not form incoherentons.
The eigenmodes in the group with the largest decay rate have no incoherenton, while in the eigenmodes belonging to the group with the smallest decay rate, all particles form incoherentons.
Furthermore, each group contains eigenmodes with various types of incoherenton.
For $N=3$, the group with the smallest decay rate consists of eigenmodes in which particles form (i) three first-order incoherentons or (ii) one first-order incoherenton and one second-order incoherenton or (iii) a single third-order incoherenton [see the bottom box of Fig.~\ref{Fig-hierarchy}(a)].
The $n$th QC gap $\Delta_{\mathrm{QC}}^{(n)}$ is defined by
\begin{equation}
	\Delta_{\mathrm{QC}}^{(n)} := \min_{\alpha, \ \beta} \big| \lambda_\alpha^{(n)} - \lambda_\beta^{(n-1)} \big|,
	\label{def_Delta_QC}
\end{equation}
where $\{ \lambda_\alpha^{(n)} \}$ are the eigenvalues with $n$ unbound pairs  [see Fig.~\ref{Fig-hierarchy}(a)].

Let us consider what would happen if the coherent tunneling amplitude $J$ is gradually increased while keeping $\gamma$ fixed (or if the dissipation strength $\gamma$ is decreased while keeping $J$ fixed).
Since $J$ represents the amplitude of the intrachain tunneling $\tilde{\mathcal{L}}_H$, the confinement length $\xi_{\mathrm{con}}$ (the QC gap $\Delta_{\mathrm{QC}}$) of an incoherenton increases (decreases) with increasing $J$.
At some critical point $J=J_c$, the confinement length $\xi_{\mathrm{con}}$ diverges, and a deconfinement transition of an incoherenton takes place.
The QC gap $\Delta_{\mathrm{QC}}$ closes at $J=J_c$.
Figure \ref{Fig-hierarchy}(b) shows the hierarchy of eigenmodes for $J>J_c$, which forms a continuum where the number of incoherentons can vary continuously with respect to the decay rate.
As in the one-particle case, the relationship between $\xi_{\mathrm{con}}$ and $\Delta_{\mathrm{QC}}$ is expected to be given by Eq.~\eqref{xi_con_Delta_QC}.

We highlight the uniqueness of our findings in the context of the segment structure of the Liouvillian spectrum for strong dissipation, demonstrated in Fig.~\ref{Fig-hierarchy}(a), which has been reported for several open quantum many-body systems in recent literature \cite{Popkov-21, Zhou-21, Wang-20}.
Firstly, Ref.~\cite{Popkov-21} primarily relies on a perturbation theory relevant only to strong dissipation.
However, the closing of the QC gap or merging of spectral bands is a nonperturbative phenomenon that cannot be adequately captured by perturbation theory.
Secondly, Ref.~\cite{Wang-20} is based on a general concept of locality and employs a randomly constructed Liouvillian. 
Consequently, it lacks an intuitive picture of the hierarchical structure of the Liouvillian spectrum. 
In contrast, our incoherenton framework offers a clear physical picture in terms of the number or order of incoherentons, which elucidates the relationship between the hierarchical structure of the spectrum and the eigenmodes associated with each spectral band.

Let us now demonstrate the scenario of Fig.~\ref{Fig-hierarchy} for the dissipative hard-core boson model introduced in Sec.~\ref{sec:hard_core_bosons}.
We consider the case of particle number $N=3$.
For each many-body eigenmode $|\rho_{\alpha})$, the number of bound pairs $N_{b,\alpha}$ is defined by Eq.~\eqref{def_N_b}.
For $J=0$, there are four highly degenerate eigenvalues $0$, $-\gamma$, $-2\gamma$, and $-3\gamma$.
The number of bound pairs and degeneracy $d$ of eigenmodes corresponding to each eigenvalue are given as follows: $N_b=3$ and $d=L(L-1)(L-2)/6$ for $\lambda=0$, $N_b=2$ and $d=L(L-1)(L-2)(L-3)/2$ for $\lambda=-\gamma$, $N_b=1$ and $d=L(L-1) \cdots (L-4)/4$ for $\lambda=-2\gamma$, and $N_b=0$ and $d=L(L-1) \cdots (L-5)/36$ for $\lambda=-3\gamma$.

Figure \ref{Fig-spec-MP}(a) shows the Liouvillian spectra with $J=0.1$, $0.15$, $0.2$, and $0.25$.
The colors of the dots represent $N_{b,\alpha}$ for the corresponding eigenmodes $|\rho_{\alpha})$.
In the presence of a nonzero but small $J$, there are four bands around $0$, $-1$, $-2$, and $-3$.
As $J$ increases, the widths of these bands increase, and at $J=J_c \simeq 0.2$, they merge almost simultaneously.
Figure \ref{Fig-spec-MP}(b) shows $\{N_{b,\alpha}\}_{\alpha=0,...,D^2-1}$ in Eq.~\eqref{def_N_b} with $L=6$ and $8$ for different values of $J$ from $0$ to $0.25$.
For $J=0$, $N_{b, \alpha}$ is degenerate at $0$, $1$, $2$, and $3$.
In the presence of nonzero $J$, since the coherent hopping mixes the eigenmodes with different $N_{b, \alpha}$, the values of $N_{b, \alpha}$ distribute over a finite width around $0$, $1$, $2$, and $3$.
The gaps between these clusters close at $J \simeq 0.2$, which is identical to the value of $J$ at which the QC gap in the Liouvillian spectrum closes.
It should be noted that the critical hopping amplitude $J_c \simeq 0.2$ is slightly shifted from that for the one-particle case $J_c=\gamma/4=0.25$ owing to the interactions among incoherentons and deconfined particles.

It is a nontrivial issue whether the critical hopping amplitude $J_c$, at which the QC gap closes, remains nonzero when the limit of infinite system size is taken at a constant density $N/L$.
In Sec.~\ref{sec:exact_many_body_solution}, we will show that a certain class of incoherentons exhibits deconfinement at a value of $J$ independent of the system size.
However, this does not mean that $J_c$ is generally independent of the system size because $J_c$ could depend on spectral bands.
Specifically, it is widely believed that an infinitesimally small integrability-breaking perturbation leads to random matrix statistics at the center of the spectrum \cite{Sa-20, Hamazaki-20, Li-21}, implying that $J_c$ becomes zero for bands located at the center of the spectrum in the thermodynamic limit.
However, even if this is the case, $J_c$ for spectral bands near the steady state may remain nonzero in the thermodynamic limit.

\begin{figure}
	\centering
	\includegraphics[width=0.45\textwidth]{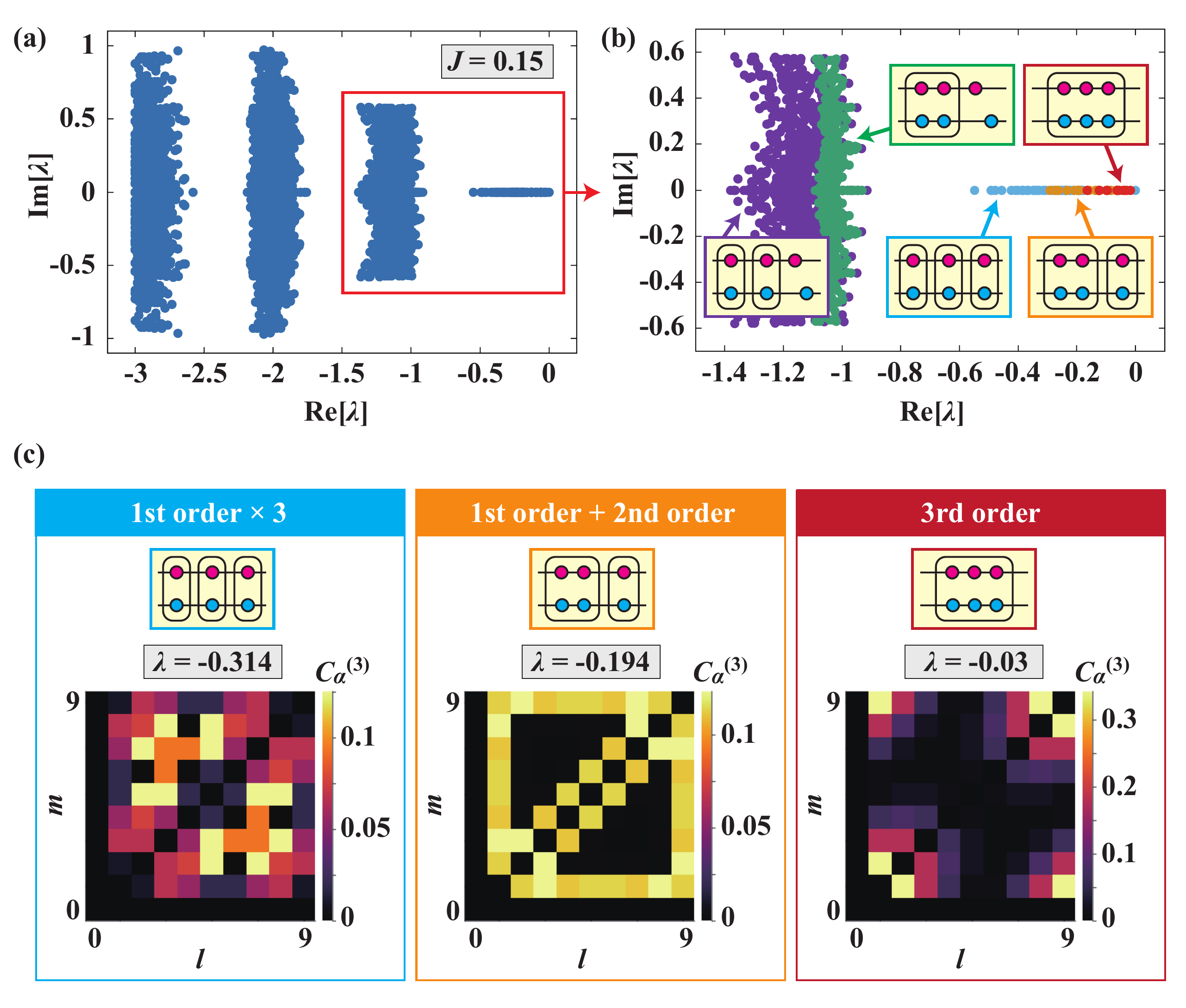}
	\caption{Classification of eigenmodes of the dephasing hard-core bosons.
	(a) Complete spectrum with $\gamma=1$ and $J=0.15$. 
	The system size is $L=10$ and the particle number is $N=3$. 
	(b) Part of the spectrum corresponding to the red square in panel (a). 
	The color differentiates between the types of eigenmodes, which are illustrated in the insets.
	(c) Color maps of $C_\alpha^{(3)}(l, m)$ for three representative eigenmodes with eigenvalues $\lambda=-0.314$, $-0.194$, and $-0.03$.
	To emphasize off-diagonal components, the values of $C_\alpha^{(3)}(l, m)$ at $l=0$, $m=0$, and $l=m$ are set to zero.
	For eigenmodes with three first-order incoherentons, $C_\alpha^{(3)}(l, m)$ is delocalized across the full range of $l$ and $m$.
	For eigenmodes with one first-order incoherenton and one second-order incoherenton, $C_\alpha^{(3)}(l, m)$ is localized at the edge of the $(l,m)$ space.
	For eigenmodes with one third-order incoherenton, $C_\alpha^{(3)}(l, m)$ is localized at the corner of the $(l,m)$ space.
	In panel (b), the eigenmodes in the band with the smallest decay rate are classified on the basis of whether the point $(l^*, m^*)$ that maximizes $C_\alpha^{(3)}(l, m)$ is located in the bulk, edge, or corners of the $(l,m)$ space.}
	\label{Fig-mode-class}
\end{figure}

Note that $N_{b, \alpha}$ counts the number of the particle pairs forming incoherentons, but does not indicate the order of incoherentons.
As illustrated in Fig.~\ref{Fig-hierarchy}, the spectral band with the smallest decay rate comprises three types of eigenmodes, which $N_{b, \alpha}$ fails to distinguish.
To quantify the number of incoherentons of each order, we introduce incoherenton correlation functions:
\begin{equation}
	C_\alpha^{(2)}(m) = \sum_{j=1}^L \frac{(\rho_\alpha| n_{j,+} n_{j,-} n_{j+m,+} n_{j+m,-} |\rho_\alpha)}{(\rho_\alpha|\rho_\alpha)},
	\label{def_inc_corr_2}
\end{equation}
\begin{equation}
	C_\alpha^{(3)}(l, m) = \sum_{j=1}^L \frac{(\rho_\alpha| n_{j,+} n_{j,-} n_{j+l,+} n_{j+l,-} n_{j+m,+} n_{j+m,-} |\rho_\alpha)}{(\rho_\alpha|\rho_\alpha)},
	\label{def_inc_corr_3}
\end{equation}
and so on.
The qualitative features of $C_\alpha^{(s)}(m_1,...,m_{s-1})$ $(s=2,3,...)$ allow the identification of the type of eigenmodes.
For instance, let us consider the spectral band with the second smallest decay rate in Fig.~\ref{Fig-hierarchy}.
It includes two types: (i) one with two first-order incoherentons and (ii) the other with a single second-order incoherenton.
While $C_\alpha^{(2)}(m)$ takes an almost constant value for type (i) eigenmodes, it exponentially decreases with respect to $m$ for type (ii) eigenmodes.
Similarly, the eigenmodes within the band with the smallest decay rate can be classified based on the behavior of $C_\alpha^{(3)}(l, m)$.
Figure \ref{Fig-mode-class}(b) shows a spectrum where color differentiates the type of eigenmodes.
In particular, color maps of $C_\alpha^{(3)}(l, m)$ for three representative eigenmodes in the band with the smallest decay rate are displayed in Fig.~\ref{Fig-mode-class}(c).
These three types of eigenmodes can be distinguished on the basis of whether $C_\alpha^{(3)}(l, m)$ is delocalized, localized on the edge, or localized at the corners.
It is worth emphasizing that quantities like $N_{b, \alpha}$ and $C_\alpha^{(2)}(m)$ can be experimentally measured in ultracold atoms on optical lattices through a process involving the interference of two system copies and the enumeration of atoms within each copy \cite{Daley-12, Islam-15}. 
Detailed experimental protocols are explained in Appendix \ref{appendix:incoherenton_correlation}.

We here remark on the universal validity of the scenario illustrated in Fig.~\ref{Fig-hierarchy}.
We expect that it is applicable to systems with local dissipation that satisfies the detailed balance condition.
In the context of Markovian open quantum systems, the detailed balance condition is expressed as $\rho_G \mathcal{L}^\dag(A)^* = \mathcal{L}(\rho_G A^*)$ for any operator $A$, where $\rho_G=e^{-\beta H}/\mathrm{Tr}[e^{-\beta H}]$ is the Gibbs state with an inverse temperature $\beta$ and $*$ denotes the complex conjugation \cite{Chetrite-12}.
If the system satisfies the detailed balance condition, it relaxes to the equilibrium state described by $\rho_G$.
The on-site dephasing $L_l=b_l^\dag b_l$ satisfies this condition with $\beta=0$.
In Appendices \ref{appendix:Bose_Hubbard_model} and \ref{appendix:hard_core_bosons_with_next_nearest_neighbor_hopping}, we present numerical results for other models, i.e., dephasing Bose-Hubbard model and dephasing hard-core bosons with next-nearest-neighbor hopping.
In the Bose-Hubbard model, unlike the hard-core model discussed above, a single lattice site can be occupied by multiple particles.
The Liouvillian eigenmodes of this model also show a hierarchical structure depending on the numbers of interchain bound states (incoherentons) and {\it intrachain} bound states, and the QC gap in the spectrum closes at a certain value of the hopping amplitude.
The results presented in Appendices \ref{appendix:Bose_Hubbard_model} and \ref{appendix:hard_core_bosons_with_next_nearest_neighbor_hopping} support the universality of our scenario.

The impact of dissipation violating the detailed balance condition on the hierarchical structure of spectra is nontrivial. 
In Sec.~\ref{sec:effect_of_particle_loss}, we explore the scenario involving particle loss and gain and demonstrate that the hierarchical structure is preserved.
However, when dissipation induces a current, the hierarchical structure depicted in Fig.~\ref{Fig-hierarchy} can undergo substantial changes.
A simple example of such dissipation is realized by the Lindblad operators $L_l=b_{l+1}^\dag b_l$, which describe stochastic hopping induced by external driving \cite{Haga-21, Eisler-11, Temme-12}.
In this case, we observe that the spectrum possesses a topologically distinct structure from the striped band structure shown in Fig.~\ref{Fig-spec-MP}, which will be elaborated in a future publication.

\section{Exact many-body solution}
\label{sec:exact_many_body_solution}

While we have shown the numerical results for the many-body system in the previous section, analytical solutions are also possible since the dissipative hard-core boson model can be exactly solved by the Bethe ansatz \cite{Medvedyeva-16}.
In the following, we derive an exact many-body solution of the model to show that some Liouvillian eigenmodes have higher-order incoherentons and that they exhibit the deconfinement transition at a critical hopping amplitude.
First, it should be noted that the hard-core bosons can be mapped to the XX chain,
\begin{equation}
	H = -\frac{J}{2} \sum_{l=1}^{L} (\sigma^x_{l} \sigma^x_{l+1} + \sigma^y_{l} \sigma^y_{l+1}),
	\label{H_XX_chain}
\end{equation}
by using the following identification between the bosonic operators $b_l$ and the Pauli matrices $\sigma_l^{\mu}$:
\begin{equation}
	b_l^{\dag} = \sigma^+_l = \frac{\sigma^x_l + i\sigma^y_l}{2}, \quad b_l = \sigma^-_l = \frac{\sigma^x_l - i\sigma^y_l}{2}.
\end{equation}
Thus, our model can be regarded as a dissipative spin chain with Lindblad operators
\begin{equation}
	L_l = \sqrt{\gamma} \sigma^+_l \sigma^-_l = \frac{\sqrt{\gamma} (\sigma^z_l + 1)}{2} \quad (l = 1, ..., L).
	\label{L_dephasing_spin}
\end{equation}
The boundary condition is set to be periodic: $\sigma_{L+1}^\mu=\sigma_1^\mu\ (\mu=x,y,z)$.
In this section, we assume that the system size $L$ is even.

Employing the Jordan-Wigner transformation, we introduce a fermion annihilation operator
\begin{equation}
	c_l := \exp \left[ \frac{i\pi}{2}\sum_{j=1}^{l-1}(\sigma_j^z+1) \right] \sigma_l^-,
\end{equation}
which satisfies the anticommutation relations $\{ c_l,c_m^\dag\}=\delta_{lm}$ and $\{ c_l,c_m \}=0$, and rewrite the Hamiltonian and the Lindblad operator as
\begin{equation}
	H = -J\sum_{l=1}^{L-1}(c_l^\dag c_{l+1}+c_{l+1}^\dag c_l)-J e^{i\pi (N+1)} (c_{L}^\dag c_1+c_1^\dag c_L),
	\label{H_nonint_fermi}
\end{equation}
and
\begin{equation}
	L_l = \sqrt{\gamma}c_l^\dag c_l \quad (l = 1, ..., L),
	\label{L_dephasing_fermi}
\end{equation}
where $N = \sum_{l=1}^L c_l^\dag c_l$ is the number of fermions.

We here use a pseudospin index $\sigma=\up,\down$ to express each chain in the ladder representation of the Liouvillian, and denote the annihilation operator of a fermion on each chain as $c_{l,\sigma}$.
Then, the ladder representation of the Liouvillian (multiplied by $i$) reads
\begin{align}
	i\tilde{\mathcal{L}} &= -\sum_{l=1}^{L-1} \sum_{\sigma=\up,\down} J_\sigma(c_{l,\sigma}^\dag c_{l+1,\sigma}+c_{l+1,\sigma}^\dag c_{l,\sigma}) \notag\\
	& \quad -\sum_{\sigma=\up,\down} J_\sigma e^{i\pi (N_{\sigma}+1)}(c_{L,\sigma}^\dag c_{1,\sigma}+c_{1,\sigma}^\dag c_{L,\sigma}) \notag\\
	& \quad +i\gamma \sum_{l=1}^L c_{l,\up}^\dag c_{l,\up}c_{l,\down}^\dag c_{l,\down} -\frac{i\gamma}{2} (N_\up + N_\down),
\end{align}
where $J_\up = -J_\down = J$, and $N_\sigma := \sum_{l=1}^L c_{l,\sigma}^\dag c_{l,\sigma}$ is the number of fermions on each chain.
The dependence of the hopping amplitude $J_\sigma$ on the pseudospin can be removed by a unitary transformation
\begin{equation}
	U^\dag c_{l,\up} U=c_{l,\up}, \quad U^\dag c_{l,\down} U=(-1)^l c_{l,\down},
	\label{U_J_flip}
\end{equation}
and the transformed Liouvillian $iU^\dag \tilde{\mathcal{L}} U$ is equivalent to the Hubbard model with an imaginary interaction strength and an imaginary chemical potential \cite{Medvedyeva-16, Ziolkowska-20}.
Since we focus on the Liouvillian $\mathcal{L}$ of an $N$-particle system, we assume $N_{\up}=N_{\down}=N$.
Note that if $N$ is odd (even), the periodic (antiperiodic) boundary condition is imposed on fermions on each chain.
For this reason, we hereafter consider the non-Hermitian Hubbard model under a flux
\begin{align}
	\mathcal{H}_{\phi} &:= -J \sum_{l=1}^L\sum_{\sigma=\up,\down} (e^{-i\phi/L}c_{l,\sigma}^\dag c_{l+1,\sigma}+e^{i\phi/L}c_{l+1,\sigma}^\dag c_{l,\sigma}) \notag\\
	& \quad +i\gamma \sum_{l=1}^L c_{l,\up}^\dag c_{l,\up}c_{l,\down}^\dag c_{l,\down} -\frac{i\gamma}{2} (N_\up + N_\down),
	\label{H_imaginary_Hubbard}
\end{align}
with the periodic boundary condition, $c_{L+1,\sigma}=c_{1,\sigma}$.
If $N$ is odd or even, $\phi$ is set to $0$ or $\pi$, respectively.
Instead of twisting the particular bond between the sites $l=1$ and $L$, we have introduced a uniform complex hopping amplitude to ensure the translation invariance of the model.
The original model with the twist at the particular bond is obtained by performing the gauge transformation $c_{l,\sigma}\to \exp(i\phi l/L)c_{l,\sigma}$ on $\mathcal{H}_{\phi}$.

The Hubbard model \eqref{H_imaginary_Hubbard} has the spin SU(2) symmetry $[\mathcal{H}_{\phi},S^\mu]=0 \ (\mu=+,-,z)$, where
\begin{gather}
	S^+ = \sum_{l=1}^L c_{l,\up}^\dag c_{l,\down}, \quad S^- = (S^+)^\dag, \\
	S^z = \frac{1}{2}\sum_{l=1}^L (c_{l,\up}^\dag c_{l,\up}-c_{l,\down}^\dag c_{l,\down}).
\end{gather}
In addition, for $\phi=0$ or $\pi$, the model \eqref{H_imaginary_Hubbard} possesses the $\eta$-SU(2) symmetry \cite{Yang-89, Yang-90} $[\mathcal{H}_{\phi},\eta_\phi^\mu]=0\ (\mu=+,-)$ and $[\mathcal{H}_{\phi},\eta^z]=0$, where
\begin{gather}
	\eta_\phi^+ = \sum_{l=1}^L(-1)^l e^{2i\phi l/L}c_{l,\up}^\dag c_{l,\down}^\dag, \quad \eta_\phi^- = (\eta_\phi^+)^\dag,\\
	\eta^z = \frac{1}{2}\sum_{l=1}^L(c_{l,\up}^\dag c_{l,\up}+c_{l,\down}^\dag c_{l,\down}-1),
	\label{eta_z}
\end{gather}
which are generalized in order to incorporate the case of antiperiodic boundary condition \cite{Nishino-04}. 
These symmetries are called the weak symmetry of the Lindblad equation and lead to a block-diagonal structure of the Liouvillian \cite{Buca-12, Albert-14}.

The one-dimensional Hubbard model \eqref{H_imaginary_Hubbard} is exactly solvable by using the Bethe ansatz method \cite{Lieb-68, Essler}. 
The Yang-Baxter integrability of the Hubbard model is preserved even when the interaction strength is complex valued \cite{Medvedyeva-16, Ziolkowska-20, Nakagawa-21}.
The Bethe equations for the Hubbard model \eqref{H_imaginary_Hubbard} are given by \cite{Lieb-68, Essler, Shastry-90}
\begin{gather}
	e^{ik_a L-i\phi}=\prod_{\alpha=1}^{N_{\down}} \frac{\Lambda_\alpha-\sin k_a-iu}{\Lambda_\alpha-\sin k_a+iu},\label{Bethe_1}\\
	\prod_{a=1}^{N_{\up}+N_{\down}} \frac{\Lambda_\alpha-\sin k_a-iu}{\Lambda_\alpha-\sin k_a+iu}=-\prod_{\beta=1}^{N_{\down}} \frac{\Lambda_\alpha-\Lambda_\beta-2iu}{\Lambda_\alpha-\Lambda_\beta+2iu},
	\label{Bethe_2}
\end{gather}
where $k_a\ (a=1, ... ,N_{\up}+N_{\down})$ is a quasimomentum, $\Lambda_\alpha\ (\alpha=1, ... ,N_{\down})$ is a spin rapidity, and $u=i\gamma/(4J)$ is the pure-imaginary dimensionless interaction strength.
An eigenvalue $\lambda$ of $\mathcal{L}$ is obtained from a solution of the Bethe equations as 
\begin{equation}
	\lambda=-\frac{\gamma(N_{\up}+N_{\down})}{2} + 2iJ\sum_{a=1}^{N_{\up}+N_{\down}} \cos k_a.
	\label{Bethe_eigenvalue}
\end{equation}

A Bethe wave function constructed from a solution of the Bethe equations \eqref{Bethe_1} and \eqref{Bethe_2} provides a Bethe eigenstate of the Hubbard model \eqref{H_imaginary_Hubbard}, which can be interpreted as a Liouvillian eigenmode in the ladder representation.
Since Bethe eigenstates satisfy the highest-weight (lowest-weight) condition of the spin SU(2) ($\eta$-SU(2)) symmetry, a general eigenstate can be obtained by acting $S^-$ or $\eta_\phi^+$ on a Bethe eigenstate \cite{Essler, Essler-91, Essler-92-1, Essler-92-2, Nishino-04}.
Noting the commutation relation $[\mathcal{H}_{\phi},\eta_\phi^+]=0$, the steady state, $\mathcal{H}_{\phi}|\Psi_0)=0$, is given by
\begin{equation}
	|\Psi_0) = (\eta_\phi^+)^N|\mathrm{v}),
\end{equation}
where $|\mathrm{v})$ is the vacuum state of fermions \cite{Medvedyeva-16}.
Since $\eta_\phi^+$ creates a bound pair of particles of the two chains, the steady state is composed of $N$ first-order incoherentons.
With the unitary transformation \eqref{U_J_flip}, the state $|\Psi_0)$ is equivalent to the infinite-temperature state of the $N$-particle sector in the original problem.
We note that an incoherenton created by $\eta_\phi^+$ is localized on a rung of the ladder, while an incoherenton in an excited eigenmode can have a nonzero confinement length.

\begin{figure}
	\centering
	\includegraphics[width=0.45\textwidth]{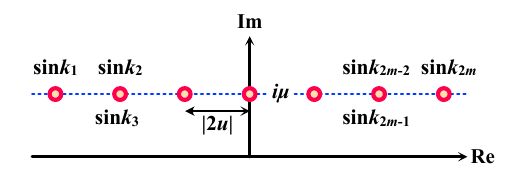}
	\caption{Schematic illustration of a $k$-$\Lambda$ string solution in the complex plane of $\sin k$.}
	\label{Fig-string-solution}
\end{figure}

The Bethe equations \eqref{Bethe_1} and \eqref{Bethe_2} for sufficiently large $L$ allow $k$-$\Lambda$ string solutions, in which a part of quasimomenta and spin rapidities forms a string pattern \cite{Essler, Takahashi-72}.
Since a $k$-$\Lambda$ string solution of length $2m$ describes a bound state made of $m$ spin-up particles and $m$ spin-down ones \cite{Essler-92-2}, it offers an $m$th-order incoherenton.
A $k$-$\Lambda$ string of length $2m$ is composed of $2m$ quasimomenta $k_{1},..., k_{2m}$ and $m$ spin rapidities $\Lambda_{1},...,\Lambda_{m}$ that satisfy 
\begin{equation}
	\begin{split}
		k_{1}&=\arcsin[i\mu+miu], \\
		k_{2}&=\pi-\arcsin[i\mu+(m-2)iu], \\
		k_{3}&=\arcsin[i\mu+(m-2)iu], \\
		&\cdots \\
		k_{2m-2}&=\pi-\arcsin[i\mu-(m-2)iu], \\
		k_{2m-1}&=\arcsin[i\mu-(m-2)iu], \\
		k_{2m}&=\pi-\arcsin[i\mu-miu],
	\end{split}
	\label{string_solution}
\end{equation}
and
\begin{equation}
	\Lambda_{j} = i\mu+(m-2j+1)iu,
\end{equation}
where $\mu \in \mathbb{R}$ is the center of the $k$-$\Lambda$ string (see Fig.~\ref{Fig-string-solution}), and we set the branch so that $-\pi/2 < \mathrm{Re}[\arcsin x] \leq \pi/2$ \cite{Medvedyeva-16, Ziolkowska-20}.

The deconfinement of incoherentons is diagnosed from the disappearance of a $k$-$\Lambda$ string solution.
Let us consider a situation in which all quasimomenta and spin rapidities form a single length-$2m$ $k$-$\Lambda$ string solution of the Bethe equations for $N_{\up}=N_{\down}=m$.
By multiplying the Bethe equations \eqref{Bethe_1} for $a=1,...,N$, we obtain
\begin{equation}
	\exp\left[i\sum_{a=1}^{2m} k_a L\right]=1,
	\label{momentum_sum_rule}
\end{equation}
where we have used Eq.~\eqref{Bethe_2} and the fact that $\phi$ is set to $0$ or $\pi$.
Since Eqs.~\eqref{string_solution} and \eqref{momentum_sum_rule} imply that $k_1+k_{2m}$ is real, we can set
\begin{equation}
	k_1 = p-i\kappa, \quad k_{2m} = q+i\kappa \quad (p,q \in \mathbb{R}, \: \kappa >0),
	\label{k_1_k_2m}
\end{equation}
without loss of generality.
From Eq.~\eqref{string_solution}, $k_1$ and $k_{2m}$ satisfy
\begin{equation}
	\sin k_1 - \sin k_{2m} = 2m iu,
\end{equation}
which leads to 
\begin{equation}
	\begin{split}
		(\sin p-\sin q)\cosh\kappa=&-2m\mathrm{Im}[u], \\
		(\cos p+\cos q)\sinh\kappa=&-2m\mathrm{Re}[u].
	\end{split}
	\label{kappa_equation}
\end{equation}
For our case with $u=i\gamma/(4J)$, the solution of Eq.~\eqref{kappa_equation} is given by
\begin{equation}
	q=p+\pi,
	\label{p_q_solution}
\end{equation}
and
\begin{equation}
	\cosh\kappa = -\frac{m\gamma}{4J\sin p}.
	\label{kappa_solution}
\end{equation}
For a given $p=\mathrm{Re}[k_1]$, the imaginary part $\kappa$ is obtained from Eq.~\eqref{kappa_solution}, and by substituting $k_1=p-i\kappa$ into $\sin k_1 = i\mu + m iu$, the center of the string can be calculated as
\begin{equation}
	\mu = - \cos p \sinh \kappa.
\end{equation}
Since $\cosh\kappa>1$, the solution of Eq.~\eqref{kappa_solution} exists only for
\begin{equation}
	-\frac{m\gamma}{4J} < \sin p < 0.
	\label{solution_condition}
\end{equation}
For $m\gamma/(4J)>1$, arbitrary $-\pi < p < 0$ satisfies the above condition.
However, for $m\gamma/(4J)\leq1$, the string solution for some $p$ around $-\pi/2$ disappears, indicating the deconfinement of the string solution at
\begin{equation}
	J_c^{(m)} = \frac{m\gamma}{4}.
	\label{J_c_MP}
\end{equation}
The eigenvalue given by Eq.~\eqref{Bethe_eigenvalue} can be calculated from Eqs.~\eqref{string_solution} and \eqref{kappa_solution} as
\begin{equation}
	\lambda=-m\gamma + \sqrt{m^2 \gamma^2 - 16 J^2 \sin^2p}.
	\label{lambda_inc_MP_p}
\end{equation}
Thus, the deconfinement of the $m$th-order incoherenton occurs at Liouvillian eigenmodes with eigenvalues $\lambda=-m\gamma$.
We here define the total momentum $K$ of the $m$th-order incoherenton by
\begin{equation}
	K = \sum_{j=1}^{2m} k_j - m\pi,
\end{equation}
where we have introduced the phase shift $m\pi$ to compensate the unitary transformation \eqref{U_J_flip}.
From Eqs.~\eqref{string_solution}, \eqref{k_1_k_2m}, and \eqref{p_q_solution}, the total momentum reads $K=2p$.
Thus, in terms of $K$, Eq.~\eqref{lambda_inc_MP_p} is rewritten as
\begin{equation}
	\lambda=-m\gamma + \sqrt{m^2 \gamma^2 - 16 J^2 \sin^2(K/2)}.
	\label{lambda_inc_MP}
\end{equation}

The $\eta$-SU(2) symmetry $[\mathcal{H}_{\phi},\eta_\phi^+]=0$ yields an eigenstate of $\mathcal{H}_{\phi}$ in the sector of $N_\up=N_\down=N\geq m$ as
\begin{equation}
	|\Psi) = (\eta_\phi^+)^{N-m} |\psi_{2m}),
\end{equation}
where $|\psi_{2m})$ is a length-$2m$ $k$-$\Lambda$ string solution of the Bethe equations.
Since $\eta_\phi^+$ generates an on-site pair of particles with opposite spins, $|\Psi)$ is interpreted as a state that involves an $m$th-order incoherenton and $N-m$ first-order incoherentons.
As the action of $\eta_\phi^+$ does not change the eigenvalue, the deconfinement transition of $|\Psi)$ occurs in Liouvillian eigenmodes with eigenvalues near $\lambda=-m\gamma$.
Thus, we conclude that the $N$-body dissipative dynamics governed by the Liouvillian $\mathcal{L}$ shows the deconfinement transition of $m$th-order incoherentons for $m=1,2,...,N$.

It is worth noting that the deconfinement of bound states does not occur in the ordinary Hermitian Hubbard model with real $u$.
In this case, the solution of Eq.~\eqref{kappa_equation} is given by
\begin{equation}
	q=p,
\end{equation}
and
\begin{equation}
	\sinh\kappa=-\frac{m\mathrm{Re}[u]}{\cos p},
\end{equation}
which can be satisfied for any value of $p$ because the range of $\sinh\kappa$ is $(-\infty, \infty)$. 
Thus, the deconfinement transition in the string solution is unique to the dissipative system which can be mapped to the non-Hermitian Hubbard model with an imaginary interaction strength.

\section{Effects of particle loss and gain}
\label{sec:effect_of_particle_loss}

In the system of hard-core bosons subject to on-site dephasing, the total number of particles is conserved during time evolution. 
A question arises as to whether the incoherenton framework is applicable to situations where particle exchange with the environment occurs. 
Such situations appear in cases like driven optical cavities \cite{Biondi-17} and exciton-polariton systems \cite{Sieberer-16}. 
In the following, we confirm that the incoherenton framework essentially holds in the presence of particle loss and gain, at least for small loss and gain rates.

\begin{figure}
	\centering
	\includegraphics[width=0.45\textwidth]{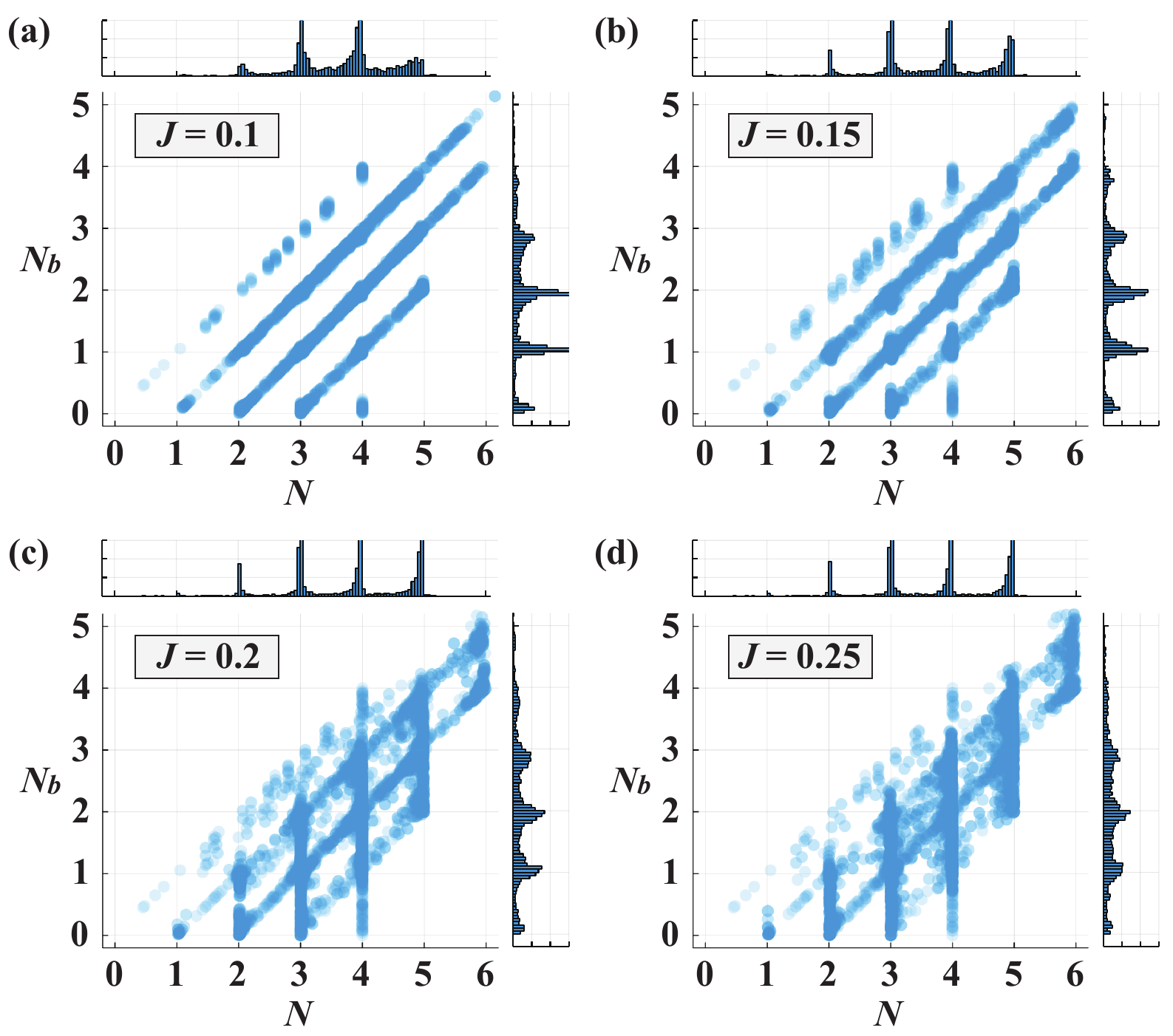}
	\caption{Scatter plots of the particle number $N$ and the number $N_b$ of interchain bound states for eigenmodes of dephasing hard-core bosons with loss and gain.
		The system size is $L=8$.
		Panels (a) through (d) represent varying hopping amplitudes: (a) $J=0.1$, (b) $J=0.15$, (c) $J=0.2$, and (d) $J = 0.25$.
		The rates of dephasing, loss, and gain are set to $\gamma=1$, $\kappa_1=0.02$, and $\kappa_2=0.01$, respectively.
		Histograms of $N$ and $N_b$ are also included.
		For smaller values of $J$, clusters are observed at integer values of $N$ and $N_b$.
		As $J$ increases, these clusters stretch along the $N_b$ axis, eventually merging around $J \simeq 0.2$.}
	\label{Fig-nb-loss}
\end{figure}

We incorporate particle loss and gain by considering the following Liouvillian:
\begin{align}
	\mathcal{L}(\rho) = &-i[H, \rho] + \sum_l \gamma \left(n_l \rho n_l - \frac{1}{2} \{ n_l^2, \rho\}\right) \nonumber \\
	&+ \sum_l \kappa_1 \left(b_l \rho b_l^\dag - \frac{1}{2} \{ b_l^\dag b_l, \rho\}\right) \nonumber \\
	&+ \sum_l \kappa_2 \left(b_l^\dag \rho b_l - \frac{1}{2} \{ b_l b_l^\dag, \rho\}\right),
	\label{L_loss_gain}
\end{align}
where $H$ is the Hamiltonian \eqref{H_hard_core_boson} of hard-core bosons, and $\kappa_1$ and $\kappa_2$ represent the rates of particle loss and gain, respectively.
When the hard-core boson model is mapped to a spin model, the loss and gain terms in Eq.~\eqref{L_loss_gain} correspond to dissipative processes that flip spins down and up at rates $\kappa_1$ and $\kappa_2$, respectively.
Note that the loss and gain terms in Eq.~\eqref{L_loss_gain} mix sectors with different particle numbers.
The ladder representation of Eq.~\eqref{L_loss_gain} can be expressed as
\begin{equation}
	\tilde{\mathcal{L}} = \tilde{\mathcal{L}}_d + \tilde{\mathcal{L}}_1 + \tilde{\mathcal{L}}_2 - \kappa_1 N - \kappa_2 (L-N),
	\label{L_loss_gain_ladder}
\end{equation}
\begin{equation}
	\tilde{\mathcal{L}}_1 := \kappa_1 \sum_l b_{l,+} b_{l,-}, \quad \tilde{\mathcal{L}}_2 := \kappa_2 \sum_l b_{l,+}^\dag b_{l,-}^\dag,
\end{equation}
where $\tilde{\mathcal{L}}_d$ is the Liouvillian \eqref{Liouvillian_dissipative_boson_ladder} of dephasing hard-core bosons, and $N=\sum_l (n_{l,+} + n_{l,-})/2$ is the total particle number.

Firstly, let us consider the scenario with particle loss but without gain, i.e., $\kappa_2=0$.
Importantly, $\tilde{\mathcal{L}}$ can be expressed in a block-upper-triangular form because $\tilde{\mathcal{L}}_1$ reduces the particle number but does not increase it.
The eigenvalues of a block-triangular matrix are given by those of its diagonal blocks.
Consequently, the spectrum of $\tilde{\mathcal{L}}$ is simply the union of spectra of each particle sector:
\begin{equation}
	\mathrm{spec}\left(\tilde{\mathcal{L}} \right) = \bigcup_{N=0}^L \mathrm{spec}\left(\tilde{\mathcal{L}}_d^{(N)}-\kappa_1 N \right),
	\label{spectra_loss}
\end{equation}
where $\tilde{\mathcal{L}}_d^{(N)}$ represents $\tilde{\mathcal{L}}_d$ in the $N$-particle sector.
This is a general property of Liouvillian with loss but without gain \cite{Torres-14, Yoshida-20, Nakagawa-21}.
Equation \eqref{spectra_loss} implies that the singular behavior of the spectra of $\tilde{\mathcal{L}}_d$, linked to the deconfinement of incoherentons, is directly transferred to the spectra with loss.
Thus, the presence of particle loss does not affect the incoherenton picture.

\begin{figure}
	\centering
	\includegraphics[width=0.45\textwidth]{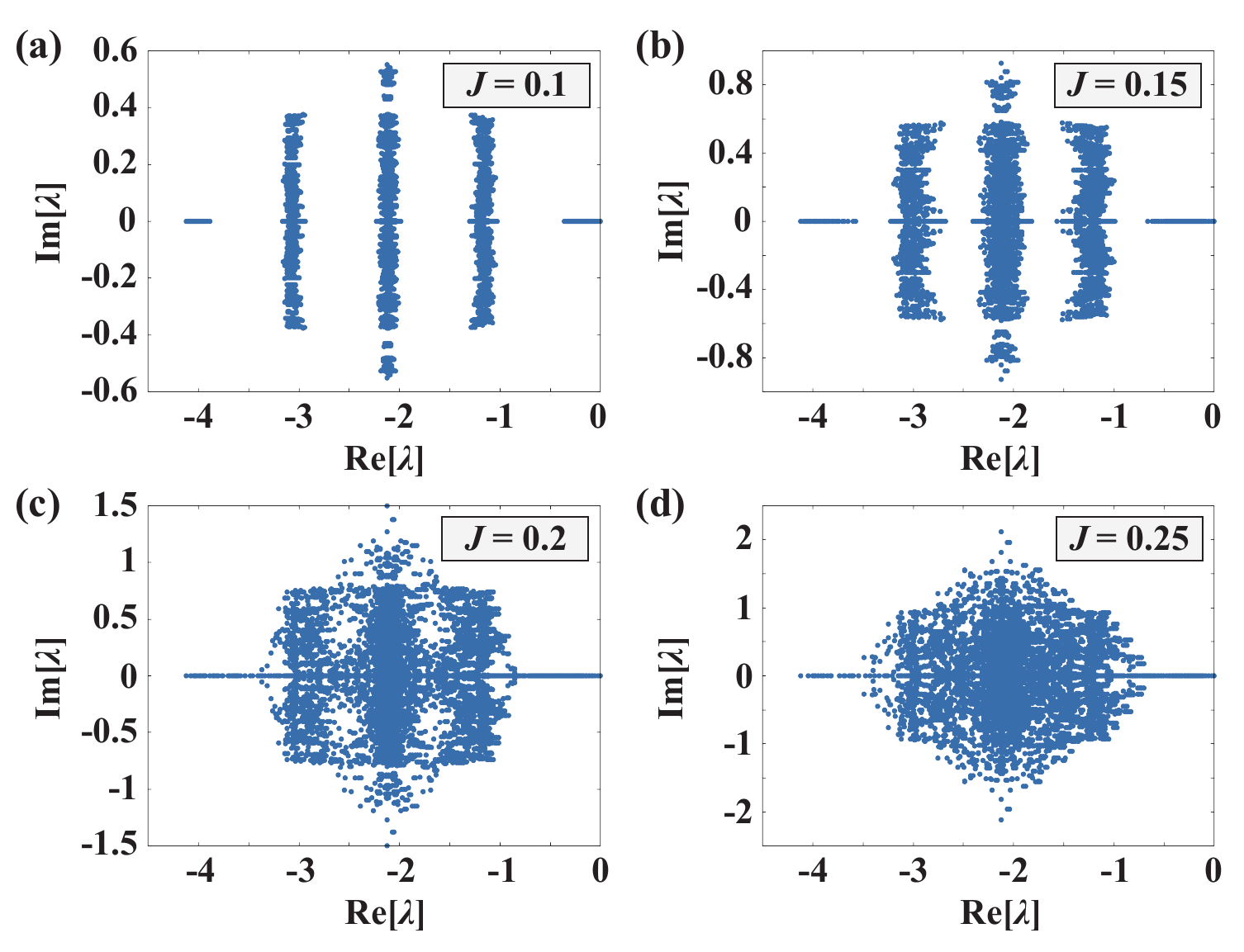}
	\caption{Liouvillian spectra of dephasing hard-core bosons with loss and gain.
		The system size is $L=8$.
		Panels (a) through (d) represent varying hopping amplitudes: (a) $J=0.1$, (b) $J=0.15$, (c) $J=0.2$, and (d) $J = 0.25$.
		The rates of dephasing, loss, and gain are set to $\gamma=1$, $\kappa_1=0.02$, and $\kappa_2=0.01$, respectively.
		The merging of spectral bands becomes apparent around $J \simeq 0.2$.}
	\label{Fig-spec-loss}
\end{figure}

Next, let us consider the situation with both particle loss and gain.
We numerically diagonalize $\tilde{\mathcal{L}}$ in the subspace with $\sum_l n_{l,+}=\sum_l n_{l,-}$.
We calculate the total particle number $N$ and the number $N_b$ of interchain bound states, as defined by Eq.~\eqref{def_N_b}, for each eigenmode.
Figure \ref{Fig-nb-loss} shows the scatter plot of $(N, N_b)$ for small loss and gain rates, $\kappa_1, \kappa_2 \ll J, \gamma$.
For smaller values of $J$, a significant number of dots cluster at integer values of $N$ and $N_b$ [see Fig.~\ref{Fig-nb-loss}(a)], as seen in sharp peaks in the histograms of $N$ and $N_b$.
As $J$ increases, these clusters stretch along the $N_b$ axis, leading to a widening of the peaks in the histogram of $N_b$.
Eventually, they merge around $J \simeq 0.2$.
Note that the sharp peaks in the histogram of $N$ remain unaffected throughout this process.
This implies that for almost eigenmodes, the mixing among sectors with different particle numbers is insignificant.
The merging of bands along the $N_b$ axis suggests the deconfinement of incoherentons [see Fig.~\ref{Fig-spec-MP}(b)].
The QC gap closing in the spectrum can also be observed in Fig.~\ref{Fig-spec-loss}.
Note that the data in Figs.~\ref{Fig-nb-loss} and \ref{Fig-spec-loss} primarily deal with situations where loss and gain rates are sufficiently smaller than the dephasing rate.
Extending the incoherenton framework to the case with strong loss and gain deserves further study.

\section{Effective description of many-body decoherence}
\label{sec:many_body_decoherence}

In Fig.~\ref{Fig-hierarchy} of Sec.~\ref{sec:hierarchy}, we have shown that the Liouvillian eigenmodes are arranged in a hierarchy characterized by incoherentons.
Each group in this hierarchy has a different decay rate and is separated from each other by the QC gap when the dissipation dominates.
In this section, we discuss the consequences of this hierarchy of Liouvillian eigenmodes for the process of quantum decoherence.
First, we introduce multi-order quantum coherence as a quantitative measure of how many incoherentons a given density matrix contains.
The time evolution of the quantum coherence is investigated for the dissipative hard-core boson model by numerically solving the master equation.
We argue that the decay process of the quantum coherence can be understood in terms of the production, diffusion, and localization of incoherentons.

\subsection{Multi-order quantum coherence}

First, the general concept of ``quantum coherence" is outlined according to Refs.~\cite{Glauber-63-1} and \cite{Glauber-63-2}.
We consider the dissipative hard-core boson model introduced in Sec.~\ref{sec:hard_core_bosons}.
Similarly to Eq.~\eqref{def_G_mP}, we define the one-particle reduced density matrix $G^{(1)}$ by
\begin{equation}
	G^{(1)}_{l_1, l_2} := \mathrm{Tr}[b_{l_1} \rho b_{l_2}^{\dag}],
	\label{def_G_1P}
\end{equation}
where $b_l^{\dag}$ and $b_l$ are the creation and annihilation operators of a boson at site $l$.
Then, the state $\rho$ is said to have the first-order coherence if $G^{(1)}$ satisfies the following relation:
\begin{equation}
	G^{(1)}_{l_1, l_2} = \left[ G^{(1)}_{l_1, l_1} G^{(1)}_{l_2, l_2} \right]^{1/2} = \left[ \langle n_{l_1} \rangle \langle n_{l_2} \rangle \right]^{1/2},
	\label{1st_coherence_condition}
\end{equation}
for any $l_1$ and $l_2$, where $n_l = b_l^{\dag} b_l$ is the number operator at site $l$.
In other words, a strong correlation between distantly separated points exists in a coherent state.
Similarly, the two-particle reduced density matrix $G^{(2)}$ is defined by
\begin{equation}
	G^{(2)}_{l_1, l_2; l_3, l_4} := \mathrm{Tr}[b_{l_1} b_{l_2} \rho b_{l_3}^{\dag} b_{l_4}^{\dag}].
\end{equation}
The state $\rho$ is said to have the second-order coherence if $G^{(1)}$ and $G^{(2)}$ satisfy Eq.~\eqref{1st_coherence_condition} and
\begin{equation}
	G^{(2)}_{l_1, l_2; l_3, l_4} = \prod_{i=1}^4 \left[ G^{(1)}_{l_i, l_i} \right]^{1/2} = \prod_{i=1}^4 \langle n_{l_i} \rangle^{1/2},
	\label{2nd_coherence_condition}
\end{equation}
for any $l_1$, $l_2$, $l_3$, and $l_4$.
The notion of the $s$th-order coherence ($s\geq3$) can also be defined from the $s$-particle reduced density matrix $G^{(s)}$ in a similar manner.

We define the amount of the first-order coherence by
\begin{equation}
	\chi_1 := \sum_{l_1, l_2} |G^{(1)}_{l_1, l_2}| (1-\delta_{l_1, l_2}),
	\label{def_chi_1}
\end{equation}
which simply measures the amount of off-diagonal components of $G^{(1)}$.
Since the steady state of the model is the infinite-temperature state $\rho_{\mathrm{ss}}=D^{-1}I$ due to dephasing, all off-diagonal components of $\rho$ vanish in the long-time limit, and then we have $\lim_{t \to \infty} \chi_1(t) = 0$.
If the state $\rho$ has the first-order coherence, e.g., a Bose-condensed pure state, the amount of the first-order coherence is given by $\chi_1 \propto NL$ because $|G^{(1)}_{l_1, l_2}| \propto N/L$ from Eq.~\eqref{1st_coherence_condition}.
It should be noted that the expectation value of an arbitrary one-body observable,
\begin{equation}
	O^{(1)} = \sum_{l_1, l_2} O^{(1)}_{l_1, l_2} b_{l_1}^{\dag} b_{l_2},
	\label{one_body_observable}
\end{equation}
can be written as
\begin{equation}
	\langle O^{(1)} \rangle = \mathrm{Tr}[O^{(1)} \rho] = \sum_{l_1, l_2} O^{(1)}_{l_1, l_2} G^{(1)}_{l_2, l_1}.
	\label{O_1_expect}
\end{equation}
In particular, $G^{(1)}$ is related to the momentum distribution of particles, which is accessible by time-of-flight experiments in ultracold atomic gases.

We also define the amount of the second-order coherence $\chi_2$ by taking the summation of $G^{(2)}_{l_1, l_2; l_3, l_4}$ over all off-diagonal indices:
\begin{align}
	\chi_2 &:= \sum_{\{ l_i \}} |G^{(2)}_{l_1, l_2; l_3, l_4}| (1-\delta_{l_1, l_3}) \notag \\
	& \quad \times (1-\delta_{l_1, l_4}) (1-\delta_{l_2, l_3}) (1-\delta_{l_2, l_4}).
	\label{def_chi_2}
\end{align}
Note that $G^{(2)}_{l, l; l_3, l_4}=G^{(2)}_{l_1, l_2; l, l}=0$ due to the hard-core condition.
If the state $\rho$ has the second-order coherence, the amount of the second-order coherence is given by $\chi_2 \propto N^2L^2$.
The amount of the higher-order coherence $\chi_s \ (s=3,4,...)$ can also be defined from the $s$-particle reduced density matrix $G^{(s)}$ in a similar manner.

When dissipation is dominant ($J \ll \gamma$), the Liouvillian eigenmodes are arranged in $N+1$ bands with different decay rates $s \gamma \ (s=0,...,N)$, as shown in Fig.~\ref{Fig-hierarchy} (a).
We denote the set of eigenmodes that belong to each band as $\{\rho_{0,\alpha}\}$, $\{\rho_{1,\alpha}\}$, ..., $\{\rho_{N,\alpha}\}$.
Each $\rho_{s,\alpha}$ involves $s$ deconfined pairs in the ladder representation.
Then, the eigenmode expansion of the density matrix can be rearranged as
\begin{equation}
	\rho = \sum_{\alpha \in \mathcal{S}_0} c_{0,\alpha} \rho_{0,\alpha} + \sum_{\alpha \in \mathcal{S}_1} c_{1,\alpha} \rho_{1,\alpha} + \cdots + \sum_{\alpha \in \mathcal{S}_N} c_{N,\alpha} \rho_{N,\alpha},
	\label{rho_hierarchical_expansion}
\end{equation}
where $\mathcal{S}_r$ denotes the set of indices for $\{ \rho_{r,\alpha} \}$.
Note that the steady state $\rho_{\mathrm{ss}}=D^{-1}I$ belongs to $\{\rho_{0,\alpha}\}$.
We refer to Eq.~\eqref{rho_hierarchical_expansion} as the hierarchical expansion of the density matrix.
The decay rate of each $\rho_{s,\alpha}$ is given by $s \gamma + O((J/\gamma)^2)$.
From the definition, the dominant contribution to the amount of the $s$th-order coherence $\chi_s$ comes from $\rho_{s,\alpha}$, because the $s$-particle reduced density matrix $G^{(s)}$ of eigenmodes with $s$ deconfined pairs has large off-diagonal components.
Thus, when $J \ll \gamma$, $\chi_s$ initially decays as
\begin{equation}
	\chi_s(t) \sim e^{-s \gamma t}.
\end{equation}
In the next subsection, it is argued that the initial decay of $\chi_s$ is due to the generation of incoherentons and that the relaxation of $\chi_s$ at long times is characterized by the localization and diffusion of incoherentons.

\subsection{Numerical results for relaxation of quantum coherence}

\begin{figure}
	\centering
	\includegraphics[width=0.45\textwidth]{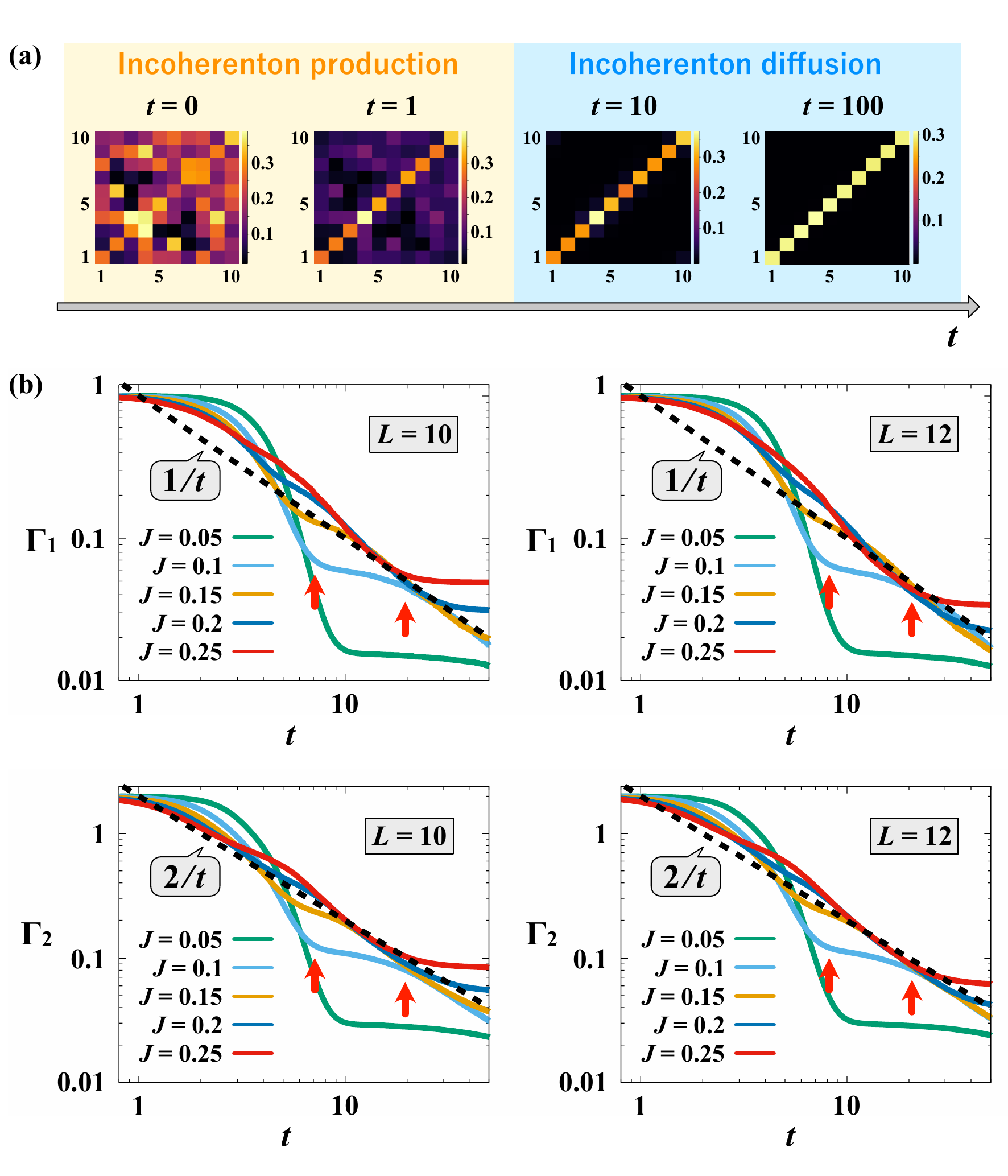}
	\caption{Relaxation of the reduced density matrix and the amount of the quantum coherence.
		(a) Time evolution of the absolute values of $G^{(1)}$ with $\gamma=1$ and $J=0.1$.
		The system size is $L=10$ and the particle number is $N=3$.
		The initial state is a random pure state.
		In the regime of production of incoherentons, a fast decay of the off-diagonal components can be seen.
		In the diffusion regime of incoherentons, on the other hand, a slow diffusion of the diagonal components occurs, leading to an infinite-temperature state in the long-time limit.
		To highlight the variation of the off-diagonal components, they are multiplied by a factor of $4$, i.e., these plots represent $(4-3\delta_{l,m})|G^{(1)}_{l,m}|$.
		(b) Time evolution of $\Gamma_1$ and $\Gamma_2$ with $\gamma=1$ and $J=0.05$, $0.1$, $0.15$, $0.2$, and $0.25$ from bottom to top at $t=10$.
		The horizontal and vertical axes are plotted on a logarithmic scale.
		$\Gamma_s$ is calculated from $\chi_s$ averaged over $100$ initial random states.
		The system sizes are $L=10$ and $12$, and the particle number is $N=3$.
		The dotted lines show $1/t$ for $\Gamma_1$ and $2/t$ for $\Gamma_2$.
		The arrows indicate the beginning and end of the localization regime ($\tau_1$ and $\tau_2$) of incoherentons for $J=0.1$.}
	\label{Fig-coherence-Gamma-numerical}
\end{figure}

By solving the quantum master equation numerically, we investigate the time evolution of $\chi_s$ for the dissipative hard-core boson model.
We take a random pure state as an initial state, i.e., $\rho_{\mathrm{ini}}=\ket{\psi_\mathrm{r}}\bra{\psi_\mathrm{r}}$ where $\ket{\psi_\mathrm{r}}$ is a normalized vector uniformly sampled from the set of unit vectors in the Hilbert space.
Figure \ref{Fig-coherence-Gamma-numerical}(a) shows the time evolution of the absolute values of the one-particle reduced density matrix $G^{(1)}$.
Two regimes can be clearly distinguished.
In the first regime, the off-diagonal components of $G^{(1)}$ decay rapidly, which implies the production of incoherentons.
In the second regime, a slow diffusion of the diagonal components is observed, and at long times they converge to $N/L$.
We refer to the first (second) regime as the incoherenton production (diffusion) regime.

It is convenient to define the time-dependent decay rate $\Gamma_s$ of the $s$th-order coherence as
\begin{equation}
	\Gamma_s(t) := - \frac{d}{dt} \ln \chi_s(t).
\end{equation}
For $J=0$, we have $\Gamma_s(t)=s\gamma$ for all $t$.
Figure \ref{Fig-coherence-Gamma-numerical}(b) shows the time evolution of $\Gamma_1$ and $\Gamma_2$ with dephasing $\gamma=1$.
For a small hopping amplitude such as $J=0.05$ or $0.1$, two plateaus and subsequent algebraic decay $\Gamma_s \sim 1/t$ are observed.
(For $J=0.1$, the beginning and end of the second plateau are indicated by arrows.)
The height of the first plateau is $s \gamma$, which is the initial decay rate of $\chi_s$.
These numerical data suggest the existence of three regimes for $J \ll \gamma$ :
\begin{equation}
	\chi_s(t) \sim 
	\begin{cases}
		e^{-s \gamma t} & (t < \tau_1); \\
		e^{-\kappa_s t} & (\tau_1 < t < \tau_2); \\
		t^{-\eta_s} & (\tau_2 < t),
	\end{cases}
	\label{chi_decay}
\end{equation}
with some intermediate decay rate $\kappa_s$ and exponent $\eta_s$.
The algebraic decay is fitted roughly by $1/t$ for $\Gamma_1$ and $2/t$ for $\Gamma_2$ [the dotted lines in Fig.~\ref{Fig-coherence-Gamma-numerical}(b)], which implies $\eta_1 \simeq 1$ and $\eta_2 \simeq 2$.
The first and third regimes of Eq.~\eqref{chi_decay} correspond to the production and diffusion regimes, respectively.
For reasons that will be explained in the next subsection, we refer to the second regime of Eq.~\eqref{chi_decay} as the localization regime of incoherentons.
As $J$ approaches the transition point $J_c \simeq 0.2$ where the QC gap closes, the second plateau shrinks and eventually disappears at $J=J_c$, leaving a small bump.

It should be noted that the power-law decay of $\chi_s$ in the diffusion regime does not last forever in a finite system.
For $t \gg \Delta_{\mathrm{L}}^{-1}$, where $\Delta_{\mathrm{L}}$ is the Liouvillian gap, the relaxation of $\chi_s$ is determined by the slowest eigenmode, and thus $\chi_s$ decays as $e^{-\Delta_{\mathrm{L}}t}$.
It is known that the Liouvillian gap closes as $\Delta_{\mathrm{L}} \sim L^{-2}$ for our model \cite{Medvedyeva-16}.
In Fig.~\ref{Fig-coherence-Gamma-numerical}(b), a third plateau of $\Gamma_s$ appears in the long-time regime (see, e.g., the data of $J=0.25$ at $t > 30$).
We can confirm that the height of this plateau scales as $L^{-2}$.
The difference in the height of the third plateau for $\Gamma_1$ and $\Gamma_2$ is due to the difference in the slowest decaying eigenmodes which contribute to $\chi_1$ and $\chi_2$.

\begin{figure}
	\centering
	\includegraphics[width=0.45\textwidth]{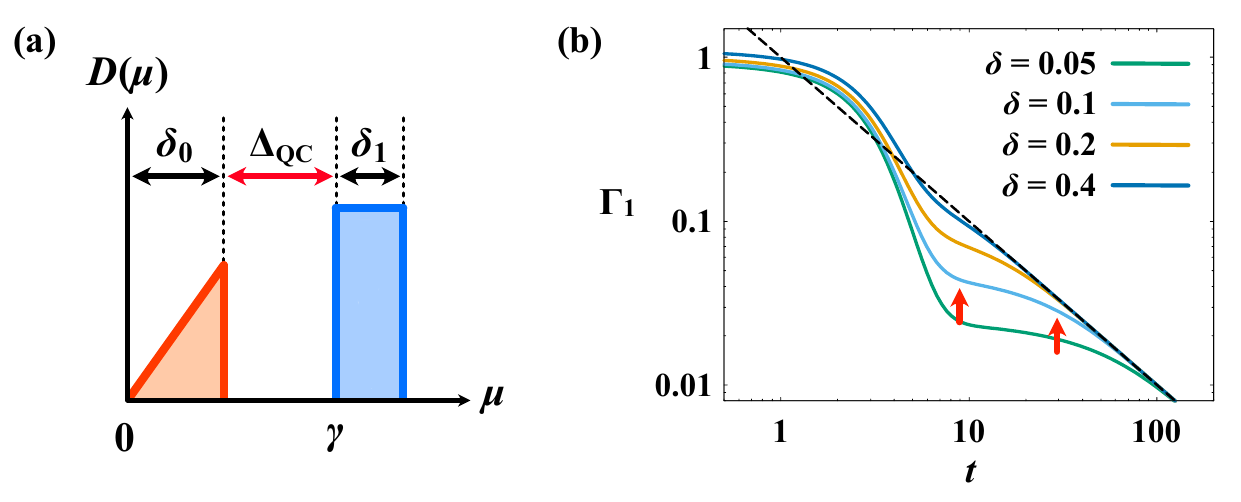}
	\caption{(a) Density of states $D_1(\mu)$ given by Eq.~\eqref{D_model} with $\eta=2$.
		(b) $\Gamma_1$ calculated from Eqs.~\eqref{chi_D} and \eqref{D_model} with $\delta:=\delta_0=\delta_1=0.05$, $0.1$, $0.2$, and $0.4$ from bottom to top.
		The parameters used are $\gamma=1$, $a_0=0.1$, $a_1=1$, and $\eta=1$.
		The dashed line denotes $1/t$.
		One can observe two plateau-like regimes for small values of $\delta$.
		The arrows indicate the beginning and end of the second plateau for $\delta=0.1$.}
	\label{Fig-coherence-Gamma-analytic}
\end{figure}

We present a simple theoretical argument showing the existence of the three relaxation regimes, i.e., regimes described by incoherenton production, localization, and diffusion.
We assume that the relaxation of $\chi_s$ is given by a superposition of exponential functions:
\begin{equation}
	\chi_s(t) = \int_0^{\infty} d\mu D_s(\mu) e^{-\mu t},
	\label{chi_D}
\end{equation}
where $D_s(\mu)$ is a weighted density of states, which expresses how many eigenmodes with decay rate $\mu$ contribute to $\chi_s$.
More precisely, Eq.~\eqref{chi_D} can be obtained by substituting the hierarchical expansion \eqref{rho_hierarchical_expansion} into the definition of $\chi_s$ and replacing the sum over eigenmodes with an integral over the decay rate.
For simplicity, we focus on the case of $s=1$.
Let us consider the following $D_1(\mu)$:
\begin{align}
	D_1(\mu) = \left\{ \begin{array}{ll}
		a_0 \mu^{\eta-1} & (0 \leq \mu \leq \delta_0); \\
		a_1 & (\gamma \leq \mu \leq \gamma+\delta_1).
	\end{array} \right.
	\label{D_model}
\end{align}
Figure \ref{Fig-coherence-Gamma-analytic}(a) shows a schematic illustration of $D_1(\mu)$.
The support of $D_1(\mu)$ near $\mu=0$ represents the contribution from incoherent eigenmodes where all particles form incoherentons.
On the other hand, the second support near $\mu=\gamma$ represents the contribution from (partially) coherent eigenmodes where only one particle pair is deconfined.
Note that the quantum coherence gap is given by $\Delta_{\mathrm{QC}}=\gamma-\delta_0$.
The exponent $\eta$ determines how the contribution from slowly-decaying eigenmodes decreases as the decay rate approaches zero.
By substituting Eq.~\eqref{D_model} into \eqref{chi_D}, one can confirm that $\chi_1 \sim t^{-\eta}$ at long times.
Figure \ref{Fig-coherence-Gamma-analytic}(b) shows $\Gamma_1$ calculated from Eq.~\eqref{D_model} with $\eta=1$ and $\delta:=\delta_0=\delta_1$.
For $\delta \ll \gamma$, one can observe two plateaus at $\Gamma_1=1$ and $\delta/2$, and for $\delta \sim \gamma$, the second plateau disappears.
The long-time behavior is given by $\Gamma_1 \simeq 1/t$.
Thus, a simple model defined by Eqs.~\eqref{chi_D} and \eqref{D_model} qualitatively reproduces the numerical results in Fig.~\ref{Fig-coherence-Gamma-numerical}(b).

\subsection{Characterization of many-body decoherence by incoherentons}

\begin{figure}
	\centering
	\includegraphics[width=0.45\textwidth]{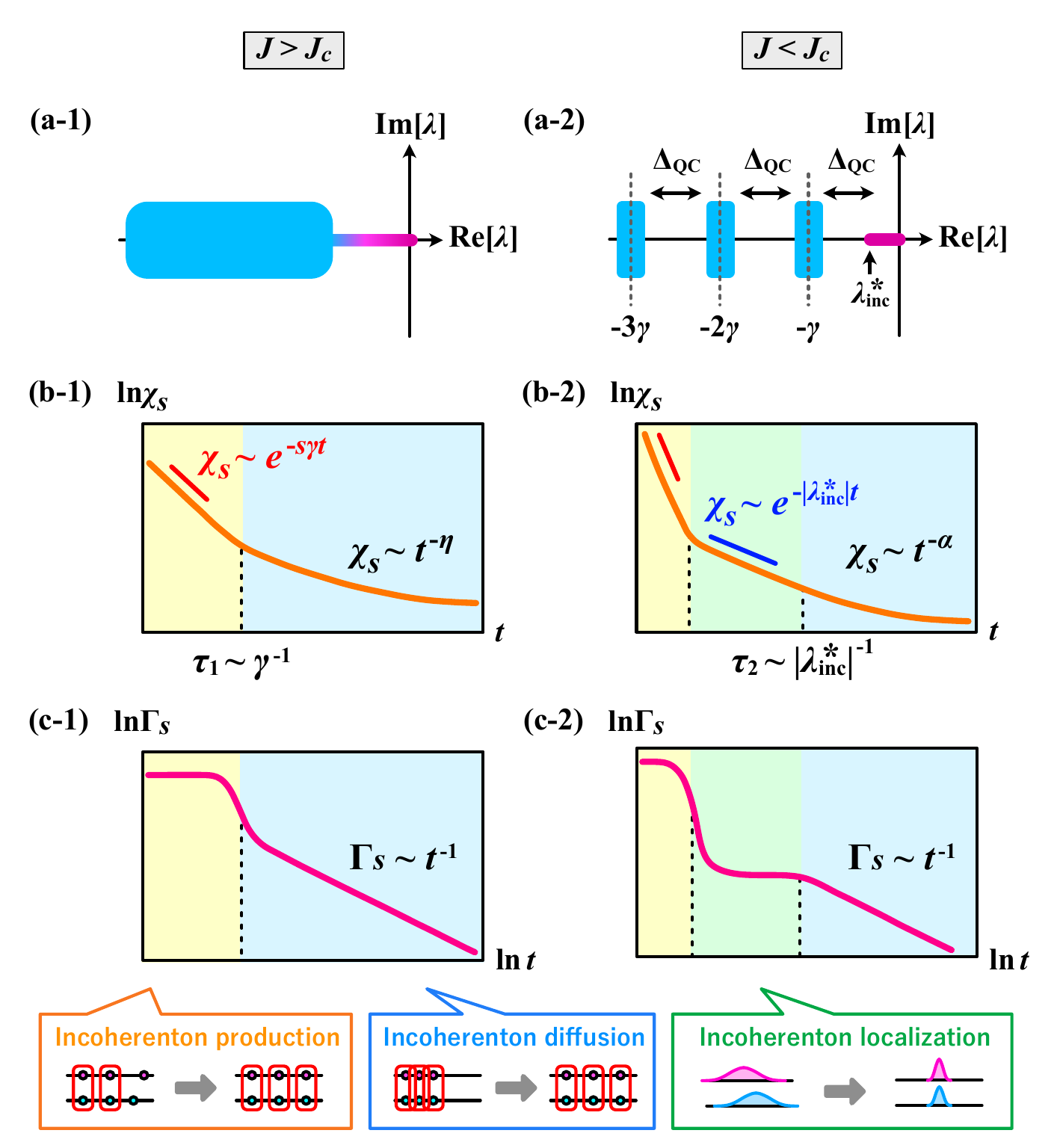}
	\caption{Summary of the effective description of many-body decoherence.
		The left panels show the case of $J > J_c$, while the right panels show the case of $J < J_c$.
		(a) Schematic illustrations of the Liouvillian spectra, where $\lambda_{\mathrm{inc}}^*$ denotes the incoherent-mode eigenvalue with the smallest real part.
		(b) $\ln \chi_s$ as a function of time $t$.
		For $J > J_c$, $\chi_s$ initially decays as $e^{-s\gamma t}$, and at long times $t \gg \gamma^{-1}$, it exhibits a power-law behavior $t^{-\eta}$.
		For $J < J_c$, there is an intermediate regime where $\chi_s$ decays as $e^{-|\lambda_{\mathrm{inc}}^*| t}$.
		(c) $\ln \Gamma_s$ as a function of $\ln t$.
		The early regime where $\chi_s$ decays exponentially and the late regime where it decays with a power-law correspond to the incoherenton-production and incoherenton-diffusion regimes, respectively.
		The intermediate regime is characterized by the localization of incoherentons.}
	\label{Fig-coherence-decay}
\end{figure}

The numerical results on the relaxation processes in the previous subsection can be explained in terms of the dynamics of incoherentons, as summarized in Fig.~\ref{Fig-coherence-decay}.
When the dissipation dominates ($J < J_c$), three distinct relaxation processes emerge:
\begin{description}
	\item[Incoherenton production] ($t < \tau_1 \sim \gamma^{-1}\ln(L\gamma/J)$)\\
	In this regime, the coherence decays exponentially as $\chi_s(t) \sim e^{-s \gamma t}$.
	Thus, $\Gamma_s$ has a plateau of height $s\gamma$ [see Fig.~\ref{Fig-coherence-decay}(c-2)].
	Since eigenmodes with a smaller number of incoherentons decay faster, the number of incoherentons increases with time in this regime.
	Let us now estimate the timescale $\tau_1$ at which the deviation from $\chi_s(t) \sim e^{-s \gamma t}$ begins.
	The magnitude of the contribution of the incoherent eigenmodes $\rho_{0,\alpha}$ to $\chi_s$ can be estimated as $L^s(J/\gamma)^s$ from perturbation theory \cite{note-1}.
	When the contributions of $\rho_{s,\alpha}$ and $\rho_{0,\alpha}$ to $\chi_s$ are comparable, the incoherenton production regime ends.
	This condition is expressed as $L^{2s} e^{-s\gamma \tau_1} \sim L^s(J/\gamma)^s$, where the factor $L^{2s}$ comes from the sum over off-diagonal components in, e.g., Eqs.~\eqref{def_chi_1} and \eqref{def_chi_2}.
	
	\item[Incoherenton localization] ($\tau_1 < t < \tau_2 \sim |\lambda_{\mathrm{inc}}^*|^{-1}$)\\
	We denote the eigenvalue with the smallest real part among the incoherent eigenmodes $\{\rho_{0,\alpha}\}$ as $\lambda_{\mathrm{inc}}^*$.
	In this regime, the relaxation of $\chi_s$ is determined by eigenmodes with decay rates of $O(|\lambda_{\mathrm{inc}}^*|)$, and thus, it decays as $\chi_s \sim e^{-\kappa_s t}$ with $\kappa_s=O(|\lambda_{\mathrm{inc}}^*|)$, which, in general, depends on $s$.
	That is, $\Gamma_s$ has a plateau of height $\kappa_s$ [see Fig.~\ref{Fig-coherence-decay}(c-2)].
	In general, eigenmodes that contain incoherentons with larger confinement length $\xi_{\mathrm{con}}$ decay faster.
	Thus, the decay of eigenmodes with eigenvalues of $O(\lambda_{\mathrm{inc}}^*)$ leads to a reduction of $\xi_{\mathrm{con}}$, i.e., the localization of incoherentons.
	
	\item[Incoherenton diffusion] ($\tau_2 < t$)\\
	In this regime, the relaxation of $\chi_s$ is determined by the incoherent eigenmodes $\{\rho_{0,\alpha}\}$ with small decay rates ($\ll |\lambda_{\mathrm{inc}}^*|$).
	In the hard-core boson model under on-site dephasing, since there exist eigenvalues arbitrarily close to $0$ for an infinitely large system, the coherence exhibits a power-law decay $\chi_s \sim t^{-\eta_s}$.
	Note that algebraic decay is a general feature of open quantum many-body systems in which the Liouvillian gap $\Delta_{\mathrm{L}}$ vanishes in the thermodynamic limit \cite{Cai-13}.
	This relaxation process proceeds by rearrangement of the positions of well-localized incoherentons, i.e., the diffusion of incoherentons.
	In a system with a nonzero $\Delta_{\mathrm{L}}$ in the thermodynamic limit, this regime is expected to be absent .
\end{description}

When the QC gap closes ($J > J_c$), we have $\tau_1 \sim \gamma^{-1} \ln L$ and $\tau_2 \sim \gamma^{-1}$, and thus, the incoherenton localization regime cannot be observed.
Instead, the incoherenton diffusion regime directly follows the incoherenton production region [see Figs.~\ref{Fig-coherence-decay}(b-1) and (c-1)].
In this case, incoherenton production and localization occur simultaneously, and these regimes cannot be clearly separated.

As mentioned above, the crossover timescale $\tau_1$ depends logarithmically with respect to the system size $L$.
This is because the sum in Eq.~\eqref{def_chi_1} or \eqref{def_chi_2} is taken over all off-diagonal components.
It may be reasonable to restrict this sum to the off-diagonal components that are close to the diagonal components.
For a typical one-body observable \eqref{one_body_observable},  $O^{(1)}_{l_1, l_2}$ rapidly decays to zero as $|l_1 - l_2|$ increases, so only off-diagonal components satisfying $l_1 \simeq l_2$ contribute to Eq.~\eqref{O_1_expect}.
Then, we can also define 
\begin{equation}
	\tilde{\chi}_1 := \sum_{l_1, l_2} |G^{(1)}_{l_1, l_2}| (1-\delta_{l_1, l_2}) e^{-c |l_1-l_2|},
	\label{chi_1_tilde}
\end{equation}
where the exponential factor with $c=O(1)$ suppresses the contribution of off-diagonal components with large $|l_1 - l_2|$.
If we focus on the relaxation of $\tilde{\chi}_1$, the crossover timescale $\tau_1$ can be given by $\gamma^{-1}\ln(\gamma/J)$, which is independent of the system size.

\section{Conclusion}
\label{sec:conclusion}

We have proposed the notion of incoherenton in open quantum many-body systems, which characterizes the hierarchical structure of Liouvillian eigenmodes and their incoherent-coherent transitions.
Under the mapping of the Liouvillian to a non-Hermitian ladder Hamiltonian, incoherentons are defined as interchain bound states.
The decay rate of each eigenmode is determined approximately by the number of incoherentons that the relevant eigenmode involves.
The quantum coherence (QC) gap is defined as the minimum difference in decay rates between eigenmodes with different numbers of incoherentons.
As the coherence parameter of the system increases, the deconfinement of an incoherenton occurs at a certain critical point, causing the QC gap closing.
For a hard-core boson system under on-site dephasing, we have demonstrated numerically and analytically the deconfinement of incoherentons.
Furthermore, the process of many-body decoherence is discussed in terms of incoherentons.
Three relaxation regimes corresponding to the production, localization, and diffusion of incoherentons are identified.

Note that our framework of incoherentons may not suffice to capture every intricate detail of the Liouvillian spectra and eigenmodes.
As highlighted in Appendix \ref{appendix:Bose_Hubbard_model}, the dephasing Bose-Hubbard model incorporates {\it intrachain} bound states, complementing the role of interchain bound states, i.e., incoherentons.
However, our primary objective is to describe and explore universal characteristics found within the complex behaviors of the Liouvillian spectra and eigenmodes, which are summarized as follows:
\begin{enumerate}
	\item Under strong dissipation, the spectrum displays multiple bands, with eigenmodes distinctly marked by both inter- and intra-chain bound states.
	\item Interchain and intrachain bound states affect decay rates and frequencies, respectively, thereby governing the temporal dynamics of eigenmodes.
	\item Decreasing dissipation leads to the merging of specific bands, signaling the deconfinement of interchain bound states.
\end{enumerate}
While we believe in the robustness of these observations, these characteristics may require refinement or further extension in more complicated situations.
The presence of additional degrees of freedom could introduce other types of bound states.
We expect that our current work provides a solid foundation for subsequent research that further refines the quasiparticle descriptions of Liouvillian eigenmodes.

In this study, we focused on systems with local bulk dissipation.
It is natural to ask whether the incoherenton picture summarized above holds for other types of dissipation as well.
When there is dissipation only at the boundary of the system \cite{Znidaric-15}, the localization of incoherentons near the boundary is expected.
The effect of nonlocal dissipation is also nontrivial, which leads to long-range interactions between the chains in the ladder representation of a Liouvillian.
The influence of long-range interactions on the formation of bound states is well studied in the context of isolated quantum many-body systems, and it has been pointed out that a new type of bound state can be realized in quantum spin chains with long-range interactions \cite{Liu-19, Lerose-19, Vovrosh-22}.
Understanding the impact of different types of dissipation on the deconfinement transition of eigenmodes and the QC gap closing in spectra deserves further study.

Quasiparticles are a key concept in many-body physics.
In isolated quantum many-body systems, the existence of well-defined quasiparticles ensures the validity of low-energy effective field theories, which describe the thermodynamic and transport properties of the system through statistical mechanics of weakly interacting quasiparticles.
Identifying the quasiparticles is to distinguish a set of relevant variables for characterizing the low-energy behavior of the system from many irrelevant variables.
It is expected that complex relaxation processes in open quantum many-body systems are described by a simple kinetic theory of various incoherentons, which should be studied in detail in future works.
A better understanding of incoherentons can provide an efficient way to predict decoherence effects in the control of large-scale quantum devices.

The formation of bound states between interacting particles is a universal phenomenon from particle physics to condensed matter physics.
Phenomena that arise from the formation of specific types of bound states include, for example, BCS-BEC crossover in interacting Fermi gases \cite{Chen-05, Giorgini-08, Strinati-18} and Efimov resonances in three-body bound states of atoms with large scattering lengths \cite{Efimov-70, Mattis-86, Braaten-06, Naidon-17}.
It is an important task to investigate the effect of the formation of various types of bound states in Lindblad ladder systems on the structure of Liouvillian spectra and dynamical features in open quantum systems.

\begin{acknowledgments}
	This work was supported by JSPS KAKENHI Grant Numbers JP22K13983, JP19J00525, and JP18H01145, and a Grant-in-Aid for Scientific Research on Innovative Areas ``Topological Materials Science'' (JSPS KAKENHI Grant Number JP15H05855).
	M. N. was supported by JSPS KAKENHI Grant Number JP20K14383.
	R. H. was supported by JSPS through Program for Leading Graduate Schools (ALPS) and JSPS fellowship (JSPS KAKENHI Grant Number JP17J03189). 
\end{acknowledgments}

\appendix

\section{General properties of Liouvillian eigenmodes}
\label{appendix:Liouvillian_eigenmodes}

The eigenmodes $\{ \rho_{\alpha} \}$ and eigenvalues $\{ \lambda_{\alpha} \}$ of a Liouvillian are defined by Eq.~\eqref{eigen_eq}.
Here, we summarize general properties of the eigenmodes and eigenvalues of the Liouvillian (see, e.g., Ref.~\cite{Minganti-18}).
\begin{enumerate}
	\item The left eigenmodes $\{ \rho_{\alpha}' \}$ are defined by
	\begin{equation}
		\mathcal{L}^{\dag}(\rho_{\alpha}')=\lambda_{\alpha}^* \rho_{\alpha}' \quad (\alpha=0, 1, ... , D^2-1),
	\end{equation}
	where $\mathcal{L}^{\dag}$ reads
	\begin{equation}
		\mathcal{L}^{\dag}(A) = -i[A, H] + \sum_{\nu} \left( L_{\nu}^{\dag} A L_{\nu} - \frac{1}{2} \{ L_{\nu}^{\dag}L_{\nu}, A \} \right).
	\end{equation}
	Since $\mathcal{L}$ is not Hermitian, $\rho_{\alpha}$ and $\rho_{\alpha}'$ are not identical.
	While the right eigenmodes are not orthogonal to each other, the right and left eigenmodes with different eigenvalues are orthogonal to each other: $\mathrm{Tr}[\rho_{\alpha}'^{\dag} \rho_{\beta}] = 0 \: (\lambda_\alpha \neq \lambda_\beta)$.
	This relation is known as the biorthogonal relation.
	
	\item The Liouvillian is diagonalizable except for a zero measure subset in the parameter space, i.e., exceptional points.
	In this case, the eigenmodes constitute a basis of the operator space, and thus, any operator $O$ can be uniquely expanded as 
	\begin{equation}
		O = \sum_{\alpha=0}^{D^2-1} c_\alpha \rho_{\alpha}, \quad c_\alpha = \frac{\mathrm{Tr}[\rho_{\alpha}'^{\dag} O]}{\mathrm{Tr}[\rho_{\alpha}'^{\dag} \rho_{\alpha}]},
	\end{equation}
	where the expression of $c_\alpha$ follows from the orthogonality between the right and left eigenmodes.
	
	\item The real parts of $\lambda_{\alpha}$ are nonpositive and there exist zero modes corresponding to the steady states.
	The negative real parts of non-steady eigenmodes ensure the relaxation to the steady state, 
	\begin{equation}
		\lim_{t \to \infty} \rho(t) = \lim_{t \to \infty} e^{\mathcal{L}t} \rho(0) = \rho_{\mathrm{ss}}.
	\end{equation}
	
	\item The Liouvillian gap is defined by $\Delta_{\mathrm{L}}=|\mathrm{Re}[\lambda_1]|$ if the steady state is unique and the eigenvalues are sorted such that $0 = |\mathrm{Re}[\lambda_0]| < |\mathrm{Re}[\lambda_1]| \leq \cdots \leq |\mathrm{Re}[\lambda_{D^2-1}]|$.
	Since $\Delta_{\mathrm{L}}$ determines the relaxation dynamics in the long-time limit, it is also referred to as the asymptotic decay rate \cite{Kessler-12}.
	
	\item $\mathrm{Tr}[\rho_{\alpha}]=0$ if $\lambda_{\alpha} \neq 0$.
	This is a consequence of the trace-preserving nature of the Liouvillian: $\mathrm{Tr}[\mathcal{L}(\rho)] = 0$.
	Thus, eigenmodes with nonzero eigenvalues are not physical states.
	
	\item If $\mathcal{L}(\rho_{\alpha}) = \lambda_{\alpha} \rho_{\alpha}$, then $\mathcal{L}(\rho_{\alpha}^{\dag}) = \lambda_{\alpha}^* \rho_{\alpha}^{\dag}$, which follows from $[\mathcal{L}(\rho_{\alpha})]^{\dag} = \mathcal{L}(\rho_{\alpha}^{\dag})$.
	This implies that (i) the Liouvillian spectrum on the complex plane is symmetric with respect to the real axis, and (ii) if $\rho_{\alpha}$ is Hermitian, the corresponding eigenvalue $\lambda_{\alpha}$ is real.
\end{enumerate}

\section{One-particle solution of Liouvillian eigenmodes}
\label{appendix:single_particle_solution}

In this Appendix, we present a detailed analysis of the Liouvillian spectrum and eigenmodes in the one-particle case without resorting to the Bethe ansatz.
In the ladder representation, the Liouvillian $\mathcal{L}$ is mapped to a non-Hermitian Hamiltonian $\tilde{\mathcal{L}}$ of a two-particle system on the ladder.
We denote by $\ket{l} \otimes \ket{m}$ the state in which each particle is located at sites $l$ and $m$ of each chain of the ladder.
Since $\tilde{\mathcal{L}}$ is translationally invariant, it is convenient to introduce a basis
\begin{equation}
	|k, l) := L^{-1/2} \sum_{m=1}^L e^{ikm} \ket{m+l} \otimes \ket{m},
	\label{momentum_base}
\end{equation}
where $k=2\pi s/L \ (s=-L/2+1,...,L/2)$ is the momentum of the center of mass and $l = -L/2+1,...,L/2$ is the relative coordinate.
We write matrix elements in this basis as
\begin{equation}
	(k, l| \tilde{\mathcal{L}} | k', l') = \delta_{kk'} \mathcal{L}(k)_{ll'},
\end{equation}
where the matrix elements between different momenta $k$ vanish owing to the translational symmetry of $\tilde{\mathcal{L}}$.
The matrix elements of $\mathcal{L}(k)$ are given by
\begin{align}
	\mathcal{L}(k)_{lm} = \left\{ \begin{array}{ll}
		- \gamma & (l=m \neq 0); \\
		i J [ 1-e^{ik(l-m)} ] & (|l-m|=1); \\
		0 & (\mathrm{otherwise}).
	\end{array} \right.
	\label{effective_tight_binding}
\end{align}
Note that the indices $l$ and $m$ satisfy the periodic boundary condition.
Equation \eqref{effective_tight_binding} defines an effective tight-binding model for the relative coordinate, which has an imaginary hopping amplitude between neighboring sites.
Let $\{ \lambda_j(k) \}_{j=1,...,L}$ be the eigenvalues of $\mathcal{L}(k)$.
For $k=0$, $\mathcal{L}(k)$ becomes diagonal for arbitrary $J$, and it has a single zero eigenvalue and an $(L-1)$-fold degenerate eigenvalue $\lambda=-\gamma$.
In particular, the zero mode is given by $\rho_{0, lm}=L^{-1/2}\delta_{lm}$.

In terms of the effective tight-binding model \eqref{effective_tight_binding}, the coherent and incoherent eigenmodes correspond to scattering and bound states, respectively.
We write an eigenmode as
\begin{eqnarray}
	\hspace{-1em} \psi_{l} = \left\{ \begin{array}{ll}
		\psi_{l}^{(+)} = c_1 e^{-\alpha l} + c_2 e^{\beta l} & (0 \leq l \leq L/2); \\
		\psi_{l}^{(-)} = \tilde{c}_1 e^{-\alpha l} + \tilde{c}_2 e^{\beta l} & (-L/2 < l < 0), \\
	\end{array} \right.
	\label{psi_ansatz}
\end{eqnarray}
where $\alpha$ and $\beta$ are complex.
The periodic boundary condition imposes the following conditions between the coefficients $c$ and $\tilde{c}$:
\begin{equation}
	\tilde{c}_1 = e^{-\alpha L} c_1, \quad \tilde{c}_2 = e^{\beta L} c_2.
	\label{c_c_tilde}
\end{equation}
Equation \eqref{c_c_tilde} and the connection condition $\psi_{0}^{(+)}=\psi_{0}^{(-)}$ at $l=0$ lead to
\begin{equation}
	c_1 + c_2 = e^{-\alpha L} c_1 + e^{\beta L} c_2.
	\label{c_1_c_2}
\end{equation}
By substituting $\psi_l$ in Eq.~\eqref{psi_ansatz} into the eigenvalue equation $\sum_m \mathcal{L}(k)_{lm} \psi_m = \lambda(k) \psi_l$ with $l \neq 0$, we have
\begin{align}
	\lambda(k) &= iJ( e^{\alpha} + e^{-\alpha} - e^{ik+\alpha} - e^{-ik-\alpha}) - \gamma \notag\\
	&= iJ( e^{-\beta} + e^{\beta} - e^{ik-\beta} - e^{-ik+\beta}) - \gamma,
	\label{lambda_analytic}
\end{align}
which leads to the following relation between $\alpha$ and $\beta$:
\begin{equation}
	\beta = \alpha + i(k-\pi).
	\label{alpha_beta}
\end{equation}
Furthermore, the eigenvalue equation for $l=0$ reads
\begin{equation}
	iJ(1-e^{ik}) \psi_{-1}^{(-)} + iJ(1-e^{-ik}) \psi_{1}^{(+)} = \lambda(k) \psi_{0}^{(+)}.
	\label{eigen_eq_analytic}
\end{equation}
From Eqs.~\eqref{c_1_c_2}--\eqref{eigen_eq_analytic}, $\alpha$, $\beta$, and $\lambda$ can be calculated as functions of $k$.

We focus on a bound state exponentially localized near $l=0$.
We assume that $\mathrm{Re}[\beta] > 0$ and $\mathrm{Re}[\alpha] > 0$.
For a sufficiently large $L$, Eq.~\eqref{c_1_c_2} reduces to
\begin{equation}
	c_2 \simeq e^{-\beta L} c_1.
	\label{c_1_c_2_large_L}
\end{equation}
Substitution of Eq.~\eqref{c_1_c_2_large_L} into Eq.~\eqref{eigen_eq_analytic} yields
\begin{equation}
	\lambda(k) = iJ(1-e^{ik}) e^{-\beta} + iJ(1-e^{-ik}) e^{-\alpha},
	\label{eigen_eq_analytic_bound}
\end{equation}
for $L \to \infty$.
From Eqs.~\eqref{lambda_analytic}, \eqref{alpha_beta}, and \eqref{eigen_eq_analytic_bound}, the eigenvalue $\lambda_{\mathrm{inc}}(k)$ associated with the bound state can be calculated as
\begin{equation}
	\lambda_{\mathrm{inc}}(k) = -\gamma + \sqrt{\gamma^2 - 8J^2(1-\cos k)},
	\label{lambda_inc_1P}
\end{equation}
which coincides with Eq.~\eqref{lambda_inc_MP} with $m=1$.

\begin{figure}
	\centering
	\includegraphics[width=0.45\textwidth]{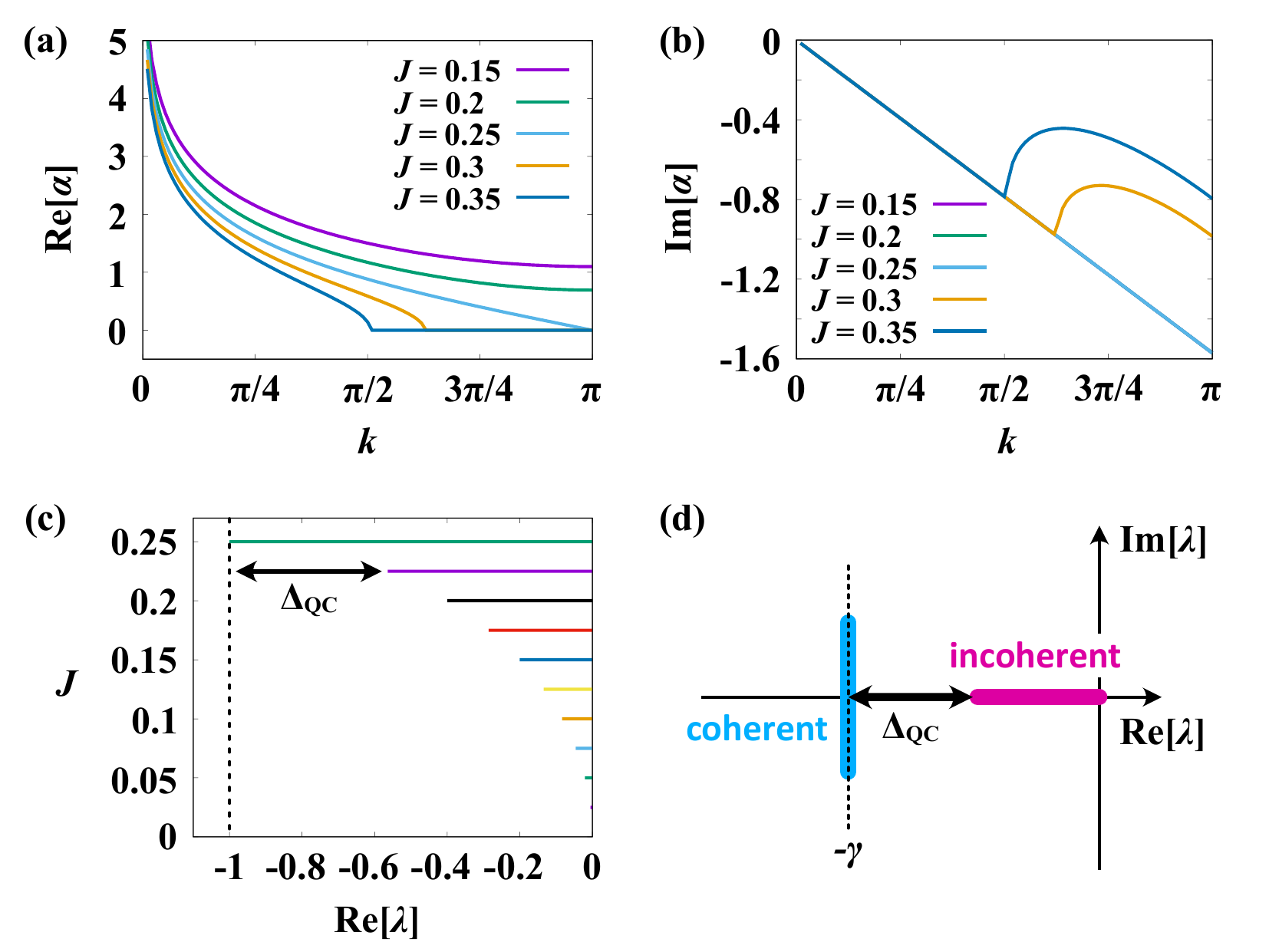}
	\caption{(a), (b) Real and imaginary parts of $\alpha$ as functions of $k$ with $\gamma=1$ for $J=0.15$, $0.2$, $0.25$, $0.3$, and $0.35$.
		(c) Trajectories of $\lambda_{\mathrm{inc}}(k)$ on the real axis as $k$ increases from $-\pi$ to $\pi$.
		$J$ increases from $0$ to $0.25$ in increments of $0.025$.
		(d) Schematic illustration of the Liouvillian spectrum in the limit of $L \to \infty$.
		The purple and blue bands represent the incoherent-mode and coherent-mode spectra, respectively.}
	\label{Fig-spec-1P-analytic}
\end{figure}

Figures \ref{Fig-spec-1P-analytic}(a) and (b) show $\mathrm{Re}[\alpha]$ and $\mathrm{Im}[\alpha]$ as functions of $k$, respectively.
Note that $\mathrm{Re}[\alpha]$ is the inverse of the confinement length $\xi_{\mathrm{con}}$.
For $k=0$, since the zero mode of $\mathcal{L}(k)$ is completely localized at $l=0$, we have $\mathrm{Re}[\alpha]=\infty$.
Figure \ref{Fig-spec-1P-analytic}(a) shows that there is a critical hopping amplitude $J_c$ below which a bound state exists ($\mathrm{Re}[\alpha]>0$) for all $k$.
On the other hand, for $J>J_c$, there is a critical wavenumber $k_c(J)$ such that the bound state disappears ($\mathrm{Re}[\alpha]=0$) for $k \geq k_{\mathrm{c}}(J)$.
From Fig.~\ref{Fig-spec-1P-analytic} (b), one finds that $\mathrm{Im}[\alpha]=-k/2$ for $J<J_c$, which is consistent with the fact that $\lambda_{\mathrm{inc}}(k)$ is real.
For $J>J_c$, $\mathrm{Im}[\alpha]$ shows a cusp at $k=k_c(J)$ owing to the disappearance of the bound state.
Figure \ref{Fig-spec-1P-analytic}(c) shows trajectories of $\lambda_{\mathrm{inc}}(k)$ for different values of $J$.
As $k$ increases from $-\pi$, $\lambda_{\mathrm{inc}}(k)$ initially moves from the left end point of the incoherent-mode spectrum to the right, reaches the origin for $k=0$, and returns to the left end point for $k=\pi$.
The quantum coherence gap $\Delta_{\mathrm{QC}}$ decreases with increasing $J$, and eventually, it closes at $J=0.25$.
Figure \ref{Fig-spec-1P-analytic}(c) is consistent with the numerical results shown in Fig.~\ref{Fig-spec-1P}(a).

Let us determine the critical hopping amplitude $J_c$ at which the confinement length $\xi_{\mathrm{con}}$ diverges.
As $J$ increases, the disappearance of the bound state firstly occurs at $k=\pi$.
Thus, from Eqs.~\eqref{lambda_analytic}, \eqref{alpha_beta}, and \eqref{eigen_eq_analytic_bound}, we have $\alpha=\beta$ and 
\begin{equation}
	\sinh \alpha = - \frac{i \gamma}{4J}.
	\label{alpha_pi}
\end{equation}
The hyperbolic sine function has the following property: if $\sinh \alpha$ is purely imaginary, $|\sinh \alpha| \leq 1$ and $|\sinh \alpha| > 1$ imply $\mathrm{Re}[\alpha] = 0$ and $\mathrm{Re}[\alpha] \neq 0$, respectively.
Thus, we have
\begin{equation}
	J_c = \frac{\gamma}{4},
\end{equation}
which agrees with Eq.~\eqref{J_c_MP} with $m=1$.
It should be noted that the incoherent-mode eigenvalue $\lambda_{\mathrm{inc}}(\pi)$ given by Eq.~\eqref{lambda_inc_1P} becomes complex for $J>J_c$.
In Fig.~\ref{Fig-spec-1P-analytic}, $J_c=0.25$ since $\gamma=1$.
From Eq.~\eqref{alpha_pi}, the confinement length $\xi_{\mathrm{con}}$ of the eigenmode with $k=\pi$ reads
\begin{equation}
	\xi_{\mathrm{con}} = \mathrm{Re}[\alpha]^{-1} \simeq \left(\frac{\gamma}{2J} - 2 \right)^{-1/2},
\end{equation}
for $J \simeq J_c$.
Since the QC gap $\Delta_{\mathrm{QC}}$ is given by Eq.~\eqref{Delta_QC_1P}, we have
\begin{equation}
	\xi_{\mathrm{con}} \simeq \frac{\gamma}{\Delta_{\mathrm{QC}}},
\end{equation}
which is consistent with Eq.~\eqref{xi_con_Delta_QC}.

Figure \ref{Fig-spec-1P-analytic}(d) shows the Liouvillian spectrum in the limit of $L \to \infty$.
It consists of the coherent-mode spectrum at $\mathrm{Re}[\lambda]=-\gamma$ parallel to the imaginary axis and the incoherent-mode spectrum on the real axis.
The reason why the real parts of the coherent-mode eigenvalues are equal to $-\gamma$ in the limit of $L \to \infty$ can be understood as follows.
Since the coherent eigenmodes are extended over the relative coordinate, $\mathrm{Re}[\alpha]$ should vanish.
By substituting $\mathrm{Re}[\alpha]=0$ into Eq.~\eqref{lambda_analytic}, we obtain $\mathrm{Re}[\lambda(k)]=-\gamma$.
By numerical diagonalization of the Liouvillian, one can also verify that the width of the coherent-mode spectrum along the real axis decreases as $L$ increases.

\section{Absence of incoherenton in continuous systems}
\label{appendix:absence_of_incoherenton}

As mentioned at the end of Sec.~\ref{sec:incoherentons}, the spatial discreteness of the lattice system is crucial for the formation of incoherentons.
In this Appendix, we show that incoherenton does not exist in systems where a free particle in continuous space undergoes dephasing.

The Hamiltonian of a free particle in one-dimensional continuous space is given by
\begin{equation}
	H = \int^{\infty}_{-\infty} dx \psi^{\dag}(x) \left( -\frac{1}{2m} \frac{\partial^2}{\partial x^2} \right) \psi(x),
	\label{Hamiltonian_continuous}
\end{equation}
where $\psi^{\dag}(x)$ and $\psi(x)$ are the creation and annihilation operators of a boson at position $x$, which satisfy the canonical commutation relations:
\begin{align}
	\begin{split}
		[\psi(x), \psi(x')] &= [\psi^{\dag}(x), \psi^{\dag}(x')] = 0, \\
		\quad [\psi(x), \psi^{\dag}(x')] &= \delta(x-x').
	\end{split}
\end{align}
The Liouvillian $\mathcal{L}$ that governs the time evolution of the density matrix $\rho$ is given by
\begin{align}
	\mathcal{L}(\rho) &= -i[H,\rho] + \int^{\infty}_{-\infty} dx \:\biggl( L(x) \rho L^{\dag}(x) \notag\\
	&\quad - \frac{1}{2} \{ L^{\dag}(x) L(x), \rho \} \biggr)\:.
	\label{Liouvillian_continuous}
\end{align}
We consider the following Lindblad operator
\begin{equation}
	L(x) = \int^{\infty}_{-\infty} dy g(x-y) \psi^{\dag}(y) \psi(y),
	\label{L_continuous}
\end{equation}
where $g(x)$ is a short-ranged function that rapidly decays for large $|x|$.
The Lindblad operator given by Eq.~\eqref{L_continuous} describes a dephasing process of a particle near position $x$.

We focus on the one-particle sector of the Hilbert space.
Let $\ket{x}=\psi^{\dag}(x)\ket{\mathrm{v}}$ ($\ket{\mathrm{v}}$: vacuum state) be the state in which a particle is located at position $x$.
Then, $\{ \ket{x} \}_{x \in (-\infty,\infty)}$ is an orthonormal basis in the one-particle sector.
In terms of this basis, the density matrix $\rho$ is written as
\begin{equation}
	\rho = \int^{\infty}_{-\infty} dx \int^{\infty}_{-\infty} dy \rho(x,y) \ket{x} \bra{y},
\end{equation}
where $\rho(x,y)$ is the matrix element of $\rho$.
In the ladder representation, $\ket{x} \bra{y}$ is mapped to a tensor-product state $\ket{x} \otimes \ket{y}$, which specifies a two-particle state in a ladder.
The Liouvillian $\mathcal{L}$ is also mapped to a non-Hermitian operator
\begin{align}
	\tilde{\mathcal{L}} &= - i H \otimes I + i I \otimes H + \int^{\infty}_{-\infty} dx \:\biggl( L(x) \otimes L(x) \notag\\
	&\quad - \frac{1}{2} L^{\dag}(x) L(x) \otimes I - \frac{1}{2} I \otimes L^{\dag}(x) L(x) \biggr)\:.
	\label{Liouvillian_ladder_continuous}
\end{align}
The matrix element of $\tilde{\mathcal{L}}$ is calculated as
\begin{align}
	&(\bra{x'} \otimes \bra{y'}) \tilde{\mathcal{L}} (\ket{x} \otimes \ket{y}) \notag\\
	& = \left[ \frac{i}{2m} \left( \frac{\partial^2}{\partial x^2}-\frac{\partial^2}{\partial y^2} \right) + \gamma(x-y)-\gamma(0) \right] \delta(x'-x) \delta(y'-y),
	\label{L_matrix_element_continuous}
\end{align}
where $\gamma(x)$ is defined by
\begin{equation}
	\gamma(x-y) := \int_{-\infty}^{\infty} dz g(x-z) g(z-y).
\end{equation}

Let $|\Phi)$ be an eigenmode of $\tilde{\mathcal{L}}$, i.e., $\tilde{\mathcal{L}} |\Phi) = \lambda |\Phi)$.
If $|\Phi)$ is written as
\begin{equation}
	|\Phi) = \int^{\infty}_{-\infty} dx \int^{\infty}_{-\infty} dy \varphi(x,y) \ket{x} \otimes \ket{y},
\end{equation}
we have an eigenvalue equation
\begin{equation}
	\left[ \frac{i}{2m} \biggl( \frac{\partial^2}{\partial x^2} - \frac{\partial^2}{\partial y^2} \biggr) + \gamma(x-y)-\gamma(0) \right] \varphi(x,y) = \lambda \varphi(x,y)
	\label{eigen_eq_continuous}
\end{equation}
from Eq.~\eqref{L_matrix_element_continuous}

The two-body wavefunction $\varphi(x,y)$ is rewritten as
\begin{equation}
	\varphi(x,y) = e^{ik(x+y)/2} \tilde{\varphi}(x-y),
	\label{phi_decouple}
\end{equation}
where $k$ is the center-of-mass momentum.
By substituting Eq.~\eqref{phi_decouple} into Eq.~\eqref{eigen_eq_continuous}, we have
\begin{equation}
	- \frac{k}{m} \tilde{\varphi}'(z) + [\gamma(z)-\gamma(0)] \tilde{\varphi}(z) = \lambda \tilde{\varphi}(z).
	\label{eigen_eq_relative_continuous}
\end{equation}

The eigenvalue equation \eqref{eigen_eq_relative_continuous} does not have a solution that is exponentially localized near $z=0$.
In fact, since $\gamma(z)$ vanishes for sufficiently large $|z|$, we have the following solution at long distances:
\begin{equation}
	\tilde{\varphi}(z) \propto \exp \left( \frac{-m[\lambda+\gamma(0)]z}{k} \right).
\end{equation}
For $\tilde{\varphi}(z)$ to be finite for $|z| \to \infty$, we must have $\mathrm{Re}[\lambda+\gamma(0)]=0$.
Thus, Eq.~\eqref{eigen_eq_relative_continuous} has only plane-wave solutions extended over the entire space.
The absence of localized solutions is a consequence of the lack of the second spatial derivative in Eq.~\eqref{eigen_eq_relative_continuous}, which is a crucial distinction from the ordinary two-body problem.
It should be noted that Eq.~\eqref{eigen_eq_relative_continuous} is equivalent to the Schr\"odinger equation for a chiral particle, whose chirality is determined from the sign of the center-of-mass momentum $k$.

\section{Measurement of incoherenton correlations}
\label{appendix:incoherenton_correlation}

In this Appendix, we discuss the measurement of incoherenton correlations defined by Eqs.~\eqref{def_inc_corr_2} and \eqref{def_inc_corr_3}.
The measurement protocol presented below is based on Refs.~\cite{Daley-12, Islam-15}, which can be implemented in ultracold atoms on optical lattices.
We denote the ladder representation of the density matrix $\rho$ as $|\rho)$. 
Let $n_{l, +(-)}$ be the density operator acting on the $l$-th site of the first (second) chain. 
We wish to measure the following quantities:
\begin{equation}
	\frac{(\rho|n_{l,+} n_{l,-}|\rho)}{(\rho|\rho)}, \quad \frac{(\rho|n_{l,+} n_{l,-}n_{m,+} n_{m,-}|\rho)}{(\rho|\rho)}.
	\label{inc_corr}
\end{equation}
In the original matrix representation, the quantities in the numerator and denominator are rewritten as
\begin{align}
	&(\rho|\rho) = \mathrm{tr}[\rho^2], \quad (\rho|n_{l,+} n_{l,-}|\rho) = \mathrm{tr}[\rho n_l \rho n_l], \nonumber \\
	&(\rho|n_{l,+} n_{l,-}n_{m,+} n_{m,-}|\rho) = \mathrm{tr}[\rho n_l n_m \rho n_l n_m],
	\label{tr_rho_n}
\end{align}
where $n_l$ is the density operator in the original system.

We prepare two copies of the original system, following the method in Refs.~\cite{Daley-12, Islam-15}. 
We define the SWAP operator $V_2$ that exchanges the state of copy 1 and copy 2 as:
\begin{equation}
	V_2 (\ket{\psi_1} \otimes \ket{\psi_2}) = \ket{\psi_2} \otimes \ket{\psi_1}.
\end{equation}
Then, the quantities in Eq.~\eqref{tr_rho_n} can be rewritten as follows:
\begin{align}
	&\mathrm{tr}[\rho^2] = \mathrm{Tr}[V_2 (\rho \otimes \rho)], \nonumber \\
	&\mathrm{tr}[\rho n_l \rho n_l] = \mathrm{Tr}[(n_l \otimes n_l) V_2 (\rho \otimes \rho)], \nonumber \\
	&\mathrm{tr}[\rho n_l n_m \rho n_l n_m] = \mathrm{Tr}[\{(n_l n_m) \otimes (n_l n_m)\} V_2 (\rho \otimes \rho)],
	\label{tr_rho_n_V2}
\end{align}
where $\mathrm{Tr}[...]$ represents the trace in the tensor product space corresponding to the two copies. 
Equations \eqref{tr_rho_n_V2} can be shown as follows.
Writing the density matrix in terms of the Fock basis ${\ket{k}}$, which specifies the occupation number at each site, as $\rho = \sum_{k_1 k_2} \rho_{k_1 k_2} \ket{k_1} \bra{k_2}$, we can transform, for instance, $\mathrm{Tr}[(n_l \otimes n_l) V_2 (\rho \otimes \rho)]$ as follows:
\begin{align}
	&\mathrm{Tr}[(n_l \otimes n_l) V_2 (\rho \otimes \rho)] \nonumber \\
	&= \sum_{\{k_i\}} \rho_{k_1 k_2} \rho_{k_3 k_4} \mathrm{Tr}[(n_l \otimes n_l) V_2 \ket{k_1} \bra{k_2} \otimes \ket{k_3} \bra{k_4}] \nonumber \\
	&= \sum_{\{k_i\}} \rho_{k_1 k_2} \rho_{k_3 k_4} \mathrm{Tr}[(n_l \otimes n_l) \ket{k_3} \bra{k_2} \otimes \ket{k_1} \bra{k_4}] \nonumber \\
	&= \sum_{\{k_i\}} \rho_{k_1 k_2} \rho_{k_3 k_4} (\bra{k_2} \otimes \bra{k_4}) (n_l \otimes n_l) (\ket{k_3} \otimes \ket{k_1}) \nonumber \\
	&= \sum_{k_1, k_2} \rho_{k_1 k_2} \bra{k_2} n_l \ket{k_2} \rho_{k_2 k_1} \bra{k_1} n_l \ket{k_1} \nonumber \\
	&= \mathrm{tr}[\rho n_l \rho n_l],
\end{align}
where we have used the fact that $n_l$ is diagonal with respect to the Fock basis $\ket{k}$.

Since the SWAP operators acting on different sites commute with each other, we will focus on a single-site system in this discussion. 
Since $(V_2)^2 = \mathrm{I}$, the eigenvalues of $V_2$ are $\pm 1$. 
We denote the symmetric eigenspace corresponding to the eigenvalue $+1$ as $\mathcal{H}_+$ and the antisymmetric eigenspace corresponding to the eigenvalue $-1$ as $\mathcal{H}_-$. 
In the case of hard-core bosons \cite{Daley-12}, $\mathcal{H}_+$ is spanned by three basis states:
\begin{equation}
	\ket{\mathrm{vac}}, \quad (a_1^\dag + a_2^\dag) \ket{\mathrm{vac}}, \quad a_1^\dag a_2^\dag \ket{\mathrm{vac}}.
	\label{H_+_basis}
\end{equation}
In contrast, $\mathcal{H}_-$ is spanned by a single basis state:
\begin{equation}
	\quad (a_1^\dag - a_2^\dag) \ket{\mathrm{vac}}.
	\label{H_-_basis}
\end{equation}
Here, $a_{1(2)}^\dag$ represents the creation operator for copy 1(2). 
To obtain the expectation value of $V_2$, it is necessary to measure the probability of the state being found in $\mathcal{H}_+$ and $\mathcal{H}_-$.
We here  introduce a ``beamsplitter operation" as follows:
\begin{equation}
	\frac{1}{\sqrt{2}} (a_1^\dag + a_2^\dag) \to a_1^\dag, \quad \frac{1}{\sqrt{2}} (a_1^\dag - a_2^\dag) \to a_2^\dag.
\end{equation}
This transformation can be implemented by introducing a weak tunnel coupling between two copies and inducing Rabi oscillations \cite{Daley-12, Islam-15}.
After this transformation, the basis of $\mathcal{H}_+$ becomes
\begin{equation}
	\ket{\mathrm{vac}}, \quad a_1^\dag \ket{\mathrm{vac}}, \quad a_1^\dag a_2^\dag \ket{\mathrm{vac}},
	\label{H_+_basis_after}
\end{equation}
and the basis of $\mathcal{H}_-$ becomes
\begin{equation}
	a_2^\dag \ket{\mathrm{vac}}.
	\label{H_-_basis_after}
\end{equation}
Note that the state $a_1^\dag a_2^\dag \ket{\mathrm{vac}}$ remains unaffected by the beamsplitter operation due to the hard-core condition suppressing tunneling between the copies.
Finally, by observing how many particles each copy contains, we can determine the probability of the state being found in $\mathcal{H}_+$ and $\mathcal{H}_-$.

We summarize the above discussion: when the preparation of copies, the beamsplitter operation, and the observation of particle number are repeated many times, we denote the probability of having no particles as $p_0^{(l)}$, having one particle in copy 1 as $p_1^{(l)}$, one particle in copy 2 as $p_2^{(l)}$, and one particle in each copy as $p_{12}^{(l)}$, where $l$ represents the site index. 
The expectation value of the physical quantity including the SWAP operator $V_2$ can be computed as follows:
\begin{equation}
	\mathrm{Tr}[V_2 (\rho \otimes \rho)] = \prod_l \left[ p_0^{(l)} + p_1^{(l)} + p_{12}^{(l)} - p_2^{(l)} \right], 
\end{equation}
\begin{equation}
	\mathrm{Tr}[(n_l \otimes n_l) V_2 (\rho \otimes \rho)] = p_{12}^{(l)} \prod_{j (\neq l)} \left[ p_0^{(j)} + p_1^{(j)} + p_{12}^{(j)} - p_2^{(j)} \right],
\end{equation}
\begin{align}
	&\mathrm{Tr}[\{(n_l n_m) \otimes (n_l n_m)\} V_2 (\rho \otimes \rho)] \nonumber \\ 
	&= p_{12}^{(l)} p_{12}^{(m)} \prod_{j (\neq l,m)} \left[ p_0^{(j)} + p_1^{(j)} + p_{12}^{(j)} - p_2^{(j)} \right].
\end{align}

\section{Dephasing Bose-Hubbard model}
\label{appendix:Bose_Hubbard_model}

\begin{figure*}
	\centering
	\includegraphics[width=\textwidth]{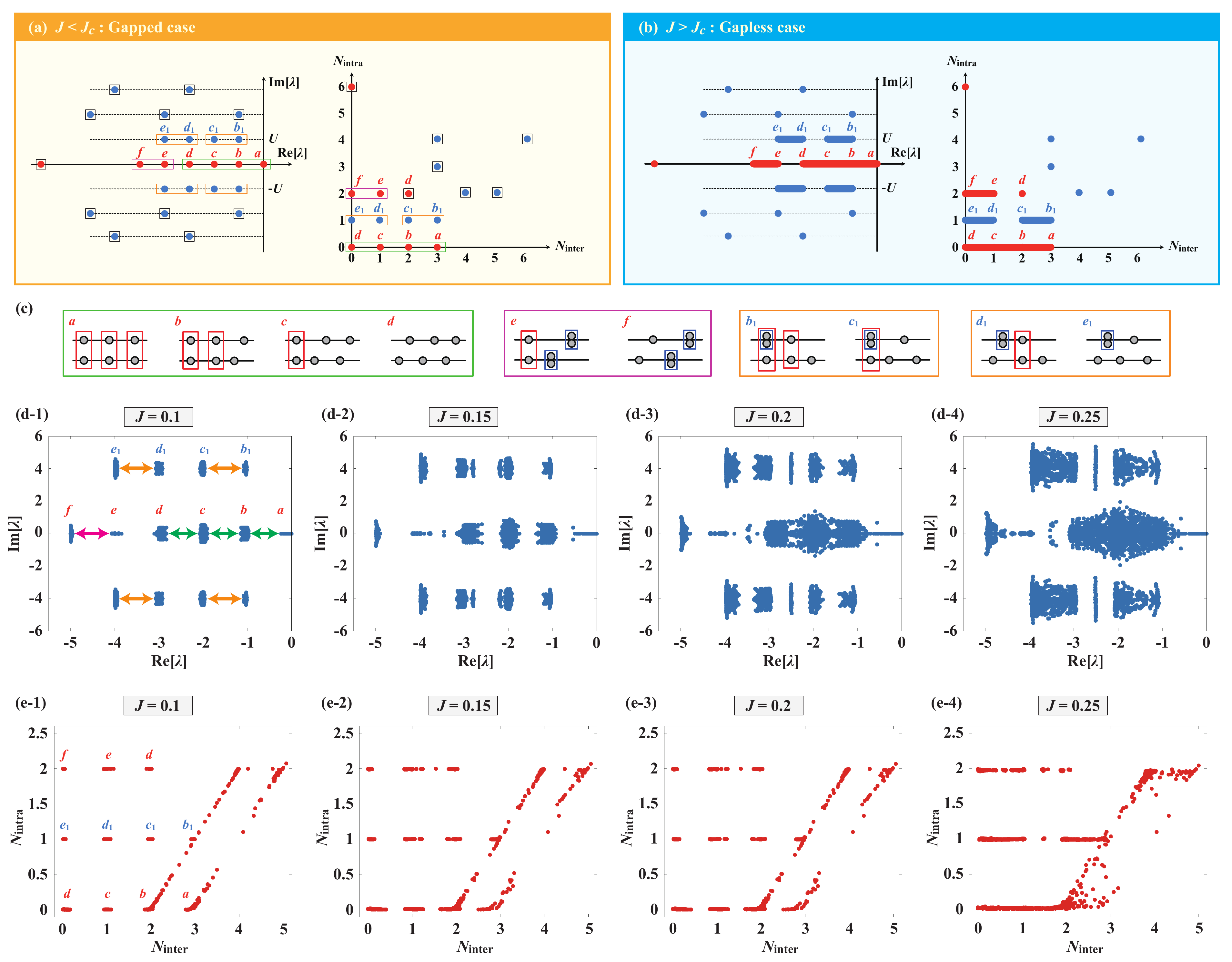}
	\caption{Three-particle Liouvillian spectra and eigenmodes of the Bose-Hubbard model under dephasing.
		Schematic illustrations of the spectra and scatter plots of $(N_\mathrm{inter}, N_\mathrm{intra})$ are shown for the gapped case (a) and the gapless case (b). 
		Panel (c) illustrates the eigenmodes indicated in panels (a) and (b), where the red (blue) squares represent an interchain (intrachain) bound state.
		In panels (a) and (b), series of spectral bands connected by breaking or creating a single first-order incoherenton are highlighted by the green, purple, orange, and black squares.
		The QC gaps are defined as the gaps between spectral bands belonging to the same series.
		In the gapless case, the QC gaps close and all bands belonging to each series merge into a single band.
		Panels (d) and (e) show the spectra and scatter plots of $(N_\mathrm{inter}, N_\mathrm{intra})$ with $J=0.1$. $0.15$, $0.2$, and $0.25$.
		The system size is $L=8$ and the particle number is $N=3$.
		The other parameters are set to $U=4$ and $\gamma=1$.
		The double arrows in panel (d-1) represents the QC gaps corresponding to each series of bands.}
	\label{Fig-BH-spec-schematic}
\end{figure*}

\begin{figure*}
	\centering
	\includegraphics[width=\textwidth]{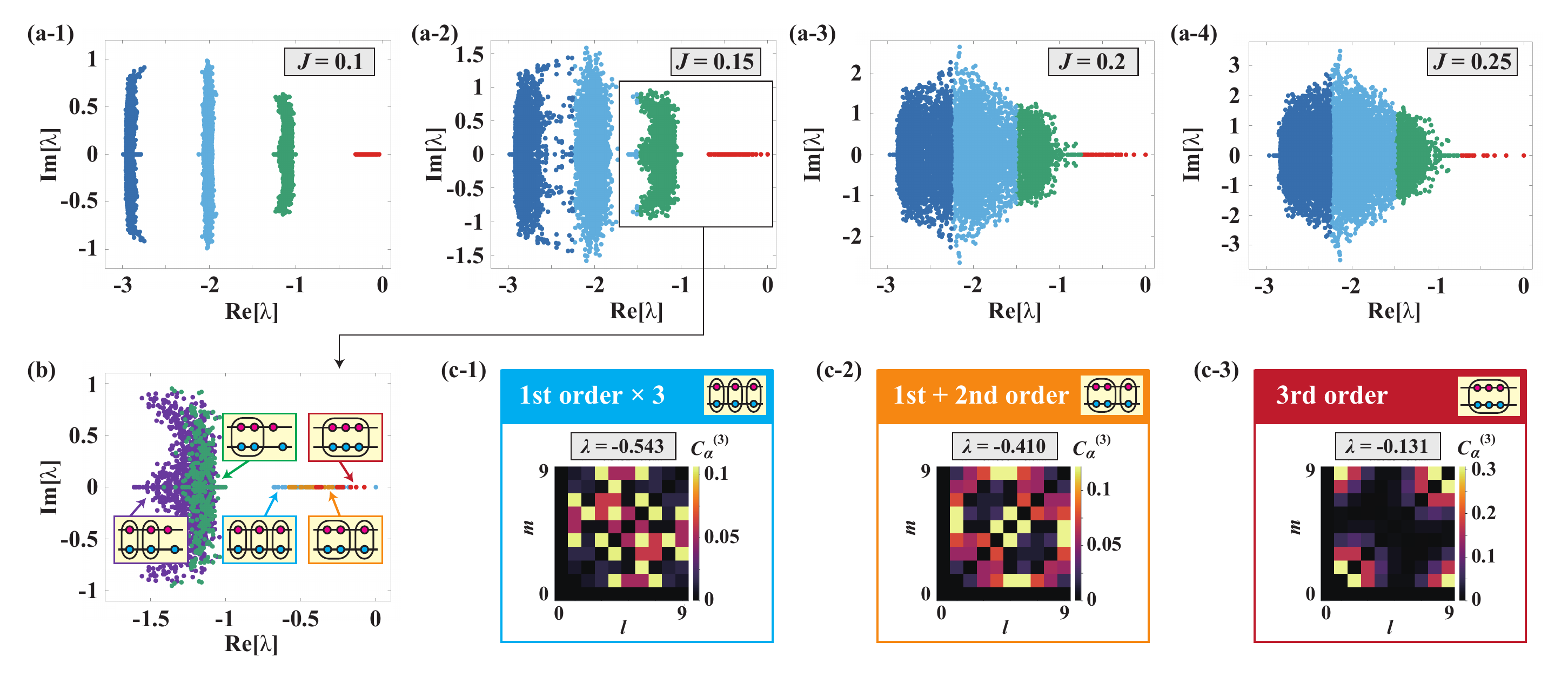}
	\caption{Liouvillian spectra for the dephasing hard-core bosons with next-nearest-neighbor hopping.
		(a) Spectra with $\gamma=1$ and $J=0.1$, $0.15$, $0.2$, and $0.25$, where $J:=J_1=J_2$.
		The systems size is $L=10$ and the particle number is $N=3$.
		The eigenvalues satisfying $0 \leq N_{b,\alpha} < 3/4$, $3/4 \leq N_{b,\alpha} < 3/2$, $3/2 \leq N_{b,\alpha} < 9/4$, and $9/4 \leq N_{b,\alpha} \leq 3$ are colored by blue, light blue, green, and red, respectively.
		(b) Part of the spectrum corresponding to the square in panel (a-2).
		The color differentiates between the types of eigenmodes, which are illustrated in the insets.
		(c) Color maps of $C_\alpha^{(3)}(l, m)$ for three representative eigenmodes with eigenvalues $\lambda=-0.543$, $-0.410$, and $-0.131$.
		To emphasize off-diagonal components, the values of $C_\alpha^{(3)}(l, m)$ at $l=0$, $m=0$, and $l=m$ are set to zero.
		For eigenmodes with three first-order incoherentons, $C_\alpha^{(3)}(l, m)$ is delocalized across the full range of $l$ and $m$.
		For eigenmodes with one first-order incoherenton and one second-order incoherenton, $C_\alpha^{(3)}(l, m)$ is localized at the edge of the $(l,m)$ space.
		For eigenmodes with one third-order incoherenton, $C_\alpha^{(3)}(l, m)$ is localized at the corner of the $(l,m)$ space.
		In panel (b), the eigenmodes in the band with the smallest decay rate are classified based on whether the point $(l^*, m^*)$ that maximizes $C_\alpha^{(3)}(l, m)$ is located in the bulk, edge, or corners of the $(l,m)$ space.}
	\label{Fig-spec-nnn-hop}
\end{figure*}

In this Appendix, we investigate the Liouvillian spectra of the Bose-Hubbard model under dephasing to substantiate the robustness of the incoherenton picture illustrated in Fig.~\ref{Fig-hierarchy}.
The Hamiltonian of the Bose-Hubbard model is given by
\begin{equation}
	H = -J \sum_{l=1}^{L} (b_{l+1}^{\dag} b_{l} + b_{l}^{\dag} b_{l+1}) + \frac{U}{2} \sum_{l=1}^L n_l (n_l-1),
	\label{Hamiltonian_BH}
\end{equation}
where $b_{l}^{\dag}$ and $b_{l}$ are the creation and annihilation operators of a boson at site $l$ and $n_l=b^{\dag}_{l}b_{l}$ is the density operator at site $l$.
Here, $J$ and $U$ represent the amplitude of coherent hopping and the strength of on-site interactions.
The Lindblad operators are given by Eq.~\eqref{L_dephasing}.
The hard-core boson model introduced in Sec.~\ref{sec:hard_core_bosons} is a special case of the Bose-Hubbard model with $U \to \infty$.
We proceed by numerically calculating the three-particle Liouvillian spectra.

A key distinction of the Bose-Hubbard model from hard-core bosons is the existence of {\it intrachain} bound states in the ladder representation, which complicates the band structure of the Liouvillian spectrum.
We will discuss that, even with the existence of intrachain bound states, the concepts of incoherenton deconfinement and QC gap closing can still be applied. 
For an eigenmode $|\rho_\alpha)$ in the ladder representation, we define the numbers of interchain and intrachain bound states as
\begin{equation}
	N_{\mathrm{inter},\alpha} := \sum_{l=1}^L \frac{(\rho_{\alpha}| n_{l,+}  n_{l,-} |\rho_{\alpha})}{(\rho_{\alpha}|\rho_{\alpha})},
\end{equation}
\begin{equation}
	N_{\mathrm{intra},\alpha} := \frac{1}{2} \sum_{\sigma=\pm} \sum_{l=1}^L \frac{(\rho_{\alpha}| n_{l,\sigma}  (n_{l,\sigma}-1) |\rho_{\alpha})}{(\rho_{\alpha}|\rho_{\alpha})},
\end{equation}
which are extensions of Eq.~\eqref{def_N_b}.

Figure \ref{Fig-BH-spec-schematic}(a) illustrates the three-particle Liouvillian spectrum and scatter plot of $\{(N_{\mathrm{inter},\alpha}, N_{\mathrm{intra},\alpha})\}_{\alpha=0,...,D^2-1}$ for a small value of $J$.
We focus on series of spectral bands connected by breaking or creating a single first-order incoherenton, which are indicated by squares in panel (a).
The structure of eigenmodes is illustrated in panel (c), where interchain and intrachain bound states are represented by red and blue squares, respectively.
As depicted, spectral bands $a$, $b$, $c$, and $d$ form a series, represented by a green square, as the eigenmodes of $b$, $c$, and $d$ can be obtained by sequentially breaking first-order incoherentons in eigenmode $a$.
Similarly, several series containing intrachain bound states can be identified, represented by purple and orange squares.
Isolated bands unconnected to any other bands are denoted by black squares.
The QC gaps are defined as the gaps between spectral bands belonging to the same series [see the double arrows in panel (d-1)].

Figure \ref{Fig-BH-spec-schematic}(b) displays the spectrum and scatter plot of $\{(N_{\mathrm{inter},\alpha}, N_{\mathrm{intra},\alpha})\}_{\alpha=0,...,D^2-1}$ for a large value of $J$.
The QC gaps between spectral bands within each series close, and larger bands emerge.
Panels (d) and (e) present the spectra and scatter plots of $\{(N_{\mathrm{inter},\alpha}, N_{\mathrm{intra},\alpha})\}_{\alpha=0,...,D^2-1}$ obtained by numerical diagonalization of the Liouvillian.
These results corroborate the scenario illustrated in panels (a) and (b).
Note that the hopping amplitude $J_c$ corresponding to the QC gap closing can vary between different spectral band series.

The scenario presented in Fig.~\ref{Fig-BH-spec-schematic} for a three-particle case can be readily extended to multiple-particle situations. 
Despite the exponential growth in spectrum complexity with particle number, series of spectral bands linked by breaking or creating a first-order incoherenton can still be identified. 
The hierarchical picture illustrated in Fig.~\ref{Fig-hierarchy} is valid for each series of spectral bands.

The description of dynamics in terms of eigenmodes with incoherentons can also be extended to that in terms of eigenmodes with intrachain bound states.
In the case of hard-core bosons, the decay rates of eigenmodes are uniquely determined by incoherentons when the spectral bands are separated by the QC gaps. 
In the case of the Bose-Hubbard model, the temporal dynamics of eigenmodes, e.g., their decay rates and frequencies, are governed by both incoherentons and intrachain bound states.
In particular, the frequencies of eigenmodes, i.e., the imaginary parts of eigenvalues, are mainly governed by the number of intrachain bound states.
This suggests a two-dimensional hierarchy encompassing the real axis (decay rates) defined by incoherentons and the imaginary axis (frequencies) specified by intrachain bound states.
While the correspondence between decay rates (frequencies) and the number of incoherentons (intrachain bound states) is not perfectly one-to-one, such a hierarchical picture provides a qualitative understanding of the spectral structure.

\section{Dephasing hard-core bosons with next-nearest-neighbor hopping}
\label{appendix:hard_core_bosons_with_next_nearest_neighbor_hopping}

In this Appendix, we present the results for dephasing hard-core bosons with next-nearest-neighbor hopping to test the robustness of the incoherenton picture to integrability-breaking perturbations.
The Hamiltonian of this system is given by
\begin{equation}
	H = -J_1 \sum_{l=1}^{L} (b_{l}^{\dag} b_{l+1} + b_{l+1}^{\dag} b_{l}) -J_2 \sum_{l=1}^{L} (b_{l}^{\dag} b_{l+2} + b_{l+2}^{\dag} b_{l}),
	\label{H_hard_core_boson_nnn_hop}
\end{equation}
where $J_1$ and $J_2$ denote the hopping amplitudes between nearest-neighbor and next-nearest-neighbor sites, respectively.
The next-nearest-neighbor hopping introduces the simplest perturbation that breaks the integrability of the original model without altering the number of spectral bands. 
In the following calculation, we assume $J:=J_1=J_2$.

Figure \ref{Fig-spec-nnn-hop}(a) presents the Liouvillian spectra for $J=0.1$, $0.15$, $0.2$, and $0.25$.
The colors of the dots represent $N_{b, \alpha}$, defined by Eq.~\eqref{def_N_b}.
With a small, nonzero $J$, four spectral bands emerge around $0$, $-1$, $-2$, and $-3$.
As $J$ increases, the widths of bands increase, and they merge almost simultaneously at $J=J_c \simeq 0.2$.
This behavior is the same as the case without next-nearest-neighbor hopping [see Fig.~\ref{Fig-spec-MP}(a)].

Each eigenmode can be classified according to the number $N_{b, \alpha}$ of incoherentons [Eq.~\eqref{def_N_b}] and the qualitative behavior of incoherenton correlations $C_\alpha^{(s)}(m_1,...,m_{s-1})$ $(s=2,3,...)$ defined by Eqs.~\eqref{def_inc_corr_2} and \eqref{def_inc_corr_3}.
Figure \ref{Fig-spec-nnn-hop}(b) depicts a spectrum where color variations differentiate eigenmode types.
Additionally, color maps of $C_\alpha^{(3)}(l, m)$ for three representative eigenmodes in the band with the smallest decay rate are shown in Fig.~\ref{Fig-spec-nnn-hop}(c).
The three types of eigenmodes can be distinguished by whether $C_\alpha^{(3)}(l, m)$ is delocalized, localized on the edge, or localized at the corners. 
All features presented in Fig.~\ref{Fig-spec-nnn-hop} are consistent with those in Fig.\ref{Fig-mode-class}, presenting further evidence that our incoherenton picture is applicable to nonintegrable systems.



\begin{thebibliography}{99}
\bibitem{note-1} When $J \ll \gamma$, coherent hopping in the Liouvillian can be considered as a perturbation.
	Recall that $\chi_s$ is a measure of the contribution from eigenmodes with $s$ deconfined particle pairs.
	To obtain such eigenmodes, the perturbation term must act $s$ times on the unperturbed state where all particles form completely localized incoherentons, and thus the factor $(J/\gamma)^s$ appears.
	Another factor $L^s$ comes from the number of configurations of $s$ incoherentons on which the perturbation term acts.
\end{thebibliography}
\end{document}